\documentclass[11pt,a4paper]{article}

\pdfoutput=1
\usepackage{jheppub}

\usepackage[english]{babel}
\usepackage{amsfonts,amsmath,amssymb,amsbsy,amsthm,bm,mathtools,fixmath}
\usepackage{comment,enumerate,footnote,graphicx,subfloat,relsize,footnote}
\usepackage{array,tabularx,tabu,multirow,framed}
\usepackage[usenames,dvipsnames]{xcolor}
\numberwithin{equation}{section}
\usepackage{cleveref}
\crefformat{pluralequation}{#2\black{eqs.~(}#1\black{)}#3}
\Crefformat{pluralequation}{#2\black{Equations~(}#1\black{)}#3}
\crefformat{pluralfigure}{#2\black{figs.~}#1#3}
\Crefformat{pluralfigure}{#2\black{Figures~}#1#3}

\usepackage{tikz}
\usetikzlibrary{arrows,shapes}
\usetikzlibrary{trees,patterns}
\usetikzlibrary{matrix,arrows} 				
\usetikzlibrary{positioning}				  
\usetikzlibrary{calc,through}				  
\usetikzlibrary{decorations.pathreplacing}  
\usepackage{pgffor}							

\usetikzlibrary{decorations.pathmorphing}	
\usetikzlibrary{decorations.markings}
\tikzset{
	>=stealth', 
    vector/.style={decorate, decoration={snake}, draw},
	provector/.style={decorate, decoration={snake,amplitude=2.5pt}, draw},
	antivector/.style={decorate, decoration={snake,amplitude=-2.5pt}, draw},
	bigvector/.style={decorate, decoration={snake,amplitude=4pt}, draw},
    fermion/.style={draw=black, postaction={decorate},
        decoration={markings,mark=at position .55 with {\arrow[draw=black]{>}}}},
    fermionbar/.style={draw=black, postaction={decorate},
        decoration={markings,mark=at position .55 with {\arrow[draw=black]{<}}}},
    fermionnoarrow/.style={draw=black},
    gluon/.style={decorate, draw=black,
        decoration={coil,amplitude=4pt, segment length=5pt}},
    scalar/.style={dashed,draw=black, postaction={decorate},
        decoration={markings,mark=at position .55 with {\arrow[draw=black]{>}}}},
    scalarbar/.style={dashed,draw=black, postaction={decorate},
        decoration={markings,mark=at position .55 with {\arrow[draw=black]{<}}}},
    scalarnoarrow/.style={dashed,draw=black},
    momentum/.style={draw=black, postaction={decorate},
        decoration={markings,mark=at position 1 with {\arrow[draw=black]{>}}}},
    antimomentum/.style={draw=black, postaction={decorate},
        decoration={markings,mark=at position 0.1 with {\arrow[draw=black]{<}}}}
}

\tikzstyle{block} = [draw, rectangle, 
    minimum height=3em, minimum width=6em]
    
\newcommand{\nc}{\newcommand}
\nc{\pd}{\partial}
\nc{\bea}{\begin{eqnarray}}
\nc{\eea}{\end{eqnarray}}
\nc{\bal}{\begin{alignedat}}
\nc{\eal}{\end{alignedat}}
\nc{\beq}{\begin{equation}}
\nc{\eeq}{\end{equation}}
\nc{\bit}{\begin{itemize}}
\nc{\eit}{\end{itemize}}
\nc{\benu}{\begin{enumerate}}
\nc{\eenu}{\end{enumerate}}
\nc{\bdes}{\begin{description}}
\nc{\edes}{\end{description}}
\nc{\bma}{\begin{pmatrix}}
\nc{\ema}{\end{pmatrix}}

\newcommand{\black}[1]	{{\color{black} 	#1}}

\nc{\nn}{\nonumber}
\nc{\hc}{\text{h.c.}}
\nc{\cc}{\text{c.c.}}

\nc{\slashed}[1]{{#1}\hspace{-2mm}/}

\nc{\abs}[1]{\left| #1 \right|}
\def\[{\left[}
\def\]{\right]}
\def\({\left(}
\def\){\right)}
\def\<{\langle}
\def\>{\rangle}

\def\g5{\gamma_{5}}

\def\GeV{{\rm GeV}}
\def\TeV{{\rm TeV}}

\def\a{\alpha}

\def\g{\gamma}
\def\d{\delta}

\def\k{\kappa}
\def\l{\lambda}
\def\m{\mu}

\def\p{\pi}

\def\s{\sigma}

\def\f{\phi}
\def\x{\chi}

\def\D{\Delta}

\def\W{\Omega}

\def\ag				{\alpha_g}
\def\agR			{\alpha_{g,\mathsmaller{\bf [R]}}}
\def\agSinglet		{\alpha_{g,\mathsmaller{\bf [1]}}}
\def\agOctet		{\alpha_{g,\mathsmaller{\bf [8]}}}
\def\as				{\alpha_s}
\def\asSinglet		{\alpha_{s,\mathsmaller{\bf [1]}}}
\def\asOctet		{\alpha_{s,\mathsmaller{\bf [8]}}}

\def\asAnn			{\alpha_s^{\rm ann}}

\def\agScatt		{\alpha_g^{\mathsmaller{S}}}
\def\agScattR		{\alpha_{g,\mathsmaller{\bf [R]}}^{\mathsmaller{S}}}
\def\agScattSinglet	{\alpha_{g,\mathsmaller{\bf [1]}}^{\mathsmaller{S}}}
\def\agScattOctet	{\alpha_{g,\mathsmaller{\bf [8]}}^{\mathsmaller{S}}}
\def\asScatt		{\alpha_s^{\mathsmaller{S}}}
\def\asScattSinglet	{\alpha_{s,\mathsmaller{\bf [1]}}^{\mathsmaller{S}}}
\def\asScattOctet	{\alpha_{s,\mathsmaller{\bf [8]}}^{\mathsmaller{S}}}

\def\agBound		{\alpha_g^{\mathsmaller{B}}}
\def\agBoundR		{\alpha_{g,\mathsmaller{\bf [R]}}^{\mathsmaller{B}}}
\def\agBoundSinglet	{\alpha_{g,\mathsmaller{\bf [1]}}^{\mathsmaller{B}}}
\def\agBoundOctet	{\alpha_{g,\mathsmaller{\bf [8]}}^{\mathsmaller{B}}}
\def\asBound		{\alpha_s^{\mathsmaller{B}}}
\def\asBoundSinglet	{\alpha_{s,\mathsmaller{\bf [1]}}^{\mathsmaller{B}}}
\def\asBoundOctet	{\alpha_{s,\mathsmaller{\bf [8]}}^{\mathsmaller{B}}}

\def\asBSF			{\alpha_{s}^{\mathsmaller{\rm BSF}}}
\def\asBSFSinglet	{\alpha_{s,\mathsmaller{\bf [1]}}^{\mathsmaller{\rm BSF}}}
\def\asBSFOctet		{\alpha_{s,\mathsmaller{\bf [8]}}^{\mathsmaller{\rm BSF}}}

\def\GammaDecSinglet{\Gamma_{\dec,\mathsmaller{\bf [1]}}}
\def\GammaDecOctet	{\Gamma_{\dec,\mathsmaller{\bf [8]}}}
\def\GammaDecR		{\Gamma_{\dec,\mathsmaller{\bf [R]}}}
\def\GammaIonSinglet{\Gamma_{\ion,\mathsmaller{\bf [1]}}}
\def\GammaIonOctet	{\Gamma_{\ion,\mathsmaller{\bf [8]}}}
\def\GammaIonR		{\Gamma_{\ion,\mathsmaller{\bf [R]}}}
\def\SigmaIonR		{\sigma_{\ion,\mathsmaller{\bf [R]}}}

\def\gR			{g_{\mathsmaller{\bf [R]}}}
\def\omegaR		{\omega_{\mathsmaller{\bf [R]}}}

\def\gh			{g_h}
\def\gs			{g_s}

\def\ah			{\alpha_h}

\def\ann		{{\rm ann}}
\def\BSF		{\mathsmaller{\rm BSF}}
\def\dec		{{\rm dec}}
\def\dh			{d_h}
\def\eff		{{\rm eff}}
\def\ion		{{\rm ion}}
\def\eq			{{\rm eq}}
\def\FO			{{\mathsmaller{\rm FO}}}

\def\gstareffsqrt	{g_{*,\rm eff}^{1/2}}
\def\gstarS		{g_{*\mathsmaller{S}}}
\def\gx			{g_\chi}
\def\gX			{g_{\mathsmaller{X}}}
\def\mh			{m_h}
\def\mpl		{m_{\mathsmaller{\rm Pl}}}
\def\mx			{m_\chi}
\def\mX			{m_{\mathsmaller{X}}}

\def\nX			{n_{\mathsmaller{X}}}

\def\sigmaeff	{\sigma_{\rm eff}}
\def\SzeroC		{S_0^{C}}
\def\Xdagger	{X^\dagger}
\def\Yx			{Y_{\chi}}
\def\YX			{Y_{X}}
\def\YXdagger	{Y_{\Xdagger}}
\def\vrel		{v_{\rm rel}}
\def\xB			{x_{\mathsmaller{B}}}
\def\xS			{x_{\mathsmaller{S}}}

\def\zetah				{\zeta_h}
\def\zetag				{\zeta_g}
\def\zetagR				{\zeta_{g, \mathsmaller{\bf [R]}}}
\def\zetagScatt			{\zeta_g^{\mathsmaller{S}}}
\def\zetagScattR		{\zeta_{g,\mathsmaller{\bf [R]}}^{\mathsmaller{S}}}

\def\zetagScattOctet	{\zeta_{g,\mathsmaller{\bf [8]}}^{\mathsmaller{S}}}

\def\zetagBoundR		{\zeta_{g,\mathsmaller{\bf [R]}}^{\mathsmaller{B}}}

\def\zetaBound			{\zeta_{\mathsmaller{B}}}
\def\zetaBoundR			{\zeta_{\mathsmaller{B,\bf [R]}}}
\def\zetaBoundSinglet	{\zeta_{\mathsmaller{B,\bf [1]}}}
\def\zetaBoundOctet		{\zeta_{\mathsmaller{B,\bf [8]}}}
\def\zetaScatt			{\zeta_{\mathsmaller{S}}}
\def\zetaScattR			{\zeta_{\mathsmaller{S, \bf [R]}}}

\def\zetaScattOctet		{\zeta_{\mathsmaller{S, \bf [8]}}}

\def\kappaR				{\kappa_{\mathsmaller{\bf [R]}}}
\def\kappaSinglet		{\kappa_{\mathsmaller{\bf [1]}}}

\def\lambdaR			{\lambda_{\mathsmaller{\bf [R]}}}

\renewcommand{\vec}{\textbf}

\begin{document}
\preprint{\parbox[t]{20ex}{TUM-HEP-1186-19 \\ Nikhef-2019-004}}

\title{Higgs-mediated bound states in dark-matter models}

\author[1,2,3]{Julia Harz}
\author[2,4]{and Kalliopi Petraki}

\affiliation[1]{Physik  Department  T70,  James-Franck-Stra\ss e,  Technische  Universit\"at  M\"unchen,  85748  Garching, Germany}
\affiliation[2]{Sorbonne Universit\'e, CNRS, Laboratoire de Physique Th\'eorique et Hautes Energies, LPTHE, F-75252 Paris, France}
\affiliation[3]{Sorbonne Universit\'e, Institut Lagrange de Paris (ILP), 98 bis Boulevard Arago, 75014 Paris, France}
\affiliation[4]{Nikhef, Science Park 105, 1098 XG Amsterdam, The Netherlands}

\emailAdd{julia.harz@tum.de}
\emailAdd{kpetraki@lpthe.jussieu.fr}


\abstract{It has been recently demonstrated that the 125~GeV Higgs boson can mediate a long-range force between TeV-scale particles, that can impact considerably their annihilation due to the Sommerfeld effect, and hence the density of thermal relic dark matter. In the presence of long-range interactions, the formation and decay of particle-antiparticle bound states can also deplete dark matter significantly.  We consider the Higgs boson as mediator in the formation of bound states, and compute the effect on the dark matter abundance. To this end, we consider a simplified model in which dark matter co-annihilates with coloured particles that have a sizeable coupling to the Higgs. The Higgs-mediated force affects the dark matter depletion via bound state formation in several ways. It enhances the capture cross-sections due to the attraction it mediates between the incoming particles, it increases the binding energy of the bound states, hence rendering their ionisation inefficient sooner in the early universe, and for large enough couplings, it can overcome the gluon repulsion of certain colour representations and give rise to additional bound states. Because it alters the momentum exchange in the bound states, the Higgs-mediated force also affects the gluon-mediated potential via the running of the strong coupling. We comment on the experimental implications and conclude that the Higgs-mediated potential must be taken into account when circumscribing the viable parameter space of related models.}

\arxivnumber{1901.10030} 

\maketitle

\clearpage
\section{Introduction \label{Sec:Intro}}

The nature of dark matter (DM) is one of the biggest open questions in modern particle physics. An intriguing possibility is that the Higgs boson participates in the physics of the dark sector. The  discovery of the Higgs and the measurement of its mass impel the thorough investigation of its implications for various DM scenarios. 

One of the most generic scenarios  for the cosmological production of DM asserts that DM arose from the thermal freeze-out of non-relativistic particles that were previously in equilibrium with the Standard Model (SM) plasma in the early universe. This possibility necessitates a sizeable coupling between the SM and DM, and is thus being probed by a variety of direct, indirect and collider experiments. The increasingly more stringent constraints have pushed the viable DM mass range in this class of models toward and beyond the TeV regime, and have given impetus to co-annihilation scenarios~\cite{Griest:1990kh,Ibarra:2015nca,Freitas:2015hsa,Baker:2015qna,ElHedri:2018atj}, which provide more flexibility in reproducing the observed DM density via freeze-out.

If the 125 GeV Higgs couples to a (multi-)TeV dark sector, then the difference between the two mass scales may give rise to long-range effects. Indeed, in models where DM or its co-annihilating partners couple to a much lighter force mediator, the resulting long-range force  distorts the wavepackets of the interacting pairs, giving rise to the well-known Sommerfeld effect~\cite{Sommerfeld:1931,Sakharov:1948yq} that alters their annihilation rate~\cite{Hisano:2002fk,Hisano:2003ec}. Long-range interactions imply also the existence of bound states; the formation and decay of particle-antiparticle bound states opens an additional two-step annihilation channel that can deplete the DM abundance~\cite{vonHarling:2014kha,Petraki:2015hla} and contribute to the DM indirect detection signals~\cite{Pospelov:2008jd,An:2016gad,An:2016kie,Petraki:2016cnz,Cirelli:2016rnw}. While such effects emanating from hidden-sector mediators, the electroweak gauge bosons, or the gluons in co-annihilation scenarios, have been considered~\cite{
Hisano:2002fk,
Hisano:2003ec,
Pospelov:2008jd,
Beneke:2014hja,
Beneke:2014gja,
Hryczuk:2010zi,
Hryczuk:2011tq,
Harz:2014gaa,
vonHarling:2014kha,
Petraki:2015hla,
Cirelli:2015bda,
An:2016kie,
Petraki:2016cnz,
ElHedri:2016onc,
Liew:2016hqo,
Asadi:2016ybp,
An:2016gad,
Kouvaris:2016ltf,
Cirelli:2016rnw,
Kim:2016kxt,
Kim:2016zyy,
Biondini:2017ufr,
Biondini:2018pwp,
Biondini:2018xor,
Biondini:2018ovz,
Baldes:2017gzw,
Baldes:2017gzu,
Pearce:2015zca,
Ellis:2018jyl,
Geller:2018biy,
Harz:2018csl,
Cirelli:2018iax,
Bhattacharya:2018ooj,
Schmiemann:2019czm}, 
only recently the long-range effect of the Higgs was pointed out~\cite{Harz:2017dlj}.

The reasoning for disregarding the latter appears to be twofold: the Higgs boson was considered too heavy to yield a long-range force, and the coupling of the Higgs to DM was assumed to be smaller than (or at best comparable to) the SM gauge couplings. However, in co-annihilation scenarios, the mass of DM or its co-annihilating partners  can be in the multi-TeV regime and their coupling to the Higgs can be sizeable, such that the range of the Higgs-mediated interaction becomes comparable to or exceeds the Bohr radius of the interacting pair of particles. Then, the interaction manifests as long-range. Reference~\cite{Harz:2017dlj} demonstrated the significant impact of the Higgs-generated potential on the annihilation cross-sections due to the Sommerfeld effect, and  consequently on the DM density. In the present paper, we investigate the impact of the Higgs-mediated force on the existence and formation of bound states, and the associated effect on the DM abundance.

For definiteness, we shall consider the simplified model of ref.~\cite{Harz:2017dlj}, where  DM is assumed to co-annihilate with a coloured scalar that transforms under the fundamental of $SU(3)_c$ and possesses a sizeable coupling to the Higgs. This scenario is encountered in the Minimal Supersymmetric SM (MSSM) --- with the coloured particle being typically the lightest stop eigenstate --- and constitutes one of the last refuges of neutralino DM; as such, it has received a lot of attention recently~\cite{Keung:2017kot,Ibarra:2015nca,Pierce:2017suq}, with the effect of radiative bound-state formation (BSF) due to gluon exchange computed in~\cite{Harz:2018csl}. We shall focus on the DM mass range (0.5 -- 5)~TeV. The lower bound of this interval largely ensures viability against current experimental constraints, while the upper bound implies that DM freezes-out after the electroweak symmetry breaking (EWSB), when the neutral component of the Higgs doublet becomes the mediator that couples to the DM co-annihilating partners. Importantly, this mass interval encompasses the range probed at the LHC and the range of highest interest for the resolution of the hierarchy problem. In fact, retaining the supersymmetry-breaking scale relatively low such that the hierarchy problem is better addressed, and reproducing the measured Higgs mass at the same time, is possible if the coupling of the stops to the Higgs is sizeable. Altogether, this scenario contains the necessary ingredients for the Higgs-mediated force to have an important effect. Similar features are present also in various Higgs-portal models (see e.g.~\cite{Freitas:2015hsa,Lopez-Honorez:2017ora}), for which we expect our findings to have important implications.

In particular, the impact of the Higgs-mediated potential on the DM abundance affects the prediction of the DM  couplings and mass. This in turn changes the interpretation of the experimental results and affects the viability of specific scenarios. The viability of models with long-range interactions with respect to indirect detection constraints is rather sensitive to the predicted DM mass, since the expected $\gamma$-ray spectrum from annihilations inside galaxies exhibits sharp parametric resonances with varying DM mass. Moreover, thermal-relic DM models are probed at precision collider experiments. The measurement of the DM abundance to an unprecedented precision, $\Omega h^2 = 0.120 \pm 0.001$, by the \emph{Planck} satellite~\cite{Aghanim:2018eyx} yields a powerful constraint that can meaningfully complement those from colliders and indirect detection. This however requires that the theoretical computation of the relic density is also sufficiently accurate. During the last couple of years, important progress has been made along this direction. While state-of-the-art numerical tools~\cite{Belanger:2013oya,Bringmann:2018lay,Ambrogi:2018jqj} have employed so far mainly leading order calculations, it has been demonstrated that higher order corrections to the annihilation processes can impact the DM density up to 20\%~\cite{Harz:2012fz,Harz:2014tma,Harz:2016dql,Baro:2007em,Schmiemann:2019czm}. The Sommerfeld effect and BSF due to the SM gauge bosons can have an even greater impact~\cite{Harz:2018csl}. As already shown in~\cite{Harz:2017dlj} and we shall see again in the following, the Higgs-mediated force can have a comparable or larger effect that far exceeds the experimental precision in the measurement of the DM density.

\smallskip

The paper is organized as follows. In \cref{Sec:Model}, we describe the simplified model, first introduced in \cite{Harz:2017dlj}, and review the calculation of the DM relic density, including the formation and decay of bound states. We then discuss the scattering and bound states in a mixed Coulomb and Yukawa potential in \cref{Sec:Wavefunctions}. In \cref{Sec:BSF}, we compute the BSF cross-sections. Since several different momentum scales enter the calculation, in \cref{sec:BSF_AlphaRunning} we discuss the running of the strong coupling and explore its effect on the cross-sections. In \cref{Sec:Results}, we present our results. We discuss the impact of the Higgs boson on the existence and formation of bound states, and on the effective annihilation cross-section, before demonstrating its effect on the predicted DM density. We conclude in \cref{Sec:Conclusions}. For easy reference, we summarize the notation used throughout the paper in \cref{tab:notation}.

\clearpage
\begin{table}[p]
\centering	
\centering
\renewcommand{\arraystretch}{1.1}

\begin{tabular}{|l|l|} 
\hline
{\bf Description} &   {\bf Symbol}
\\ \hline  \hline 
DM particle & $\x$
\\ \hline
DM mass 
& $\mx$
\\ \hline
Coloured coannihilating partners of DM
& \parbox{0.4\textwidth}{\smallskip
$X$ and $X^\dagger$ \\ 
{}[scalars, ${\bf 3}$ and $\bar{\bf 3}$ under $SU(3)_c$]}
\\[1.5ex] \hline
Mass of coannihilating partners 
& $\mX$
\\ \hline
Total mass of pair of coannihilating partners 
& $M \equiv 2\mX$
\\ \hline
Reduced mass of pair of coannihilating partners 
& $\mu \equiv \mX/2$
\\ \hline
\parbox{0.5\textwidth}{\smallskip
Relative mass difference between \\ 
DM and coannihilating partners} 
& $\Delta \equiv (\mX - \mx)/\mx$
\\[1.5ex] \hline 
Mass of Higgs-like boson $h$
& $\mh = 125~\GeV$
\\ \hline \hline
$X, X^\dagger$ coupling to the Higgs  
& $\ah=g_h^2/(16 \pi)$ 
\\ \hline 
Strong coupling~~[$Q$: momentum transfer]
& $\as(Q) \equiv \gs^2(Q)/(4\pi)$
\\
in scattering and bound states (singlet, octet)
& $\asScatt$~~~and~~~$\asBoundSinglet,~~\asBoundOctet$
\\ 
in annihilation and BSF vertices (singlet, octet)
& $\asAnn$~~~and~~~$\asBSFSinglet,~~\asBSFOctet$
\\ \hline
Gluon-generated Coulomb potential coupling &
\\
in the scattering states (singlet, octet)
& $\agScattSinglet \equiv (4/3)\asScatt, \qquad 
   \agScattOctet   \equiv -\asScatt/6$
\\
in the bound states (singlet, octet)
& $\agBoundSinglet \equiv (4/3)\asBoundSinglet, \quad 
   \agBoundOctet   \equiv -\asBoundOctet/6$
\\ \hline
Bohr momentum of the bound states ({\bf R =1,8})
& $\kappaR  \equiv \mu(\agBoundR + \ah)$
 \\ \hline
\parbox{0.5\textwidth}{\smallskip
Expectation value of relative velocity of \\
interacting particles in the scattering state} 
& $\vec{v}_{\rm rel}, \quad \vrel = |\vec{v}_{\rm rel}|$
\\[1.5ex] \hline
\parbox{0.5\textwidth}{\smallskip
Expectation value of momentum \\ 
in the scattering state in the CM frame}
& $\vec{k}  \equiv \mu \vec{v}_{\rm rel}, \quad k \equiv |\vec{k}|$
\\[1.5ex] \hline \hline 
\multirow{7}{*}{
\parbox{0.5\textwidth}{
Dimensionless variables that \\
parametrise the long-range effects and \\
the parts of the annihilation and BSF \\ 
cross-sections that are evaluated numerically \\
(colour representation {\bf R =1,8})
}}
& $\dh \equiv \mu \ah / \mh$
\\ \cline{2-2}
& $\lambdaR \equiv \agBoundR /\ah$
\\ \cline{2-2}
& $\zetah  \equiv \ah / \vrel$
\\ \cline{2-2}
& $\zetagScattR  \equiv \agScattR / \vrel$
\\ 
& $\zetagBoundR  \equiv \agBoundR / \vrel$
\\ \cline{2-2}
& $\zetaScattR  \equiv (\ah + \agScattR) / \vrel$
\\ 
& $\zetaBoundR  \equiv (\ah + \agBoundR) / \vrel$
\\ \hline
Binding energy of $n\ell m$ bound state
&${\cal E}_{n\ell} = -  \gamma_{n\ell}^2(\lambda,\dh) \times \kappa^2 / (2\mu)$
\\[1mm] \hline
Kinetic energy of scattering state in CM frame
&${\cal E}_{\vec k} = \vec k^2/(2\mu)$
\\ \hline
Wavefunction of $n\ell m$ bound state
&$\psi_{n\ell m} (\vec r) = \kappa^{3/2} 
\left[\dfrac{\x_{n\ell}^{}(\kappa r)}{\kappa r}\right] 
Y_{\ell m}(\W_{\vec r})$
\\[2ex] \hline 
Wavefunction of scattering state
&
$
\phi_{\vec k} (\vec r) \! = \! \displaystyle \sum_{\ell=0}^\infty (2\ell+1) 
\! \left[ \dfrac{\x_{|\vec k|,\ell}^{}(k r)}{k r} \right] 
\! P_{\ell}(\hat{\vec k}\cdot\hat{\vec r})
$
\\[2ex] \hline
\parbox{0.5\textwidth}{\smallskip
Dimensionless radial space coordinates \\ 
for bound and scattering states}
&$\xB \equiv \kappa r$~~and~~$\xS \equiv kr$
\\[2ex] \hline
Dimensionless time parameters for freeze-out
& $x \equiv \mx / T, \quad \tilde{x} \equiv  \mX / T$
\\ \hline
\end{tabular}
\caption{\label{tab:notation} 
Summary of notation. For the running of the strong coupling, see \cref{tab:MomentumTransfers}.}
\addcontentsline{toc}{subsection}{Summary of notation}	
\end{table}

\clearpage
\section{Simplified model and relic density \label{Sec:Model}}

\subsection{Simplified model \label{sec:Model_Lagrangian}}

We assume that DM is a Majorana fermion $\chi$ with mass $m_\chi$, that co-annihilates with a complex scalar $X$ with mass $\mX$. $X$ is a triplet under $SU(3)_c$  and couples to a real scalar $h$ of mass $m_h=125~\GeV$, that aims to resemble the SM Higgs boson. Both $\x$ and $X$ are odd under a $\mathbb{Z}_2$ symmetry, and assumed to be the lightest (LP) and next-to-lightest particles (NLP). We denote the LP--NLP relative mass splitting as
\beq
\D \equiv (\mX - \mx)/\mx,
\label{eq:Delta}
\eeq
and define the total and the reduced mass of an $XX^\dagger$ pair,
\beq
M = 2\mX \quad \text{and} \quad \mu = \mX/2 .
\label{eq:Masses_TotalReduced}
\eeq

If $\Delta \ll1$, the DM abundance is determined not only by the $\x-\x$ annihilations, but also by the $\x-X$, $\x-X^\dagger$ and $X-X^\dagger$ (co)annihilation processes. We shall assume that the $X-X^\dagger$ annihilation processes dominate the total depletion rate of the $\mathbb{Z}_2$-odd particles, and neglect all other contributions. This is a reasonable approximation, since the  former involve the strong coupling and the coupling to the Higgs, which can be quite significant, while the latter involve the Weak coupling and are typically less efficient. Moreover, our goal here is to investigate, within a self-consistent framework, the long-range effect of the Higgs on the formation of bound states by particles charged under $SU(3)_c$. Including the $\chi-\chi$,  $\chi-X$ and $\chi-X^\dagger$ processes in the computation of the DM density, would necessitate further model specifications and result in loss of generality. We note though, that a scenario where the depletion of DM is dominated by the annihilation of its coloured co-annihilating partners is realistic. For instance, within the MSSM it is possible to find viable stop-neutralino co-annihilation scenarios where, for small relative mass differences, the $\tilde{t}-\tilde{t}^\dagger$ annihilation contributes about $75\%-80\%$ to the neutralino depletion rate~\cite{Harz:2014gaa,Schmiemann:2019czm}.

Implicit in our computations is the hypothesis that the LP and the NLP stay in chemical equilibrium, i.e.~their interconversion rate exceeds the Hubble expansion rate,
\begin{align}
\Gamma(X + \mathrm{SM} \leftrightarrow \chi + \mathrm{SM}) \gg H\,,
\end{align}
at least until the decoupling of the $X-X^\dagger$ annihilation processes. This implies a lower limit on the interconversion cross-section,  
$\sigma(X+\mathrm{SM} \leftrightarrow \chi+ {\rm SM}) \sim 
\Gamma(X+ \mathrm{SM} \leftrightarrow \chi+ {\rm SM}) / n_{\mathsmaller{\rm SM}}^{\mathrm{eq}}$, where  $n_{\mathsmaller{\rm SM}}^{\rm eq}$ stands for the number density of the SM particles participating in these processes. With the latter being relativistic during the DM freeze-out, $n_{\mathsmaller{\rm SM}}^{\mathrm{eq}} \sim T^3$, and $H\sim 17~T^2/M_{\mathrm{Pl}}$, we find the condition
\begin{align}
\sigma(X + \mathrm{SM} \leftrightarrow \chi + \mathrm{SM}) 
\gg \frac{17~x_{\rm dec}}{m_\chi M_{\mathrm{Pl}}} 
\sim 6 \times 10^{-11}~{\rm pb} \(\frac{\TeV}{m_\chi}\) \,,
\label{eq:ChemicalEq_Condition}
\end{align}
where we set $x_{\rm dec} \equiv {m_\chi} / T_{\rm dec} \sim 100$ to encompass even $X-X^\dagger$ annihilations that occur somewhat after the DM freeze-out and can be significant in scenarios with light force mediators. 
Evidently, the required cross-section is many orders of magnitude below the Weak scale.  
How constraining the assumption of LP-NLP chemical equilibrium in reality is, depends on the model and the specific processes under consideration. 
For instance, in the MSSM, the interconversion between stops and neutralinos is mediated by SM particles, and the rate comfortably satisfies the condition \eqref{eq:ChemicalEq_Condition}. On the other hand, the gluino-neutralino interconversion is usually mediated by a heavy squark, and the condition of chemical equilibrium sets an upper limit on its mass, of the order of 100~TeV~\cite{Nagata:2015hha}. 
We discuss this and other caveats of our calculations further in  \cref{sec:Results_Caveats}.

\bigskip

Given the above considerations, the Lagrangian terms relevant for our purposes are~\cite{Harz:2017dlj}
\beq
{\d \cal L} = 
(D_{\m,ij} X_j)^\dagger \, (D_{ij'}^\m X_{j'}) 
- \mX^2 \, \Xdagger_j X_j^{} 
+ \frac{1}{2}(\partial_\m h) (\partial^\m h) - \frac12 \mh^2 h^2
- \gh^{} \mX \ h \, \Xdagger_j X_j^{}  ,
\label{eq:Lagrangian}
\eeq
where $D_{\m,ij} = \d_{ij} \partial_\mu + i \gs \, G_\m^a T^a_{ij}$ is the covariant derivative with $G_\m^a$ being the gluon fields and $T^a$ the corresponding generators. We define the couplings that will appear in the non-relativistic potential in \cref{sec:Model_NRpotential}, as 
\begin{align}
\as &\equiv \gs^2 / (4 \pi)  , \label{eq:alpha_s_def}
\\
\ah &\equiv \gh^2 / (16 \pi) . \label{eq:alpha_h_def}
\end{align}

Let us briefly comment on the origin of the trilinear term $h |X|^2$. In the corresponding dimensionful coupling, we have factored out $\mX$ for convenience. This factorization does \emph{not} imply that the coupling originates solely from the quartic term $|H|^2 |X|^2 \supset |H^0|^2 |X|^2$, with $H$ and $H^0$ being the Higgs doublet and its neutral component, whose radial excitation after the  EWSB is $h$. The $h|X|^2$ term may be also sourced by a gauge-invariant trilinear coupling $H X_1^\dagger X_2$, where $X_1$ and $X_2$ have different $SU(2)_L \times U(1)_{Y}$ charges and mix after the EWSB to generate the mass eigenstate $X$. This dynamics is in fact encountered in the MSSM, where  $X_1$ and $X_2$ would be the two stop fields that couple to the Higgs via a SUSY-breaking $A$ term.  
Moreover, the $X$, $X_1$, $X_2$ fields -- being scalars -- may also have $SU(2)_L \times U(1)_Y$ invariant mass terms.  In order to remain agnostic about the underlying UV complete model, we shall treat $\gh$ and $\mX$ as independent parameters that are not delimited by one another, or by the Higgs vacuum expectation value, $v_{\mathsmaller{H}} \simeq 246~\GeV$.

Here, we are interested in the long-range effect of the Higgs, which emanates from the $h|X|^2$ coupling. Thus for simplicity, we do not include quartic terms in the scalar potential, or the couplings of the Higgs to the SM particles.

\subsection{Boltzmann equation for the relic density \label{sec:Model_BoltzmannEq}}

Following \cite{Edsjo:1997bg}, we define $\tilde{Y}$ to be the sum over the yields of all (co)annihilating particles,
\beq 
\tilde{Y} \equiv \Yx + \YX^{} + \YXdagger  =   \Yx + 2\YX^{} ,
\label{eq:Ytilde} 
\eeq
where $Y_i$ is the ratio of the number density $n_i$ over the entropy density of the universe $s \equiv (2\pi^2/45) \, \gstarS^{} \, T^3$. We denote by  $\gstarS^{}$ and $g_*$ the entropy and energy degrees of freedom respectively, and define 
\begin{align}
\gstareffsqrt &\equiv \frac{\gstarS}{\sqrt{g_*}}
\(1 + \frac{T}{3g_{*\mathsmaller{S}}} \frac{dg_{*\mathsmaller{S}}}{dT} \) .
\label{eq:gstareff}
\end{align}
Under the assumption of chemical equilibrium between the LP and NLP, the evolution of the DM density can be described by a single Boltzmann equation, 
\beq
\frac{d\tilde{Y}}{dx} 
= - \sqrt{\frac{\pi}{45}} 
\, \frac{\mpl \mx \, \gstareffsqrt}{x^2}
\, \<\sigmaeff \, \vrel \>
\, (\tilde{Y}^2 - \tilde{Y}_\eq^2)  .
\label{eq:Boltzmann}
\eeq
As usual, we have defined the dimensionless time parameter $x \equiv \mx / T$. The yields in equilibrium of $\x$, $X$ and $X^\dagger$ are ($\gx = 2$, $\gX^{} = 3$)
\begin{subequations}
\label{eq:Y_equil}
\label[pluralequation]{eqs:Y_equil}
\begin{align}
\Yx^\eq &= 
\frac{90}{(2\pi)^{7/2}} \ \frac{\gx}{\gstarS} \ x^{3/2} \ e^{-x} , 
\label{eq:Yx_equil}
\\
\YX^\eq = \YXdagger^\eq &= 
\frac{90}{(2\pi)^{7/2}} \ \frac{\gX}{\gstarS} \ [(1+\D)x]^{3/2} \ e^{-(1+\D)x} . \label{eq:YX_equil}
\end{align}
\end{subequations}

While $\<\sigmaeff \, \vrel \>$ in general denotes the effective, thermally averaged cross-section that includes all annihilation and co-annihilation processes, each weighted by the number densities of the interacting species, we assume for the purpose of our study that the NLP self-annihilation is the dominant contribution. Then, the effective cross-section becomes
\beq
\<\sigmaeff \, \vrel \> = 
\frac{2\YX^\eq \YXdagger^\eq \, \<\s_{\mathsmaller{X\Xdagger}} \, \vrel\> }{\tilde{Y}_\eq^2} 
= \< \s_{\mathsmaller{X\Xdagger}} \, \vrel\> 
\( \frac{2 \gX^2 (1+\D)^3 \, e^{-2 x \, \D}}{\[\gx + 2\gX (1+\D)^{3/2} \, e^{- x \, \D }\]^2} \) ,
\label{eq:sigma_eff}
\eeq
where
\beq
\< \sigma_{XX^\dagger} \,\vrel \> \equiv 
\< \sigma_{\mathrm{ann}}  \,\vrel \> +
\< \sigma_{\BSF} \,\vrel \>_{\eff} 
\label{eq:sigma_XXdagger}
\eeq
includes all processes that deplete the $XX^\dagger$ pairs. It comprises not only of the direct annihilation channels, but also of the BSF processes, whose cross-sections must be weighted by the fraction of bound states that decay rather than being ionised, as we discuss below.

\subsection{Non-relativistic potential \label{sec:Model_NRpotential}}

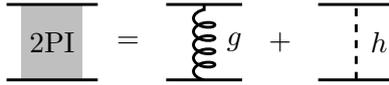
\begin{figure}[t!]
\centering
\begin{tikzpicture}[line width=1.1pt, scale=1]
\begin{scope}[shift={(-1,0)}]
\draw[fill=lightgray,draw=none]    (-0.4,0) -- (0.4,0) -- (0.4,1) -- (-0.4,1) -- cycle;
\draw[fermionnoarrow] (-0.6,1) -- (0.6,1);
\draw[fermionnoarrow] (-0.6,0) -- (0.6,0);
\node at (0,0.5) {2PI};
\end{scope}
\node at (0,0.5) {$=$};
\begin{scope}[shift={(1,0)}]
\draw[fermionnoarrow] (-0.5,1) -- (0.5,1);
\draw[fermionnoarrow] (-0.5,0) -- (0.5,0);
\draw[gluon]          (0,0) -- (0,1);
\node at (0.4,0.5) {$g$};
\end{scope}
\node at (2,0.5) {$+$};
\begin{scope}[shift={(3,0)}]
\draw[fermionnoarrow] (-0.5,1) -- (0.5,1);
\draw[fermionnoarrow] (-0.5,0) -- (0.5,0);
\draw[scalarnoarrow]  (0,0) -- (0,1);
\node at (0.3,0.5) {$h$};
\end{scope}
\end{tikzpicture}
\caption[]{\label{fig:2PI} 
The gluon and Higgs exchange are the leading order contributions to the 2-particle-irreducible (2PI) kernel that gives rise to the long-range interaction between $X$ and $X^\dagger$.}
\end{figure}

In the presence of a long-range interaction between two particles, the asymptotic states are not well approximated by plane waves. In order to determine their properties, we must resum the two-particle-irreducible (2PI) diagrams (see e.g.~\cite{Petraki:2015hla}). In our model, the $X, X^\dagger$ pairs interact via gluon and Higgs exchange, both of which may result in a sizeable long-range effect~\cite{Harz:2017dlj}. In the non-relativistic and weak coupling regime, the resummation of the gluon and Higgs exchange diagrams shown in \cref{fig:2PI} amounts to solving the Schr\"odinger equation with the mixed Coulomb and Yukawa potential
\beq 
V(r) = - \frac{\ag}{r} - \frac{\ah}{r} \ e^{-\mh r} ,
\label{eq:Potential}
\eeq 
where $\ah$ has been defined in \cref{eq:alpha_h_def} and derived in ref.~\cite{Petraki:2016cnz}. The coupling $\ag$ is related to $\as$ defined in \cref{eq:alpha_s_def} but depends also on the colour representation of the $XX^\dagger$ state. The product of a colour triplet and an anti-triplet decomposes into a singlet and an octet state, ${\bf 3 \otimes  \bar{3}} = {\bf 1 \oplus 8 }$, such that the coupling $\ag$ amounts to~\cite{Kats:2009bv}
\begin{subequations}
\label{eq:alpha_g}
\label[pluralequation]{eqs:alpha_g}
\begin{align}
\agSinglet &\equiv \asSinglet \times C_F = (4/3)\asSinglet ,
\label{eq:alpha_g_Singlet}
\\
\agOctet &\equiv \asOctet \times (C_F-C_A/2) = -\asOctet/6,
\label{eq:alpha_g_Octet}
\end{align}
\end{subequations}
for the singlet and the octet configurations, where $C_F=4/3$ and $C_A=3$ are the quadratic Casimir invariants of the fundamental and adjoint representations of $SU(3)$. In \cref{eqs:alpha_g}, we have differentiated between the strong coupling in the singlet and octet states, $\asSinglet$ and $\asOctet$ respectively, to accommodate the possibly different momentum transfer along the 2PI diagrams, which affects the value of $\as$ due its running.

For the massive mediator $h$, the interaction manifests as long-range if the range of the potential $\mh^{-1}$ is comparable to or larger than the corresponding Bohr radius  $(\mu \ah)^{-1}$. To parametrise the range of the interaction we use the dimensionless ratio
\beq
\label{eq:dh}
d_h \equiv \mu\ah / \mh .	
\eeq
The Sommerfeld effect and BSF become significant when the average momentum transfer between the scattering particles, $\mu \vrel$, is comparable to or lower than the Bohr momentum, $\mu \alpha$. To compare the relevant scales in our set-up, we introduce the parameters
\begin{subequations}
\label{eq:zeta}
\label[pluralequation]{eqs:zeta}
\begin{align}
\zetagR		&\equiv  \agR/\vrel  ,	\label{eq:zeta_g} \\
\zetah		&\equiv  \ah/\vrel  ,	\label{eq:zeta_h}
\end{align}
\end{subequations}
where in \cref{eq:zeta_g}, ${\bf R=1,8}$ denotes the colour representation. Because the average momentum transfer along the 2PI diagrams is different in the scattering and the bound states, we shall also discern between $\zetagScattR$ and $\zetagBoundR$, for which we evaluate the strong coupling at the respective scale, as we shall see in the following (cf.~\cref{sec:BSF_AlphaRunning} for a summary). However, since the running of $\ah$ depends on the underlying UV model, we neglect it in the following. For convenience, we define
\begin{subequations}
\label{eq:zeta_SB}
\label[pluralequation]{eqs:zeta_SB}
\begin{align}
\zetaScattR &\equiv \zetagScattR + \zetah = (\agScattR +\ah)/\vrel, \label{eq:zetaScatt} \\
\zetaBoundR &\equiv \zetagBoundR + \zetah = (\agBoundR +\ah)/\vrel. \label{eq:zetaBound}
\end{align}
\end{subequations}
Finally, for the bound states, it will be useful to define the parameter
\begin{align}
\lambdaR &\equiv \agBoundR/\ah  = \zetagBoundR / \zetah .
\label{eq:lambda}
\end{align}

The dimensionless variables defined in \cref{eq:dh,eq:zeta,eq:zeta_SB,eq:lambda} suffice to parametrise the pieces of the annihilation and BSF cross-sections that need to be evaluated numerically. Generically, the Sommerfeld effect and the capture into bound states are expected to be significant for $|\zeta_g|\gtrsim {\cal O}(1)$ and $\zeta_h, d_h \gtrsim {\cal O}(1)$, for the Coulomb and Yukawa potentials, respectively. However, in ref.~\cite{Harz:2017dlj} it was shown that  the Coulomb potential affects how long-range the Yukawa potential manifests. In \cref{Sec:Wavefunctions}, we revisit this point and determine the conditions for the existence of bound states and their properties.

\paragraph{Thermal effects.}

The coupling of the mediators --- the gluons and the Higgs --- to the relativistic SM plasma modifies the zero-temperature potential \eqref{eq:Potential} in several ways. On one hand, it generates thermal contributions to the mediator masses that screen the long-range effect.  On the other hand, it gives rise to frequent scatterings between $X,X^\dagger$ and the relativistic plasma that allow for non-radiative dissipation of energy from an $XX^\dagger$ pair, and thus ensure equilibrium between bound and scattering states. In fact, the $XX^\dagger$ spectral function morphs into a continuum that encompasses energies both above and below $2\mX$~\cite{Kim:2016zyy,Kim:2016kxt}.  These thermal effects have been considered in DM freeze-out computations using linear response theory, first in refs.~\cite{Biondini:2017ufr,Biondini:2018pwp,Biondini:2018xor}, where radiative capture processes were neglected, and subsequently in ref.~\cite{Biondini:2018ovz} which incorporated BSF via gluon emission. Recently, ref.~\cite{Binder:2018znk} derived \emph{ab initio} a formalism for the description of DM long-range interactions in a plasma background, starting from non-equilibrium quantum field theory.

In our analysis, we neglect thermal effects. In the scattering states, the screening of the gluon-mediated force due to the thermal gluon masses is negligible, because of the fairly large average velocity of the $X,X^\dagger$ particles during DM freeze-out, which ensures that the average momentum transfer in an $XX^\dagger$ pair is larger than the thermal gluon  mass. The radiative BSF processes we consider are sufficiently rapid to bring bound states in equilibrium at early times. In equilibrium, and at temperatures much higher than the binding energy, the DM depletion via bound-state decay becomes independent of the BSF rate, whether this is due to radiative and/or scattering processes; we return to this point in \cref{sec:Model_BoundStates}. More importantly, we find that the impact of the bound states on the DM relic density occurs mostly at later times, when the plasma temperature approaches or falls below the binding energy. In this regime, the thermal corrections become less important, and the radiative capture is typically the dominant BSF process.

\subsection{Annihilation \label{sec:Model_Annihilation}}

\begin{figure}[t!]
\centering
\begin{tikzpicture}[line width=1.1pt, scale=1]

\begin{scope}[shift={(0,0)}]
\begin{scope}[shift={(-2.5,0)}]
\node at (-1.25, 1) {$X^{}$};
\node at (-1.25, 0) {$X^\dagger$};
\draw[fermion]    (-1,1) -- (0,1);
\draw[fermion]    ( 0,1) -- (0,0);
\draw[fermionbar] (-1,0) -- (0,0);
\draw[gluon] (0,0) -- (1,0);
\draw[gluon] (0,1) -- (1,1);
\node at (1.2,1) {$g$};
\node at (1.2,0) {$g$};
\end{scope}
\begin{scope}[shift={(0,0)}]
\node at (-0.25, 1) {$X^{}$};
\node at (-0.25, 0) {$X^\dagger$};
\draw[fermion]    ( 0,1) -- (0.5,0.5);
\draw[fermionbar] ( 0,0) -- (0.5,0.5);
\draw[gluon] (0.5,0.5) -- (1,0);
\draw[gluon] (0.5,0.5) -- (1,1);
\node at (1.2,1) {$g$};
\node at (1.2,0) {$g$};
\end{scope}
\begin{scope}[shift={(2.5,0)}]
\node at (-0.25, 1) {$X^{}$};
\node at (-0.25, 0) {$X^\dagger$};
\draw[fermion]    ( 0,1) -- (0.5,0.5);
\draw[fermionbar] ( 0,0) -- (0.5,0.5);
\draw[gluon] (0.5,0.5) -- (1,0.5);
\draw[gluon] (1,0.5)   -- (1.5,0);
\draw[gluon] (1,0.5)   -- (1.5,1);
\node at (1.7,1) {$g$};
\node at (1.7,0) {$g$};
\end{scope}
\begin{scope}[shift={(6.5,0)}]
\node at (-1.25, 1) {$X^{}$};
\node at (-1.25, 0) {$X^\dagger$};
\draw[fermion]    (-1,1) -- (0,1);
\draw[fermion]    ( 0,1) -- (0,0);
\draw[fermionbar] (-1,0) -- (0,0);
\draw[scalarnoarrow] (0,0) -- (1,0);
\draw[scalarnoarrow] (0,1) -- (1,1);
\node at (1.2,1) {$h$};
\node at (1.2,0) {$h$};
\end{scope}
\end{scope}


\end{tikzpicture}
\caption[Tree-level diagrams of particle-antiparticle annihilation into pairs of gluons or Higgses]{\label{fig:Annihilation_Perturb}
Tree-level diagrams contributing to the $X-X^\dagger$ annihilation into gluons and into Higgs bosons. The $u$-channel diagram that match the corresponding $t$-channel diagram  (\emph{first diagram})  is not shown. The $s$-channel annihilation into gluons (\emph{third}) yields a $p$-wave contribution such that we neglect this process in our computations. }
\bigskip
\begin{tikzpicture}[line width=1.1pt, scale=1.1]

\begin{scope}[shift={(2.9,0)}]
\node at (-2.25, 1) {$X^{}$};
\node at (-2.25, 0) {$X^\dagger$};
\draw[fill=lightgray,draw=none]    (-1.9,0) -- (-1.9,1) -- (-1.4,1) -- (-1.4,0) -- cycle;
\node at (-1.65,0.5) {\scriptsize 2PI};
\node at (-1.1,0.5) {$\cdots$};
\draw[fill=lightgray,draw=none]    (-0.8,0) -- (-0.8,1) -- (-0.3,1) -- (-0.3,0) -- cycle;
\node at (-0.55,0.5) {\scriptsize 2PI};
\draw[fermion]    (-2,1) -- (  0,1);
\draw[fermion]    ( 0,1) -- (0.5,0.5);
\draw[fermionbar] ( 0,0) -- (0.5,0.5);
\draw[fermionbar] (-2,0) -- (  0,0);
\filldraw (0.5,0.5) circle (5pt);
\draw[fermionnoarrow] (0.5,0.5) -- (1,0);
\draw[fermionnoarrow] (0.5,0.5) -- (1,1);
\end{scope}

\end{tikzpicture}
\caption[]{\label{fig:Annihilation_SE} 
The annihilation processes are influenced by the Sommerfeld effect, a non-perturbative phenomenon that corresponds diagrammatically to the resummation of the infinite ladder of 2-particle-irreducible (2PI) diagrams. The 2PI kernel for our model is shown in \cref{fig:2PI} and includes both possible long-range interactions: the gluon and the Higgs exchange (left). The black blob represents the perturbative part of the various annihilation channels, shown in \cref{fig:Annihilation_Perturb}.}
\end{figure}

The dominant contributions to the direct annihilation of $XX^\dagger$ pairs arise from the $XX^\dagger \rightarrow gg$ and $XX^\dagger \rightarrow hh$ channels, shown in \cref{fig:Annihilation_Perturb}. We neglect the $p$-wave suppressed contributions $XX^\dagger \rightarrow q \bar{q}, gh$. (Similarly we neglect the $p$-wave suppressed contribution of the $s$-channel diagram of $XX^\dagger \rightarrow gg$). To leading order, the colour-averaged perturbative cross-sections are~\cite{Harz:2017dlj}
\begin{subequations}
\label{eq:AnnihilationCrossSections_tree}
\label{eqs:AnnihilationCrossSections_tree}
\begin{align}
(\s \vrel)_{X \Xdagger \to gg}^{\mathrm{perturb}} &= 
\frac{14}{27} \frac{\pi (\asAnn)^2}{\mX^2} ,
\label{eq:sigma_XXbarTogg_tree}
\\
(\s \vrel)_{X \Xdagger \to hh}^{\mathrm{perturb}} &= 
\frac{4\pi \ah^2}{3\mX^2} \ \frac{(1-\mh^2/\mX^2)^{1/2}}{[1-\mh^2/(2\mX^2)]^2} ,
\label{eq:sigma_XXbarTohh_tree}
\end{align}
\end{subequations}
where $\asAnn \equiv \as(Q=\mX)$ arises from the gluon emission vertices in the hard annihilation process and must be evaluated at momentum transfer $Q=\mX$. Both the colour-singlet and octet scattering states contribute in \cref{eq:sigma_XXbarTogg_tree}, while only the singlet can annihilate into $hh$. Due to their different gluon-mediated potential the two states are affected differently by the Sommerfeld effect. The full cross-sections are~\cite{Harz:2017dlj} 
\begin{subequations}
\label{eq:AnnihilationCrossSections}
\label{eqs:AnnihilationCrossSections} 
\begin{align}
(\s \vrel)_{X \Xdagger \to gg} &= 
(\s \vrel)_{X \Xdagger \to gg}^{\mathrm{perturb}} \times 
\( \frac27 S_{0, {\bf [1]}} + \frac57 S_{0, {\bf [8]}} \) ,
\label{eq:sigma_XXbarTogg}
\\
(\s \vrel)_{X \Xdagger \to hh} &= 
(\s \vrel)_{X \Xdagger \to hh}^{\mathrm{perturb}}
\times S_{0,{\bf [1]}}  ,
\label{eq:sigma_XXbarTohh}
\end{align}
\end{subequations}
where $S_{0, {\bf [1]}}$ and $S_{0, {\bf [8]}}$ indicate the $s$-wave Sommerfeld factors for the colour-singlet and octet states respectively. For the mixed Coulomb and Yukawa potential \eqref{eq:Potential}, the $s$-wave Sommerfeld factor depends on the parameters defined in \cref{eq:zeta,eq:dh}, i.e.~$S_0 = S_0 (\zetagScatt, \zetah, \dh)$, as we discuss in  \cref{sec:Wavefunctions_ScatteringStates}. Thus, taking \cref{eqs:alpha_g} into account,
\begin{subequations}
\label{eq:Annihilation_S0}	
\label[pluralequation]{eqs:Annihilation_S0}	
\begin{align}
S_{0, {\bf [1]}} &= S_0 
\(\frac{4\asScatt}{3\vrel},~\frac{\ah}{\vrel},~\frac{\mu \ah}{\mh} \) ,
\\
S_{0, {\bf [8]}} &= S_0 \(-\frac{\asScatt}{6\vrel},~\frac{\ah}{\vrel},~\frac{\mu \ah}{\mh} \) .
\end{align}
\end{subequations}
Here, $\asScatt \equiv \as (Q =\mu \vrel) = \asScattSinglet = \asScattOctet$ denotes the strong coupling evaluated at the average momentum transfer in the scattering states, $Q = \mu \vrel$, which is independent of the colour representation. The thermally-averaged total annihilation cross-section is
\beq
\< \sigma_{\ann} \vrel \> = 
\(\frac{\mu}{2\pi T}\)^{3/2}
\int d^3 \vrel 
\ \exp \(-\frac{\mu \vrel^2}{2T}\) 
\[ (\s \vrel)_{X \Xdagger \to gg} +(\s \vrel)_{X \Xdagger \to hh} \] .
\label{eq:sigmaANN_averaged}
\eeq

\subsection{Bound-state formation, ionisation and decay \label{sec:Model_BoundStates}}

\paragraph{Formation.}

Attractive long-range interactions imply the existence of bound states that may form from the scattering states via dissipation of energy.  In the non-relativistic regime, the energy available to be dissipated is the difference between the initial and final states, i.e.~the kinetic energy of the scattering state in the centre-of-momentum (CM) frame ${\cal E}_{\bf k} = {\bf k}^2/(2\mu) = \mu \vrel^2/2$ minus the binding energy of the bound state ${\cal E}_{n\ell} = -\gamma^2_{n\ell}\times \kappa^2/(2 \mu)$, where $\kappa \equiv \mu (\agBound+\ah)$ is the generalised Bohr momentum of the bound state and $\gamma_{n\ell}$ parametrises the departure of ${\cal E}_{n\ell}$ from its Coulomb limit $\dh \to \infty$ (cf.~\cref{sec:Wavefunctions_BoundStates}). For capture into the ground state, the dissipated energy is
\beq
\omega \simeq {\cal E}_{\bf k} - {\cal E}_{10} = \frac{\mu}{2} [\vrel^2 + (\agBound+\ah)^2\gamma_{10}^2] .
\label{eq:omega}
\eeq

\begin{figure}[t!]
\centering    
\begin{tikzpicture}[line width=1.1pt, scale=1.1]

\begin{scope}[shift={(0,0)}]
\node at (-2.75, 1) {$X$};
\node at (-2.75, 0) {$X^\dagger$};
\draw[fill=lightgray,draw=none]    (-2.4,0) -- (-2.4,1) -- (-1.9,1) -- (-1.9,0) -- cycle;
\node at (-2.15,0.5) {\scriptsize 2PI};
\node at (-1.6,0.5) {$\cdots$};
\draw[fill=lightgray,draw=none]    (-1.3,0) -- (-1.3,1) -- (-0.8,1) -- (-0.8,0) -- cycle;
\node at (-1.05,0.5) {\scriptsize 2PI};
\draw[fermion] 		(-2.5,1) -- (-0.5,1);
\draw[fermion] 		(-0.5,1) -- (-0.05,0.55);
\draw[fermionbar] 	(-0.5,0) -- (-0.05,0.45);
\draw[fermionbar] 	(-2.5,0) -- (-0.5,0);
\draw[fill=lightgray,draw=none]    (2.4,0) -- (2.4,1) -- (1.9,1) -- (1.9,0) -- cycle;
\node at (2.15,0.5) {\scriptsize 2PI};
\node at (1.6,0.5) {$\cdots$};
\draw[fill=lightgray,draw=none]    (1.3,0) -- (1.3,1) -- (0.8,1) -- (0.8,0) -- cycle;
\node at (1.05,0.5) {\scriptsize 2PI};
\draw[fermionbar] 	(2.5,1) -- (0.5,1);
\draw[fermionbar] 	(0.5,1) -- (0.1,0.6);
\draw[fermion]		(0.5,0) -- (0.1,0.4);
\draw[fermion]	 	(2.5,0) -- (0.5,0);
\draw[line width=0.7pt,gray] (1.6,0.5) ellipse (1cm and 0.9cm);
\node at (2.85,0.5) {${\cal B}$};
\draw[gluon] (0,1.5) -- (0,0.75);
\node at (0.35,1.3) {$g$};
\filldraw[lightgray]  (0,0.5) circle (7pt);
\node at (0,0.5) {${\cal C}$};
\end{scope}

\end{tikzpicture}
\caption[Feynman diagram representation of bound-state formation via gluon emission]{\label{fig:FeynmanDiagram_BSF} 
The amplitude for the radiative capture into bound states consists of the (non-perturbative) initial and final state wavefunctions, and the perturbative 5-point function that includes the radiative vertices. The wavefunctions (cf.~\cref{Sec:Wavefunctions}) are determined by the resummation of the  2-particle-irreducible (2PI) diagrams. The 2PI kernel is shown in \cref{fig:2PI}.
}
\medskip
\begin{tikzpicture}[line width=1.1pt, scale=1.1]
\node at (-3,0.5) {${\cal C} = $};
\begin{scope}[shift={(-1.2,0)}]
\draw[fermion] 		(-0.8,1) -- (0,1);
\draw[fermionbar] 	(-0.8,0) -- (0,0);
\draw[gluon] (0,0)   -- (0,0.5);
\draw[gluon] (0,1)   -- (0,0.5);
\draw[gluon] (0.6,0.5) -- (0,0.5);
\draw[fermion] 		(0,1) -- (0.8,1);
\draw[fermionbar] 	(0,0) -- (0.8,0);
\end{scope}
\node at (0, 0.5) {$+$};
\begin{scope}[shift={( 2.4,0)}]
\begin{scope}[shift={(-1.2,0)}]
\draw[fermion] 		(-0.8,1) -- (0,1);
\draw[fermionbar] 	(-0.8,0) -- (0.8,0);
\draw[gluon] (0,1) -- (0,1.6);
\draw[fermion] (0,1) -- (0.8,1);
\end{scope}
\node at (0, 0.5) {$+$};
\begin{scope}[shift={(1.2,0)}]
\draw[fermion] 		(-0.8,1) -- (0.8,1);
\draw[fermionbar] 	(-0.8,0) -- (0,0);
\draw[gluon] (0,-0.6) -- (0,0);
\draw[fermionbar] 	(0,0) -- (0.8,0);
\end{scope}
\end{scope}
\end{tikzpicture}
\caption[Gluon emission vertices of the radiative capture into bound states]{\label{fig:FeynmanDiagrams_BSF_RadiativeVertices} 
Leading order contributions to the radiative part of the capture into bound states via gluon emission (cf.~\cref{Sec:BSF}).}
\end{figure}
%
%
The capture into bound states via emission of an ultrasoft gluon is depicted in  \cref{fig:FeynmanDiagram_BSF,fig:FeynmanDiagrams_BSF_RadiativeVertices}. 
The decompositions ${\bf 3 \otimes  \bar{3}} = {\bf 1 \oplus 8 }$  and
${\bf 8} \otimes {\bf 8} = {\bf 1}_S + {\bf 8}_A + {\bf 8}_S + {\bf 10}_A + \overline{\bf 10}_A + {\bf 27}_S$ imply that the allowed transitions are
\begin{subequations}
\label{eq:BSF_gluons}
\label[pluralequation]{eqs:BSF_gluons}
\begin{align}
(X + \Xdagger)_{\bf [8]} \ &\to \ 
{\cal B} (X\Xdagger)_{\bf [1]} + g_{\bf [8]}	, 
\label{eq:BSF_8to1}  
\\ 
(X + \Xdagger)_{\bf [1]} \ &\to \ 
\left\{ {\cal B} (X\Xdagger)_{\bf [8]} + g_{\bf [8]} \right\}_{\bf [1_S]}	, 
\label{eq:BSF_1to8} 
\\ 
(X + \Xdagger)_{\bf [8]} \ &\to \ 
\left\{ {\cal B} (X\Xdagger)_{\bf [8]} + g_{\bf [8]} \right\}_{{\bf [8_S]} \text{ or } {\bf [8_A]}}	. 
\label{eq:BSF_8to8} 
\end{align}
\end{subequations}
In~\eqref{eq:BSF_1to8} and \eqref{eq:BSF_8to8}, the indices outside the curly brackets denote the colour representation of the entire final state (bound state plus gluon). When considering the gluon potential solely, only the colour-singlet $XX^\dagger$ combination supports a bound state, thus \eqref{eq:BSF_8to1} is the sole capture process via gluon emission. However, if the coupling to the Higgs is sufficiently strong, the Higgs exchange can overcome the gluon-mediated repulsion in the octet state and give rise to octet bound states. Then, the processes~\eqref{eq:BSF_1to8} and \eqref{eq:BSF_8to8} may be possible. In \cref{sec:Wavefunctions_BoundStates}, we determine the condition for the existence of bound states in a mixed repulsive Coulomb plus attractive Yukawa potential, and in \cref{sec:BSF_AlphaRunning} we adapt this condition to the colour-octet states in our model, taking into account the $\as$ running.

If the energy \eqref{eq:omega} available to be dissipated exceeds the Higgs mass, then bound states may form also via Higgs emission, according to 
\begin{subequations}
\label{eq:BSF_Higgs}
\label[pluralequation]{eqs:BSF_Higgs}
\begin{align}
(X + \Xdagger)_{\bf [1]} \ &\to \ {\cal B} (X\Xdagger)_{\bf [1]} + h	, 
\label{eq:BSFh 1to1}  \\
(X + \Xdagger)_{\bf [8]} \ &\to \ {\cal B} (X\Xdagger)_{\bf [8]} + h	. 
\label{eq:BSFh 8to8} 
\end{align}
\end{subequations}
However, the capture of particle-antiparticle pairs via scalar emission is subject to cancellations that suppress the processes \eqref{eq:BSF_Higgs} by higher orders in $\as$ and $\ah$ with respect to \eqref{eq:AnnihilationCrossSections} and \eqref{eq:BSF_gluons}~\cite[eqs.~(4.10)]{Petraki:2016cnz}. Although other couplings in the scalar potential, such as the quartic $ h^2|X|^2$ term, can enhance and even dominate the cross-sections for the processes \eqref{eq:BSF_Higgs}~\cite{Oncala:2018bvl}, these cross-sections remain subdominant to the direct annihilation and/or BSF via gluon emission. In the following, we thus neglect BSF via Higgs emission.

The thermally-averaged BSF cross-section is
\beq
\< \sigma_{\BSF} \vrel \> = 
\(\frac{\mu}{2\pi T}\)^{3/2}
\int d^3 \vrel 
\ \exp \(-\frac{\mu \vrel^2}{2T}\) 
[1+f_g(\omega)]
\ \sigma_{\BSF} \vrel ,
\label{eq:sigmaBSF_averaged}
\eeq
where $f_g(\omega) = 1/(e^{\omega/T} - 1)$ is the gluon occupation number, with $\omega$ being the energy of the emitted gluon, given by \cref{eq:omega}. The factor $1+f_g(\omega)$ accounts for the Bose enhancement due to the final-state gluon, and is necessary to ensure the detailed balance between the bound-state formation and ionisation processes at $T \gtrsim \omega$, which encompasses a significant temperature range that is relevant to the DM freeze-out~\cite{vonHarling:2014kha}.\footnote{
As is standard, we have omitted the Bose enhancement factor in the thermal averaging of the annihilation cross-sections \eqref{eq:sigmaANN_averaged}, since the energy of the annihilation products, $E = \mX$, far exceeds the temperature of the relativistic bath at the temperature range relevant for freeze-out, $T \lesssim \mX /20$.}

\smallskip 

Once bound states form, they may either be ionised back into their constituents by the ambient radiation, or decay into radiation, as we now discuss.

\paragraph{Ionisation.}

The ionisation cross-section is related to the BSF cross-section via the Milne relation (cf.~e.g.~\cite[appendix D]{Harz:2018csl}),
\beq
\SigmaIonR =  \frac{\gX^2}{g_g \, \gR^{}} 
\( \frac{\mu^2\vrel^2}{\omegaR^2}\)  
\sum_{\bf R_\textsl{S}}
\sigma_{\BSF}^{\mathsmaller{\bf [R_\textsl{S}]\to [R]}} ,
\label{eq:sigma_ion}
\eeq
where ${\bf R}_S$, ${\bf R}$ denote the colour representations of the scattering and bound $XX^\dagger$ states, with $g_{\mathsmaller{\bf [1]}} = 1$ and $g_{\mathsmaller{\bf [8]}} = 8$ being the colour-singlet and octet degrees of freedom respectively, and $g_g = 8$ being the gluon degrees of freedom. In \cref{eq:sigma_ion}, we have made explicit the dependence of the gluon energy \eqref{eq:omega} on the colour representation of the bound state. Note that for the ionisation cross-section of the colour-octet bound states -- if they exist -- we must include the contributions from both processes \eqref{eq:BSF_1to8} and \eqref{eq:BSF_8to8}, i.e.~sum over ${\bf R}_S = {\bf 1,8}$ as indicated in \cref{eq:sigma_ion}.  The ionisation rate of a bound state is then
\beq
\GammaIonR = 
g_g \ \int_{\omegaR^{\min}}^\infty \dfrac{d\omegaR}{2\pi^2} 
\dfrac{\omegaR^2}{e^{\omegaR / T}-1} \ \SigmaIonR ,
\nn
\eeq
where $\omegaR^{\min}$ is the minimum energy required for ionisation, i.e.~the binding energy, recovered from \cref{eq:omega} for $\vrel=0$. Using \cref{eq:sigma_ion,eq:omega}, we arrive at
\beq
\GammaIonR = \dfrac{9 \mu^3}{2\pi^2 \gR^{}}
\int_0^\infty d\vrel \ 
\dfrac{\vrel^2}{
\exp\left\{\dfrac{\mu [(\agBoundR+\ah)^2\gamma_{1,0}^2+ \vrel^2]}{2T} \right\} 
- 1} 
\ \sum_{\mathsmaller{{\bf R}_S}} \sigma_{\BSF}^{\mathsmaller{\bf [R_\textsl{S}] \to [R]}} \vrel .
\label{eq:GammaIon}
\eeq
Note that in \cref{eq:GammaIon}, $\gamma_{1,0}^{}$ also depends on ${\bf R}$ even though it is not explicitly indicated. 
From \cref{eq:sigmaBSF_averaged,eq:GammaIon}, we can explicitly verify the principle of detailed balance,
\beq
(n_{\mathsmaller{X}}^{\eq})^2 
\sum_{\mathsmaller{ {\bf R}_S}}
\<\sigma_{\BSF}^{\mathsmaller{\bf [R_\textsl{S}] \to [R]}} \vrel \> 
= n_{\mathsmaller{{\cal B}, \bf [R]}}^{\eq} \, \GammaIonR ,
\label{eq:DetailedBalance}
\eeq
where $n_{\mathsmaller{X}}^{\eq}$ and $n_{\mathsmaller{{\cal B}, \bf [R]}}^{\eq}$ are the equilibrium densities of $X$ and of $XX^\dagger$ bound states of colour representation ${\bf [R]}$, respectively. Note that \cref{eq:DetailedBalance} is more general than our derivation here may suggest; it holds true independently of what processes -- radiative or scattering -- may contribute to the capture into and the ionisation of bound states.
It is clear from \eqref{eq:GammaIon} and \eqref{eq:DetailedBalance}, that $\Gamma_{\ion}$ decreases exponentially when $T$ drops below the binding energy and most particles in the plasma do not have enough energy to dissociate the bound states.

\paragraph{Decay.}

The decay rate of the $\ell=0$ bound states is related to the perturbative $s$-wave annihilation cross-section times relative velocity as (see e.g.~\cite{Petraki:2015hla})
\beq
\GammaDecR =
(\sigma_{\ann, \mathsmaller{\bf [R]}}^{s-\rm wave} \vrel )
\ |\psi_{n 00}^{\mathsmaller{\bf [R]}}(0)|^2 ,
\label{eq:GammaDecay_def}  
\eeq
where $\psi_{n 00}^{\mathsmaller{\bf [R]}}(0)$ indicates the $n$-th level $s$-wave bound-state wavefunction of the colour representation ${\bf R}$, evaluated at the origin, and will be discussed in \cref{sec:Wavefunctions_BoundStates}. The cross-section $\sigma_{\ann, \mathsmaller{\bf [R]}}^{s-\rm wave} \vrel$ corresponds to the colour configuration of the bound state and should be averaged over the bound-state colour degrees of freedom only, rather than those of an unbound $XX^\dagger$ pair. Using the perturbative annihilation cross-sections \eqref{eq:AnnihilationCrossSections_tree} and taking into account the colour decomposition in \cref{eq:sigma_XXbarTogg}, we obtain for the ground states
\begin{subequations}
\label{eq:GammaDecay}
\label[pluralequation]{eqs:GammaDecay}
\begin{align}
\GammaDecSinglet &=
9\( 
\frac{4\pi (\asAnn)^2}{27\mX^2} + 
\frac{4\pi \ah^2}{3\mX^2} \ \frac{(1-\mh^2/\mX^2)^{1/2}}{[1-\mh^2/(2\mX^2)]^2}
\) 
|\psi_{100}^{\bf [1]}(0)|^2
, 
\label{eq:GammaDecaySinglet}  
\\
\GammaDecOctet &= \frac{9}{8}
\( \frac{10\pi (\asAnn)^2}{27\mX^2} \) 
|\psi_{100}^{\bf [8]}(0)|^2  . 
\label{eq:GammaDecayOctet} 
\end{align}
\end{subequations}

\paragraph{Effective bound-state formation cross-section.}

The effect of unstable bound states on the DM relic abundance is described by a system of coupled Boltzmann equations for the bound and unbound particles that describe the interplay between bound-state formation, ionisation and decay~\cite{vonHarling:2014kha}. Only the bound states that decay into radiation before being ionised contribute to the depletion of the DM abundance. It is thus possible to describe the impact of bound states using a single Boltzmann equation for the unbound particles and an effective BSF cross-section that incorporates the branching ratio of the bound states that decay rather than being ionised. For our model, this is
\begin{align}
\< \sigma_{\BSF} \vrel\>_{\eff} 
&= 
\< \sigma_{\BSF}^{\mathsmaller{\bf [8] \to [1]}} \vrel \>
\times \(\frac{\GammaDecSinglet}{\GammaDecSinglet + \GammaIonSinglet} \) 
\nn \\
&+
\( \< \sigma_{\BSF}^{\mathsmaller{\bf [1] \to [8]}} \vrel \> 
+  \< \sigma_{\BSF}^{\mathsmaller{\bf [8] \to [8]}} \vrel \> \)
\times \(\frac{\GammaDecOctet}{\GammaDecOctet + \GammaIonOctet} \) 
.
\label{eq:sigmaBSFeff}
\end{align}
The decay becomes faster than ionisation at temperatures comparable to or smaller than the binding energy, when most particles in the thermal bath do not possess enough energy to disassociate the bound states and $\Gamma_{\ion}$ decreases exponentially with decreasing temperature.

The effective BSF cross-section \eqref{eq:sigmaBSFeff} together with the thermally averaged annihilation cross-sections \eqref{eqs:AnnihilationCrossSections} yield the total effective annihilation cross-section \eqref{eq:sigma_XXdagger} that determines the DM relic density according to the Boltzmann \cref{eq:Boltzmann}.

\paragraph{Ionisation equilibrium.} 

If the BSF and ionisation cross-sections are sufficiently large to set $\Gamma_{\ion} \gg \Gamma_{\dec}$ at high temperatures, then the \emph{effective} BSF cross-section \eqref{eq:sigmaBSFeff} becomes independent of the actual BSF cross-section during that time. Instead, it is proportional to the annihilation cross-section, which determines the bound-state decay rate. Indeed, using the detailed balance \cref{eq:DetailedBalance} and the decay rate~\eqref{eq:GammaDecay_def}, we find 
\begin{align}
\< \sigma_{\BSF} \vrel\>_{\eff} 
\stackrel{{\rm large}~T}{\simeq} 
&\sum_{  \mathsmaller{\bf R_{\textsl{S}}, R}  }
\< \sigma_{\BSF}^{\mathsmaller{\bf [R_{\textsl{S}}] \to [R]}} \vrel\> 
\(\frac{\GammaDecR}{\GammaIonR}\)
= \sum_{  \mathsmaller{\bf R}  }
\frac{  n_{ \mathsmaller{{\cal B}, {\bf [R]}} }^\eq  } {(\nX^\eq)^2}
\GammaDecR
\nn \\
\stackrel{\phantom{{\rm large}~T}}{\simeq} 
&\sum_{  \mathsmaller{\bf R}  } 
\frac{\gR}{\gX^2}
\ \frac{\exp \( |{\cal E}_{10}^{\mathsmaller{\bf [R]}}|/T \)} {(\pi \mX T)^{3/2}}
\ (\sigma_{\ann, \mathsmaller{\bf [R]}}^{s-\rm wave} \vrel )
\ |\psi_{100}^{\mathsmaller{\bf [R]}}(0)|^2 ,
\label{eq:sigmaBSFeff_highT}
\end{align}
where we included the ground-state contributions only, and recall that ${\cal E}_{10}^{\mathsmaller{\bf [R]}}$ are their binding energies. \Cref{eq:sigmaBSFeff_highT} implies that at $T~\gg~|{\cal E}_{10}^{\mathsmaller{\bf [R]}}|$, the bound states contribute to the Sommerfeld factor of the colour configuration ${\bf [R]}$, by
\beq
S_{\rm bound, \mathsmaller{\bf [R]}} 
\simeq
|\psi_{100}^{\mathsmaller{\bf [R]}}(0)|^2  / (\pi \mX T)^{3/2} .
\label{eq:SBSF_largeT}
\eeq

To assess the importance of \cref{eq:SBSF_largeT}, it is informative to consider a simple attractive Coulomb potential of strength $\alpha$, for which $|\psi_{1,0,0}(0)|^2 =  (\mu \alpha)^3 / \pi$. Then, \cref{eq:SBSF_largeT} yields
$\pi^{-5/2} [\mX \alpha^2 / (4T)]^{3/2} \ll 1$, 
where the square bracket contains the binding-energy-to-temperature ratio, $|{\cal E}_{10}| / T$. This contribution can be compared with the thermally averaged Sommerfeld factor from the scattering states, which in this regime is 
$1+ (4\pi)^{1/2} [\mX \alpha^2 / (4T)]^{1/2}$. 
Clearly, the enhancement due to the bound states is subdominant to that due to the scattering states. Including excited states in \cref{eq:sigmaBSFeff_highT} would somewhat increase the contribution of the bound levels, but not change this conclusion.

The depletion of DM via BSF becomes more significant as the temperature of the universe lowers and approaches the binding energy~\cite{vonHarling:2014kha}; this is in part manifested by the exponential factor in \cref{eq:sigmaBSFeff_highT}. In this regime, the bound-state decay becomes comparable to or faster than ionisation, which is now exponentially suppressed, [cf.~\cref{eq:GammaIon}], and $\<\sigma_{\BSF} \vrel \>_{\eff}$ rapidly saturates to $\<\sigma_{\BSF} \vrel \>$ [cf.~\cref{eq:sigmaBSFeff}]. 
It is important to note that if $\sigma_{\BSF}$ is comparable to or exceeds  $\sigma_{\ann}$, then it is possible that the DM depletion via BSF contributes significantly to the effective annihilation rate even starting from temperatures that are larger than the binding energy by a factor of a few (cf.~refs.~\cite{vonHarling:2014kha,Harz:2018csl} and \cref{Sec:Results}).


\clearpage
\section{Wavefunctions and overlap integrals \label{Sec:Wavefunctions}}

\subsection{Scattering states \label{sec:Wavefunctions_ScatteringStates}}

The non-relativistic potential due to gluon and Higgs exchange for the scattering states is
\beq
V_{\mathsmaller{S}}(r) = - \dfrac{\agScatt}{r} - \dfrac{\ah}{r} \ e^{-\mh r} ,
\label{eq:ScatPotential2}
\eeq
where we will consider both positive and negative $\agScatt$, but only $\ah \geqslant 0$.

\paragraph{Wavefunctions.} 
The scattering states are described by a wavefunction $\f_{\vec k}(\vec r)$, parametrised by the continuous quantum number
\beq 
\vec{k} = \mu \vec{v}_{\rm rel} ,
\label{eq:ScattMomentumTransfer}
\eeq 
where $\vec{k}$ and $\vec{v}_{\rm rel}$ are the expectation values of the momenta of the interacting particles in the CM frame, and of their relative velocity. The strong coupling $\asScatt$ that determines
$\agScatt$ in \cref{eq:ScatPotential2} is evaluated at $Q=|{\bf k}|$.
The scattering state wavefunction obeys the Schr\"odinger equation
\beq 
\[-\frac{\nabla^2}{2\mu} + V_{\mathsmaller{S}} (\vec r)\] \f_{\vec k}(\vec r) 
= {\cal E}_{\vec k} \, \f_{\vec k}(\vec r) ,
\label{eq:SchrEq_Scatt}
\eeq 
with
\beq
{\cal E}_{\vec{k}} \equiv \frac{\vec{k}^2}{2\mu}  
= \frac{\mu \vrel^2}{2}> 0 .
\label{eq:EnergyScatt}
\eeq 
They are normalised according to
\beq
\int d^3r \: \f_{\vec k}^*(\vec r) 
\, \f_{\vec k'}(\vec r) 
= (2\p)^3 \d^3(\vec k - \vec k') . 
\label{eq:phi norm}
\eeq
To solve the Schr\"odinger equation numerically, we perform the separation of variables 
\beq
\f_{\vec k}(\vec r) = 
\sum_{\ell = 0}^\infty (2\ell+1) 
\left[ \frac{\x_{|\vec k|,\ell}^{} (k r)}{k r} \right]
\: P_{\ell} (\hat{\vec k} \cdot\hat{\vec r}) ,
\label{eq:phi var separ}
\eeq
and introduce the dimensionless radial space coordinate 
\beq 
\xS \equiv k r ,
\label{eq:xS}
\eeq 
such that the radial Schr\"{o}dinger equation reads
\beq
\x_{|\vec k|,\ell}''(\xS) + 
\left\{1 -\frac{\ell(\ell+1)}{\xS^2} 
+\frac{2}{\xS} \[\zetagScatt +\zetah \, e^{-(\zetah/\dh)\,\xS} \]  \right\} \x_{|\vec k|,\ell}^{} (\xS) = 0 ,
\label{eq:chi scatt diff}
\eeq
where we have used the parameters $\zetagScatt \equiv \agScatt/\vrel$, $\zetah \equiv \ah/\vrel$ and $\dh\equiv\mu \ah/\mh$ defined in \cref{eq:zeta,eq:dh}.  
At $\xS\to 0$, and for $\ell>0$,  the second term of \cref{eq:chi scatt diff} is dominated by the centrifugal contribution. In this region, the two independent solutions of \cref{eq:chi scatt diff}  scale as $\xS^{\ell+1}$ (regular) and $\xS^{-\ell}$ (irregular). We are interested in the regular solutions, which imply the boundary condition
\beq
\lim_{x\to 0} \x_{|\vec k|,\ell}'(\xS) = (\ell+1) \lim_{\xS\to 0} \, [\x_{|\vec k|,\ell}^{}(\xS)/\xS] \, .
\label{eq:BC derivative@0}
\eeq
The condition~\eqref{eq:BC derivative@0} will be valid also for $\ell = 0$. To fully specify the wavefunction $\x_{|\vec k|,\ell}^{}$, we shall also use the asymptotic behaviour at $\xS \to \infty$. At large $\xS$, the wavefunction $\x_{|\vec k|, \ell}^{}$ behaves as (see e.g.~ref.~\cite[chapter 7]{Sakurai_QMbook})
\beq
\x_{|\vec k|,\ell}^{} (\xS) 
\ \stackrel{\xS \to \infty}{\longrightarrow} \ 
\frac{1}{2i}\[e^{i (\xS+\d_\ell)} - e^{-i(\xS -\ell \pi)}\] \, ,
\nn 
\eeq
where the phase shifts $\delta_\ell$ depend on $\zetagScatt$, $\zetah$ and $\dh$. This implies that 
\beq
\left|\x_{|\vec k|,\ell}^{} (\xS) \right|^2 + 
\left|\x_{|\vec k|,\ell}^{} (\xS -\p/2) \right|^2 
= 1 \, .
\label{eq:SS WF large x norm}
\eeq

\paragraph{Coulomb limit.} 
In the limit $\dh \to \infty$, the Schr\"odinger \cref{eq:chi scatt diff} depends only on $\zetaScatt = \zetagScatt + \zetah$, and can be solved analytically,
\begin{subequations}
\beq
\chi_{|\vec k|,\ell}^C (\xS) 
= \sqrt{\SzeroC (\zetaScatt)} 
\ \frac{2^\ell}{(2\ell+1)! }
\ \frac{\Gamma(1+\ell-i\zetaScatt)}{\Gamma(1-i\zetaScatt)}
\ \xS^{1+\ell} 
\ e^{-i \xS}
\ {}_1F_1 (1+\ell+i\zetaScatt; \ 2\ell+2; \ 2i \xS) ,
\label{eq:chi_k Coul}
\eeq
where ${}_1F_1$ is the confluent hypergeometric function of the first kind, and 
\beq
\SzeroC (\zeta) \equiv \frac{2\pi\zeta}{1-e^{-2\pi\zeta}} .
\label{eq:S0Coul}
\eeq
The sum over the $\ell$ modes in \cref{eq:phi var separ} can be expressed in closed form such that the total scattering-state wave function in the Coulomb limit reads
\beq
\f_{\vec k}^C (\vec r) =  \lim_{\dh \to \infty} \f_{\vec k} (\vec r) =  
\ \sqrt{\SzeroC (\zetaScatt)} 
\ {}_1F_1 [i\zetaScatt;\ 1; \ i(kr - \vec k \cdot \vec r)] 
\ e^{i \vec k \cdot \vec r} . 
\label{eq:phi_Coul}
\eeq
\end{subequations}
Note that the limit $\dh \to 0$ (pure gluon-mediated potential) can be also obtained from the above with the substitution $\zetaScatt \to \zetagScatt$.

\paragraph{Sommerfeld factor $\boldsymbol{S_{0}}$ for $\boldsymbol{s}$-wave annihilation.} 
This is (see e.g.~\cite{Cassel:2009wt})
\beq
S_0 (\zetag,\zetah,\dh) = |\f_{\vec{k}} (r=0)|^2 = 
\lim_{\xS \to 0} \left|\frac{\x_{|\vec{k}|,0}^{}(\xS)}{\xS}\right|^2 .
\label{eq:S0 def}
\eeq
In the Coulomb limits $\dh \to 0$ and $\dh \to \infty$, the Sommerfeld factor reduces to
\begin{align} 
\lim_{\dh^{} \to 0} S_0 (\zetagScatt,\zetah,\dh) &= \SzeroC (\zetagScatt) ,
\\
\lim_{\dh^{} \to \infty} S_0 (\zetagScatt,\zetah,\dh) &= \SzeroC (\zetagScatt+\zetah) ,
\end{align} 
with $\SzeroC(\zeta)$ given in \cref{eq:S0Coul}. The features of $S_0$ and the impact of Higgs enhancement on the relic density have been discussed in detail in ref.~\cite{Harz:2017dlj} (see in particular figs.~2~and~3 therein).

\subsection{Bound states \label{sec:Wavefunctions_BoundStates}}

The non-relativistic potential for the bound states, due to gluon and Higgs exchange, is
\beq
V_{\mathsmaller{B}}(r) = - \dfrac{\agBound}{r} - \dfrac{\ah}{r} \ e^{-\mh r} .
\label{eq:V(r)_bound}
\eeq
Note the difference in $\ag$ with respect to \cref{eq:ScatPotential2}. While $\agScatt$ in the scattering potential is to be evaluated at the scale $Q = \mu \vrel$, $\agBound$ in the bound state potential is to be taken at the Bohr momentum scale, as will be discussed below.

\paragraph{Wavefunctions.} 
The Schr\"{o}dinger equation for bound states is
\beq
\[-\frac{\nabla^2}{2\mu} + V_{\mathsmaller{B}}(\vec r)\] \psi_{n\ell m}(\vec r) 
= {\cal E}_{n\ell} \, \psi_{n\ell m}(\vec r) \, ,
\label{eq:SchrEq Bound}
\eeq
where $\{n \ell m\}$ are the standard principal and angular momentum quantum numbers, and the wavefunctions are normalised as
\beq
\int d^3r \: \psi_{n\ell m}^*(\vec r) \, \psi_{n'\ell' m'}^{} (\vec r) 
= \d_{nn'} \d_{\ell\ell'} \d_{m m'} . \label{eq:psi norm}
\eeq
We define the generalised Bohr momentum
\beq
\kappa \equiv \mu (\agBound+\ah)  ,
\label{eq:kappa}
\eeq
such that the discrete binding energy levels of the system are
\beq
{\cal E}_{n\ell} 
\equiv - \gamma_{n\ell}^2 \times \frac{\kappa^2}{2\mu} 
= - \frac{1}{2} \mu \, (\agBound + \ah)^2 \, \gamma_{n\ell}^2 < 0,
\label{eq:E_nl}
\eeq
where the factor $\gamma_{n\ell}$ is determined numerically (see below) and parametrises the departure from the Coulomb limit $\dh\to \infty$ where $\gamma_{n\ell} = 1/n$. We perform the separation of variables 
\beq
\psi_{n\ell m}(\vec r) 
= \k^{3/2} \[ \frac{\x_{n\ell}^{}(\k r)}{\k r} \] Y_{\ell m} (\W_{\vec r}) \, .
\label{eq:psi var separ}
\eeq
Defining the dimensionless space coordinate 
\beq 
\xB \equiv \kappa r , 
\label{eq:xB}
\eeq 
we find that $\x_{n\ell}^{}$ is normalised as
\beq
\int_0^\infty d\xB \: |\x_{n\ell}^{}(\xB)|^2 = 1 ,
\label{eq:chi bound norm}
\eeq
and obeys the radial Schr\"{o}dinger equation
\beq
\x_{n\ell}''(\xB) + 
\left\{
- \gamma_{n\ell}^2  -\frac{\ell(\ell+1)}{\xB^2}
+\frac{2}{(1+\lambda)\xB} \, \( \lambda + \exp\[-\frac{\xB}{(1+\l)\dh}\]   \)
\right\} 
\x_{n\ell}^{} (\xB) = 0 \, .
\label{eq:chi bound diff}
\eeq
We recall that the dimensionless parameters $\lambda \equiv \agBound/\ah$ and $\dh \equiv \mu \ah / \mh$ have been defined in \cref{eq:dh,eq:lambda}. It follows that the bound-state wavefunctions $\chi_{n\ell}$ and the (normalised) energy eigenvalues $\gamma_{n\ell}$ depend on $\lambda$ and $\dh$ only. Assuming $\ah>0$, the necessary condition $\kappa > 0$ for bound states to exist implies $\lambda >-1$. This condition is necessary but not sufficient, as we shall see below.

\begin{figure}[th!]
\centering
\includegraphics[width=0.45\textwidth]{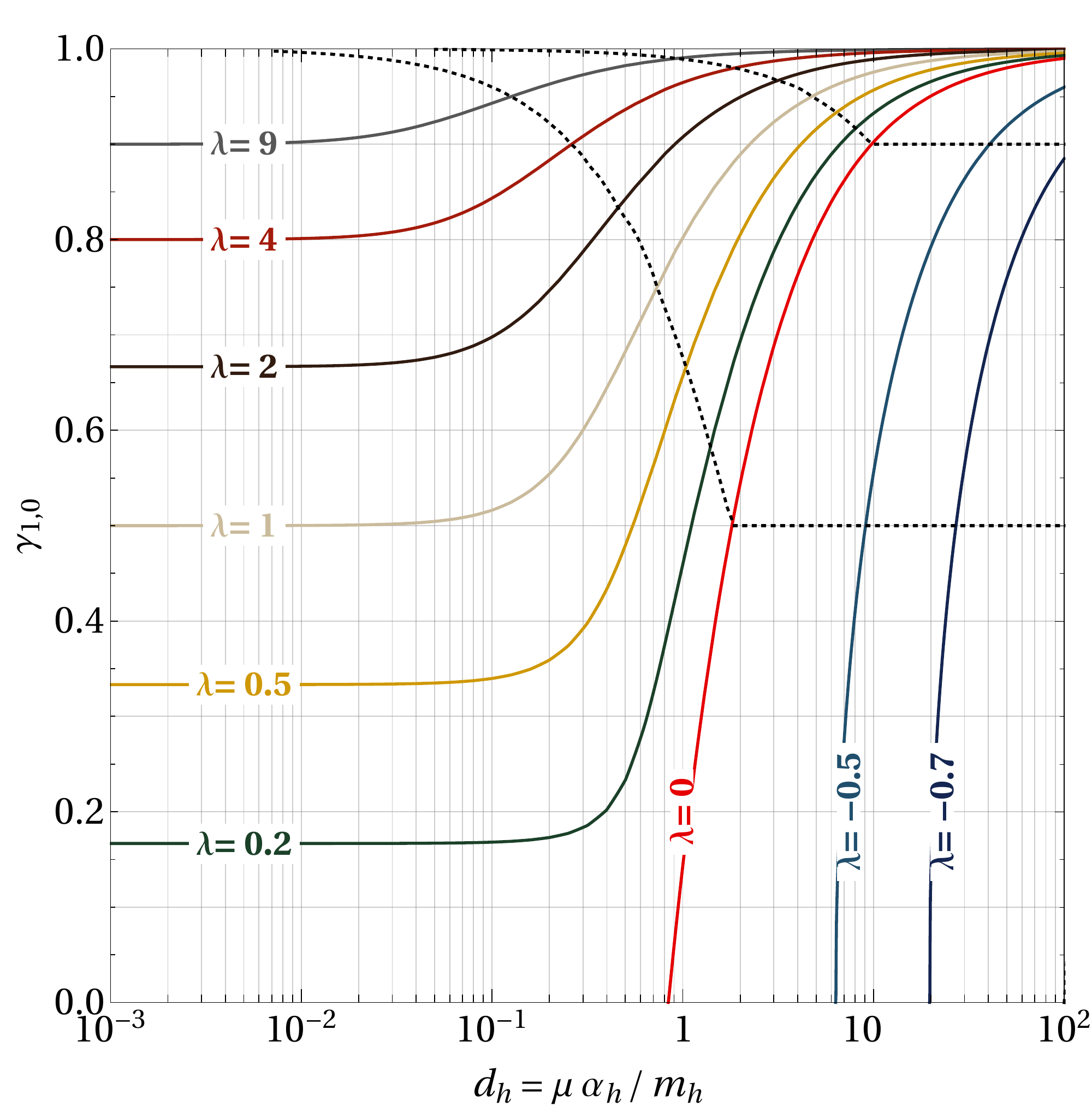}~~~~
\includegraphics[width=0.45\textwidth]{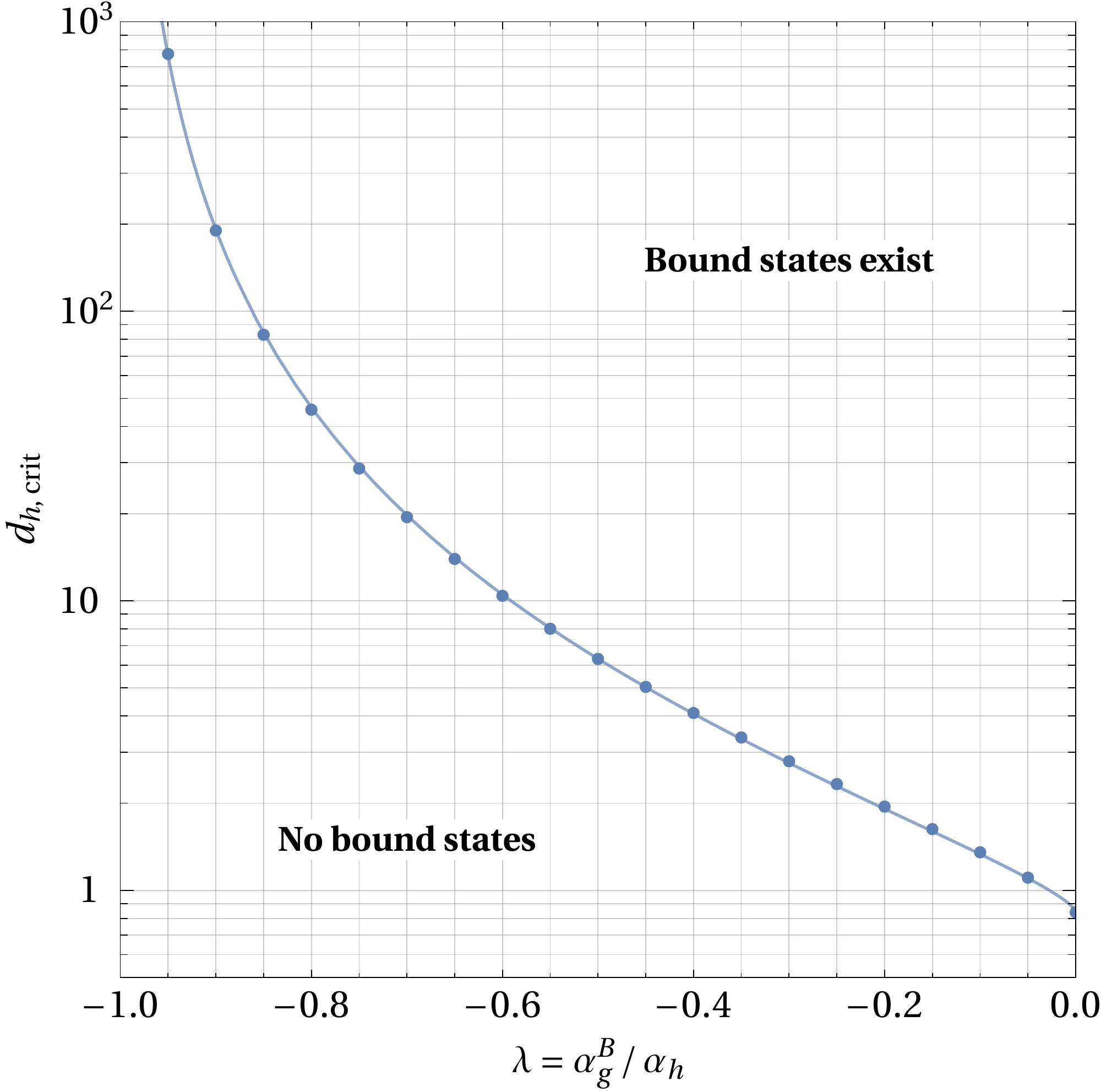}
\caption[]{\label{fig:gamma}
\emph{Left:} $\gamma_{1,0}(\lambda,\dh)$ parametrises the ground-state binding energy, ${\cal E}_{1,0} = -\gamma_{1,0}^2 \: \kappa^2/(2\mu)$, where $\kappa \equiv \mu (\agBound + \ah)$. The two \emph{dotted black lines} connect the points for which the binding energy is 50\% and 90\% between the two Coulomb limits, $\dh\to 0$ and $\dh\to\infty$. Comparing with a simple attractive Yukawa potential ($\lambda = 0$, \emph{red line}), the superposition of an attractive $(\lambda>0)$ or repulsive ($\lambda<0$) Coulomb component results in the Yukawa potential having a long-range effect for smaller and larger $\dh$ respectively. This is analogous to the features exhibited by the scattering states (see the $s$-wave Sommerfeld factor for a mixed Coulomb and Yukawa potential in ref.~\cite[fig.~2]{Harz:2017dlj}).
\\[1ex]
\emph{Right:} The critical values $d_{h, \rm crit}$ above which bound states exist in the mixed repulsive Coulomb and attractive Yukawa potential. The points are the numerical results, while the line shows the analytical fit \eqref{eq:dh_critical}.}
\end{figure}

As in the case of the scattering states, we are interested in the regular solutions of \cref{eq:chi bound diff}, which scale as
\beq
\lim_{\xB\to 0} \x_{n\ell}'(\xB) = (\ell+1) \lim_{\xB\to 0} \, [\x_{n\ell}(\xB)/\xB] .
\label{eq:BC derivative@0 bound}
\eeq
The discrete spectrum of eigenvalues $\gamma_{n\ell}^{} = \g_{n\ell} (\lambda, \dh)$ is then determined by requiring that $\x_{n\ell}^{}$ vanish at infinity,
\beq
\lim_{\xB\to\infty} \x_{n\ell}^{}(\xB) = 0 .
\label{eq:BC 0@infinity}
\eeq
We present some numerical results in \cref{fig:gamma} and discuss them next.

\paragraph{Coulomb limits.} 
In the left panel of \cref{fig:gamma}, the two plateaus indicate the two Coulomb limits, $\dh \to 0$ (pure gluon-mediated potential) and $\dh \to \infty$ (Coulomb limit of the Higgs-mediated potential), which allow for analytic solutions of \cref{eq:chi bound diff}. 
\begin{itemize}
\item
In the limit $\dh \to \infty$, the spectrum of eigenvalues and the radial wavefunction are 
\beq
\g_{n\ell}^C = \lim_{\dh \to \infty} \g_{n\ell} = 1/n \, ,
\label{eq:gamma nl Coul}
\eeq
\beq
\x_{n\ell}^C (\xB) 
= \frac{1}{n} \[\frac{(n-\ell-1)!}{(n+\ell)!}\]^{1/2}
\ e^{-\xB/n} \ (2\xB/n)^{\ell+1} \ L_{n-\ell-1}^{(2\ell+1)} (2\xB/n) ,
\label{eq:chi_bound_Coul}
\eeq
where $L_n^a$ are the generalised Laguerre polynomials of degree $n$, and we assume the normalisation condition $\int_0^\infty z^a e^{-z} L_n^{(a)}(z) L_m^{(a)}(z) dz = [\Gamma(n+a+1)/n!] \, \d_{n,m}$.

\item
In the limit $\dh \to 0$, bound states exist only for $\l>0$ ($\agBound>0$). In this case, the binding energy takes the Coulomb value for the gluon-generated potential only, 
\beq
\lim_{\dh\to 0} {\cal E}_{n\ell} = \frac{1}{n^2}\frac{\k_g^2}{2\mu},
\label{eq:Enl_CoulGluon}
\eeq
where $\kappa_g \equiv \mu \agBound$. This implies
\beq
\lim_{\dh \to 0} \g_{n\ell} = \frac{\l}{n(1+\l)} .
\label{eq:gamma_nl_CoulGluon}
\eeq

\end{itemize}

\paragraph{Existence condition.} 
If the Coulomb component of the potential is not attractive, $\l \leqslant 0$ (equivalently, $\agBound \leqslant 0$), the Yukawa potential has to be sufficiently long-range for bound states to exist, i.e.~$\dh$ has to be sufficiently large. In the right panel of \cref{fig:gamma}, we show the critical value $d_{h,\rm crit}(\lambda)$ such that for $\dh \geqslant d_{h,\rm crit}(\lambda)$ at least one bound level exists.  We find that in the range $-0.96 \leqslant \l \leqslant 0$, $d_{h,\rm crit}$ can be fitted by the analytical expression
\beq
d_{h, \rm crit} (\l) \simeq \frac{d_0 \ \exp \[(-\l)^{0.59}\]}{(1+\l)^{1.95}} ,
\label{eq:dh_critical}
\eeq
with an accuracy of better than $3\%$. The value $d_0 = d_{h, \rm crit} (\l=0) \simeq 0.84$ corresponds to a Yukawa potential only and is in agreement with the results of ref.~\cite[appendix A2]{Petraki:2016cnz}.

\subsection{Overlap integrals \label{sec:Wavefunctions_OverlapIntegrals}}

In order to evaluate the amplitudes for the BSF processes of interest, we need the following overlap integrals of the scattering and bound state wavefunctions~\cite{Harz:2018csl}
\begin{subequations}
\label{eq:OverlapIntegrals_def}
\label[pluralequation]{eqs:OverlapIntegrals_def}
\begin{align}
\boldsymbol{{\cal J}}_{\vec{k}, \{n\ell m\}}
(\vec{b}) 
&\equiv
\int \frac{d^3 p}{(2\pi)^3} \ \vec{p}
\ \tilde{\phi}_{\vec{k}} (\vec{p}+\vec{b}) 
\ \tilde{\psi}_{n\ell m}^* (\vec{p})
\nn \\
&=
i \int d^3 r 
\ \phi_{\vec{k}} (\vec{r}) 
\ [\nabla \psi_{n\ell m}^* (\vec r)] 
\ e^{-i \vec{b} \vec{r}} ,
\label{eq:Jcal_CoordSpace}
\\
\boldsymbol{{\cal Y}}_{\vec{k}, \{n\ell m\}}
&\equiv
8\pi\mu\asBound
\int \frac{d^3 p}{(2\pi)^3} \frac{d^3 q}{(2\pi)^3} 
\frac{\vec{q}-\vec{p}}{(\vec{q}-\vec{p})^4} 
\, \tilde{\phi}_{\vec{k}}  (\vec{q})
\, \tilde{\psi}_{n\ell m}^*(\vec{p}) 
\nn \\
&=
-i \mu \asBound \int d^3 r \, \psi_{n\ell m}^* (\vec r) \, \phi_{\vec{k}} (\vec{r}) \, \hat{\vec{r}} ,
\label{eq:Ycal_CoordSpace}
\end{align}
\end{subequations}
where we used the following Fourier transforms
\begin{subequations}
\label{eq:FourierTransformsWavefunctions}
\label[pluralequation]{eqs:FourierTransformsWavefunctions}
\begin{align}
\tilde{\phi}_{\vec{k}}^{}(\vec{q}) &= \int d^3 r \ \phi_{\vec{k}}^{}(\vec{r}) \ e^{-i\vec{q} \, \vec{r}} ,&
\phi_{\vec{k}}^{}(\vec{r}) &= \int \frac{d^3 q}{(2\pi)^3} \ \tilde{\phi}_{\vec{k}}^{}(\vec{q}) \ e^{i\vec{q} \, \vec{r}} ,
\\ 
\tilde{\psi}_{n\ell m}^{}(\vec{q}) &= \int d^3 r \ \psi_{n\ell m}^{}(\vec{r}) \ e^{-i\vec{q} \, \vec{r}} ,&
\psi_{n\ell m}^{}(\vec{r}) &= \int \frac{d^3 q}{(2\pi)^3} \ \tilde{\psi}_{n\ell m}^{}(\vec{q}) \ e^{i\vec{q} \, \vec{r}} .
\end{align}
\end{subequations} 
Note that  the integrals \eqref{eq:OverlapIntegrals_def} depend on the representations of the scattering and bound states. For simplicity we leave this dependence implicit in this section, but will denote it explicitly in \cref{Sec:BSF}, as 
$\boldsymbol{{\cal J}}_{\vec{k}, \{n\ell m\}}
^{\mathsmaller{\bf [R_\textsl{S},R_\textsl{B}] }}$
and
$\boldsymbol{{\cal Y}}_{\vec{k}, \{n\ell m\}}
^{\mathsmaller{\bf [R_\textsl{S},R_\textsl{B}] }}$,
where ${\bf R_\textsl{S}}$ and  ${\bf R_\textsl{B}}$ stand for the scattering-state and bound-state colour representations respectively.

For the numerical evaluation of the integrals \eqref{eq:OverlapIntegrals_def}, we express them in terms of the radial wavefunctions $\chi_{\vec{k},\ell}$ and $\chi_{n\ell}$ introduced in \cref{sec:Wavefunctions_ScatteringStates,sec:Wavefunctions_BoundStates}. The integrals $\boldsymbol{{\cal J}}_{\vec{k}, \{n\ell m\}}$ have been previously analysed in ref.~\cite[appendix B]{Petraki:2016cnz}. The leading order contribution to $\boldsymbol{{\cal J}}_{\vec{k}, \{n\ell m\}} (\vec{b})$ with $\vec{b} \propto \vec{P}_g$ is independent of $\vec{b}$~\cite{Petraki:2015hla,Petraki:2016cnz} such that in the following we denote $\boldsymbol{{\cal J}}_{\vec{k}, \{n\ell m\}} = \boldsymbol{{\cal J}}_{\vec{k}, \{n\ell m\}} ({\bf b}=0)$. Adapting the result to our notation for capture into the ground state, $\{n\ell m\} = \{100\}$, we arrive at\footnote{
Note that in ref.~\cite{Petraki:2016cnz}, a factor of $(-i)$ has been missed in going from eq.~(B.8b) to eq.~(B.9b). While this factor is not important for the computations of ref.~\cite{Petraki:2016cnz}, it is important here, since it determines the relative phase of $\boldsymbol{{\cal J}}_{\vec{k}, \{100\}}$ and $\boldsymbol{{\cal Y}}_{\vec{k}, \{100\}}$.
} 
%
\begin{subequations}
\label{eq:IntegralsCal_NonCoulomb}
\label[pluralequation]{eqs:IntegralsCal_NonCoulomb}
\beq
\boldsymbol{{\cal J}}_{\vec{k}, \{100\}}
\simeq  -i \, \hat{\vec{k}}  
\ \frac{\kappa}{k} 
\ \sqrt{\frac{4\pi}{\kappa}}
\int_0^\infty d\xB 
\[ \frac{\chi_{1,0}^* (\xB)}{\xB} - \frac{d\chi_{1,0}^* (\xB)}{d\xB} \] 
\ \chi_{|\vec{k}|,1}^{} (\xB / \zetaBound)  ,
\label{eq:Jcal_LO}
\eeq
where we recall from \cref{eq:zetaBound} that $\zetaBound = \kappa / k$. Similarly, the integral \eqref{eq:Ycal_CoordSpace} becomes
\beq
\boldsymbol{{\cal Y}}_{\vec{k}, \{100\}} = 
-i \, \frac{\mu \asBound}{k \sqrt{4\pi \kappa}}
\ \sum_{\ell = 0}^{\infty} \ (2\ell+1)
\ \int_0^\infty d\xB \ \chi_{1,0}^*(\xB) 
\ \chi_{|\vec{k}|,\ell}^{}(\xB / \zetaBound)
\ \int d\W \ \hat{\vec{r}} \ P_\ell ( \hat{\vec{k}} \cdot \hat{\vec{r}}) .
\nn 
\eeq
The angular integral above is (see e.g.~ref.~\cite[eq.~(B.7b) with $\vec{b} \to 0$]{Petraki:2016cnz})
\beq
\int d\W \ \hat{\vec{r}} \ P_\ell ( \hat{\vec{k}} \cdot \hat{\vec{r}}) =
\delta_{\ell,1} \ (4\pi/3) \ \hat{\vec{k}}  .
\nn
\eeq
Thus, we arrive at the final expression
\beq
\boldsymbol{{\cal Y}}_{\vec{k}, \{100\}} = 
-i \, \hat{\vec{k}} \ \frac{\mu \asBound}{k}
\sqrt{\frac{4\pi}{\kappa}}
\ \int_0^\infty d\xB \ \chi_{1,0}^*(\xB) 
\ \chi_{|\vec{k}|,1}^{}(\xB / \zetaBound)
.
\label{eq:Ycal_LO}
\eeq
\end{subequations}
We shall use \cref{eqs:IntegralsCal_NonCoulomb} to evaluate the BSF cross-sections in \cref{Sec:BSF}.

\paragraph{Coulomb limit.} 

In the limit $\dh \to \infty$, the scattering and bound state wavefunctions, \eqref{eq:phi_Coul} and \eqref{eq:chi_bound_Coul} are
\begin{subequations}
\label{eq:Wavefunctions_Coulomb}
\label[pluralequation]{eqs:Wavefunctions_Coulomb}
\begin{align}
\phi_{\vec k}^C (\vec r) &=  
\ \sqrt{\SzeroC (\zetaScatt)} 
\ {}_1F_1 [i\zetaScatt; \ 1; \ i(kr - \vec k \cdot \vec r)] 
\ e^{i \vec k \cdot \vec r} ,
\\	
\psi_{\{100\}}^C (\vec r) &=  \sqrt{\frac{\kappa^3}{\pi}} \ e^{-\kappa r} ,
\end{align}
\end{subequations}
where $\zetaScatt \equiv \zetagScatt + \zetah$ [cf.~\cref{eq:zetaScatt}], 
$\SzeroC (\zetaScatt) \equiv 2\pi \zetaScatt/(1-e^{-2\pi\zetaScatt})$ [cf.~\cref{eq:S0Coul}], 
and ${}_1F_1$ is the confluent hypergeometric function of the first kind. 
Following refs.~\cite{Pearce:2013ola,Petraki:2015hla,Petraki:2016cnz,Harz:2018csl}, we derive the Coulomb limit of the overlap integrals using the identity~\cite{AkhiezerMerenkov_sigmaHydrogen}
\beq
\int d^3r 
\ \frac{ e^{ i(\vec{k}-\vec{b}) \cdot \vec{r} - \kappa r }  }{4\pi r}
\ {}_1F_1 [ i \zetaScatt, 1, i (kr - \vec{k} \cdot \vec{r}) ] 
= \dfrac{[\vec{b}^2 + (\kappa - i k)^2]^{-i \zetaScatt}}
{[(\vec{k}-\vec{b})^2 +\kappa^2]^{1-i\zetaScatt}}
\equiv f_{\vec{k},\vec{b}} (\kappa) .
\label{eq:Identity_HyperGeo}
\eeq
\Cref{eqs:OverlapIntegrals_def} become~\cite[appendix~B]{Harz:2018csl}
\begin{subequations}
\label{eq:IntegralsCal_Coulomb}
\label[pluralequation]{eqs:IntegralsCal_Coulomb}
\begin{align}
\boldsymbol{{\cal J}}_{\vec{k}, \{100\}}^C (\vec{b})
&= \kappa
\sqrt{16 \pi \kappa^3 \, \SzeroC (\zetaScatt)} 
\ [ \nabla_{\vec{b}} f_{\vec{k},\vec{b}}(\kappa) ]
\label{eq:Jcal_Coulomb} ,
\\
\boldsymbol{{\cal Y}}_{\vec{k}, \{100\}}^C 
&= \mu \asBound 
\sqrt{16 \pi \kappa^3 \, \SzeroC (\zetaScatt)} 
\ [ \nabla_{\vec{b}} f_{\vec{k},\vec{b}}(\kappa) ]_{\vec{b}=0} 
\label{eq:Ycal_Coulomb} ,
\end{align}
where
\beq
\[\nabla_{\vec b} f_{\vec k, \vec b}(\kappa)\]_{\vec b = \vec 0} 
= \hat{\vec{k}} \ \frac{2(1-i\,\zetaScatt)}{k^3} \ 
\frac{\exp\[-2 \zetaScatt \, {\rm arccot} (\zetaBound) \]}{(1+\zetaBound^2)^2} ,
\label{eq:df/db@b=0}
\eeq
\end{subequations}
and we used again $\kappa/k = \zetaBound$. As noted before, it suffices to evaluate $\boldsymbol{{\cal J}}_{\vec{k}, \{100\}} (\vec{b})$ at $\vec{b}=\vec{0}$~\cite{Petraki:2015hla,Petraki:2016cnz} such that $\boldsymbol{{\cal Y}}_{\vec{k}, \{100\}}$ and $\boldsymbol{{\cal J}}_{\vec{k}, \{100\}}$ are related as
\beq
\dfrac{ \boldsymbol{{\cal Y}}_{\vec{k}, \{100\}}^C }
{ \boldsymbol{{\cal J}}_{\vec{k}, \{100\}}^C }
= \dfrac{\asBound}{\ah+\agBound} .
\label{eq:YtoJratio_Coul}
\eeq
with the squared amplitude of $\boldsymbol{{\cal J}}_{\vec{k}, \{100\}}^C$ being
\beq
|\boldsymbol{{\cal J}}_{\vec{k}, \{100\}}^C|^2 = 
\frac{2^6 \pi}{\kappa} 
\ \SzeroC (\zetaScatt) \, (1+\zetaScatt^2) 
\ \frac{\zetaBound^6 \,  \exp\[-4 \zetaScatt \, {\rm arccot} (\zetaBound) \]}{(1+\zetaBound^2)^4} .
\label{eq:JcalSquare_Coul}
\eeq
In \cref{sec:BSF_CrossSections_Coulomb}, we shall use \cref{eq:YtoJratio_Coul,eq:JcalSquare_Coul} to obtain analytical expressions for the BSF cross-sections in the Coulomb limit.

\clearpage
\section{Bound-state formation cross-sections \label{Sec:BSF} }

Having outlined how to evaluate the scattering and bound state wavefunctions, we have the tools at hand to calculate the cross-sections for the BSF processes discussed in \cref{sec:Model_BoundStates}. 

\subsection{Amplitudes \label{sec:BSF_Amplitude}}

The radiative capture via gluon emission has been computed in ref.~\cite{Harz:2018csl} for general groups and representations, in terms of the overlap integrals \eqref{eq:OverlapIntegrals_def}. Adapting the results to our model, we find the following colour-averaged squared amplitudes:
\begin{subequations}
\label{eq:MatrixElements}
\label[pluralequation]{eqs:MatrixElements}
\begin{align}
(X + X^\dagger)_{\bf [8]} \to {\cal B}(X X^\dagger)_{\bf [1]} + g:& \nn
\\[1ex]
\frac{1}{3^2}
\abs{ \boldsymbol{{\cal M}}_{\vec{k}\to \{n\ell m\}}^{\bf [8] \to [1]} }^2 
=& \( \frac{2^{7}\pi\asBSF \, M^2}{3^3 \mu} \) \times 
\abs{
\boldsymbol{{\cal J}}_{\vec{k}, \{n\ell m\}}^{\bf [8,1]}  +
\frac{3}{2} \ \boldsymbol{{\cal Y}}_{\vec{k}, \{n\ell m\}}^{\bf [8,1]}
}^2 ,
\label{eq:MBSFsquared_BoundStateSinglet}
\\[2ex]
(X + X^\dagger)_{\bf [1]} \to {\cal B}(X X^\dagger)_{\bf [8]} + g:& \nn
\\[1ex]
\frac{1}{3^2}
\abs{ \boldsymbol{{\cal M}}_{\vec{k}\to \{n\ell m\}}^{\bf [1] \to [8]} }^2 
=& \( \frac{2^{7}\pi\asBSF \, M^2}{3^3\mu} \) \times 
\abs{\boldsymbol{{\cal J}}_{\vec{k}, \{n\ell m\}}^{\bf [1,8]}
-\frac{3}{2} \ \boldsymbol{{\cal Y}}_{\vec{k}, \{n\ell m\}}^{\bf [1,8]} }^2  ,
\label{eq:MBSFsquared_ScatteringStateSinglet}
\\[2ex]
(X + X^\dagger)_{\bf [8]} \to {\cal B}(X X^\dagger)_{\bf [8]} + g:& \nn
\\[1ex]
\frac{1}{3^2}
\abs{ \boldsymbol{{\cal M}}_{\vec{k}\to \{n\ell m\}}^{\bf [8] \to [8]}  }^2 
=& \( \frac{2^{7} \pi\asBSF \, M^2}{3^3\mu} \) \times
\frac{5}{2}
\abs{ \boldsymbol{{\cal J}}_{\vec{k}, \{n\ell m\}}^{\bf [8,8]} }^2 ,
\label{eq:MBSFsquared_Total}
\end{align}
\end{subequations}
where $\asBSF \equiv \as (Q = |\vec{P}_g|)$ is the strong coupling evaluated at momentum transfer equal to the momentum of the emitted gluon; this factor  arises from the gluon emission vertices in the diagrams of \cref{fig:FeynmanDiagrams_BSF_RadiativeVertices}. $|\vec{P}_g|$ is found from energy-momentum conservation to be 
$|\vec{P}_{g}| = {\cal E}_{\vec{k}} - {\cal E}_{n\ell} 
=  (\mu/2) [ (\ah+\agBound)^2 \gamma_{n\ell}^2 + \vrel^2 ]$. We recall that in \cref{eqs:MatrixElements}, the superscripts on the overlap integrals denote, in order, the scattering-state and bound-state colour representations.

\subsection{Cross-sections for capture into the ground state \label{sec:BSF_CrossSections}}

The differential cross-section for the radiative capture into bound states is
\beq
\vrel \frac{d \sigma_{\vec{k}\to \{n\ell m\}} }{d\Omega} = 
\frac{|\vec{P}_{g}|}{64 \pi^2 M^2 \mu}
\( 
|\boldsymbol{{\cal M}}_{\vec{k}\to \{n \ell m\}}|^2 
- |\hat{\vec{P}}_g \cdot \boldsymbol{{\cal M}}_{\vec{k}\to \{n \ell m\}}|^2
\) .
\label{eq:sigmaBSF_differential}
\eeq
For capture into the ground state, the leading-order amplitude is 
$\boldsymbol{{\cal M}}_{\vec{k}\to \{100\}} \propto \vec{k}$, as seen from the expressions \eqref{eq:IntegralsCal_NonCoulomb} for the overlap integrals. Thus, carrying out the angular integration, we arrive at 
\beq
\sigma_{\vec{k}\to \{100\}} \vrel = 
\frac{(1+\zetaBound^2 \g_{1,0}^2)\vrel^2}{48 \pi M^2}
\ |\boldsymbol{{\cal M}}_{\vec{k}\to \{100\}}|^2 ,
\label{eq:sigmaBSF_general}
\eeq
where we recall that $\zetaBound \equiv (\ah + \agBound)/\vrel$ [cf.~\cref{eq:zetaBound}].
Using the amplitudes \eqref{eq:MatrixElements} and the overlap integrals \eqref{eq:IntegralsCal_NonCoulomb}, we find from \cref{eq:sigmaBSF_general} the coloured-averaged radiative capture cross-sections to be
\begin{subequations}
\label{eq:sigmaBSF_GeneralGroup}
\label[pluralequation]{eqs:sigmaBSF_GeneralGroup}
\beq
\sigma_{\vec{k}\to \{100\}} \vrel = 
\frac{\pi \asBSF (\ah +\agBound)}{\mX^2}
\ \frac{2^7}{3^4}
\times ( 1 + \zetaBound^2 \gamma_{1,0}^2 )
\ {\cal H}  ,
\tag{\ref{eq:sigmaBSF_GeneralGroup}}
\eeq
with
\begin{align} 
{\cal H}_{\mathsmaller{\bf [8] \to [1]}} &\equiv 
\left |
\int_0^\infty d\xB 
\ \chi_{\vec{k},1}^{\mathsmaller{\bf [8]}} \(\frac{\xB }{\zetaBoundSinglet}\)	
\[ 
\dfrac{ \chi_{1,0}^{\mathsmaller{\bf [1]}} (\xB) }{\xB} 
-\dfrac{d\chi_{1,0}^{\mathsmaller{\bf [1]}} (\xB)}{d\xB}
+\dfrac{3}{2} \( \dfrac{\asBoundSinglet}{\ah + \agBoundSinglet} \) 
\chi_{1,0}^{\mathsmaller{\bf [1]}} (\xB) 
\]^* 
\right|^2,
\label{eq:H_OctetToSinglet} \\
{\cal H}_{\mathsmaller{\bf [1] \to [8]}} &\equiv 
\left| \int_0^\infty d\xB 
\ \chi_{\vec{k},1}^{\mathsmaller{\bf [1]}} 
\(\frac{\xB}{\zetaBoundOctet}\)
\[ 
\dfrac{ \chi_{1,0}^{\mathsmaller{\bf [8]}} (\xB) }{\xB} 
- \dfrac{d\chi_{1,0}^{\mathsmaller{\bf [8]}} (\xB)}{d\xB} 
- \dfrac{3}{2} \( \dfrac{\asBoundOctet}{\ah + \agBoundOctet} \) 
\chi_{1,0}^{\mathsmaller{\bf [8]}} (\xB) \]^* 
\right|^2 ,
\label{eq:H_SingletToOctet} \\
{\cal H}_{\mathsmaller{\bf [8] \to [8]}}
&\equiv  \frac{5}{2}
\abs{\int_0^\infty d\xB 
\ \chi_{\vec{k},1}^{\mathsmaller{\bf [8]}} \(\frac{\xB}{\zetaBoundOctet}\)
\[ \dfrac{\chi_{1,0}^{\mathsmaller{\bf [8]}} (\xB)}{\xB} 
- \dfrac{d\chi_{1,0}^{\mathsmaller{\bf [8]}} (\xB)}{d\xB} \]^* 
}^2 .
\label{eq:H_OctetToOctet}
\end{align}
\end{subequations}
We use \cref{eqs:sigmaBSF_GeneralGroup} to evaluate numerically the BSF cross-sections.

\subsection{Coulomb limit\label{sec:BSF_CrossSections_Coulomb}}

In the limit $\dh \to \infty$, we may use the analytical expressions \eqref{eq:YtoJratio_Coul} and \eqref{eq:JcalSquare_Coul} for the overlap integrals. We find the colour-averaged BSF cross-sections to be
\begin{subequations}
\label{eq:sigmaBSF_Coulomb_All}
\label[pluralequation]{eqs:sigmaBSF_Coulomb_All}
\beq
\sigma_{\BSF}^C \vrel = \frac{\pi \asBSF (\ah +\agBound)}{\mX^2}
\ \frac{2^{11}}{3^4}  
\ f_c \times S_{\BSF}^C  (\zetaScatt, \zetaBound) ,
\label{eq:sigmaBSF_Coulomb}
\eeq
where $f_c$ is a numerical factor that depends on the transition,
\beq
f_c = \left\{
\begin{alignedat}{100}
&\[1 + \frac{3}{2} \(\frac{\asBoundSinglet}{\ah + 4\asBoundSinglet/3} \) \]^2,&
\quad
&{\bf [8] \to [1],}& 
\\
&\[1 - \frac{3}{2} \(\frac{\asBoundOctet}{\ah -\asBoundOctet/6} \) \]^2,&
\quad
&{\bf [1] \to [8],}& 
\\
&5/2&
\quad
&{\bf [8] \to [8],}& 
\end{alignedat}
\right.
\label{eq:fc}
\eeq
and
\beq
S_{\BSF}^C (\zetaScatt, \zetaBound) \equiv
\( \frac{2\pi \zetaScatt}{1-e^{-2\pi \zetaScatt}} \)
(1+\zetaScatt^2)
\[ \frac{\zetaBound^4 \ \exp \[ - 4 \, \zetaScatt \ {\rm arccot} (\zetaBound) \]}{(1+\zetaBound^2)^3}  \] .
\label{eq:SBSF_Coulomb}
\eeq
\end{subequations}
We emphasise that, even when not explicitly denoted above, the couplings $\asBSF$, $\asBound$, $\agBound$, $\agScatt$, and therefore also 
$\zetaScatt \equiv (\agScatt + \ah)/\vrel$ and 
$\zetaBound \equiv (\agBound + \ah)/\vrel$, depend on the colour representations of the initial and final states, and are thus different for every transition.

The function $S_{\BSF}^C  (\zetaScatt, \zetaBound)$ encapsulates all the velocity dependence of $\sigma_{\BSF}^C \vrel$. The first two factors in \cref{eq:SBSF_Coulomb} arise solely from the scattering state wavefunction and correspond to the Sommerfeld effect on $p$-wave processes in the Coulomb limit,  $S_1^C (\zetaScatt) = [2\pi \zetaScatt/ (1-e^{-2\pi \zetaScatt})] (1+\zetaScatt^2)$. The factors inside the square brackets in \cref{eq:SBSF_Coulomb} arise from the convolution of the scattering and bound state wavefunctions with the radiative vertices. Comparing \cref{eq:sigmaBSF_GeneralGroup,eq:sigmaBSF_Coulomb_All}, yields $S_{\BSF}^C (\zetaScatt,\zetaBound) = [(1+\zetaBound^2) / (16 f_c)] 
\, \lim_{\dh\to \infty} {\cal H}$.

Two limiting cases of \cref{eqs:sigmaBSF_Coulomb_All} are of particular interest:
\bit
\itemsep 0pt
\item
For $\ah \ll \asBound$, only colour-singlet bound states exist, hence the transition ${\bf [8] \to [1]}$ is the only possible, with $f_c = 289/64$. The corresponding cross-section is
\beq
\sigma_{\BSF}^{\bf [8] \to [1]} \vrel = 
\ \frac{2^7 17^2}{3^5} 
\ \frac{\pi \asBSFSinglet \asBoundSinglet}{\mX^2}
\times S_{\BSF}^C  (\zetaScattOctet , \zetaBoundSinglet) .
\label{eq:sigmaBSF_alphahSmall}
\eeq

\item
For $\ah \gg \asBound$, the colour factors \eqref{eq:fc} reduce to
\beq
f_c \simeq \left\{
\begin{alignedat}{100}
&1,&
\qquad
&{\bf [8] \to [1],}& 
\\
&1,&
\qquad
&{\bf [1] \to [8],}& 
\\
& 5/2,&
\qquad
&{\bf [8] \to [8].}&
\end{alignedat}
\right.
\label{eq:fc_SU3fundamental_alphahLarge}
\eeq

\eit

\subsection[The running of $\as$]{The running of $\boldsymbol{\as}$ \label{sec:BSF_AlphaRunning}}

\begin{figure}[t!]
\centering
\includegraphics[height=0.44\textwidth]{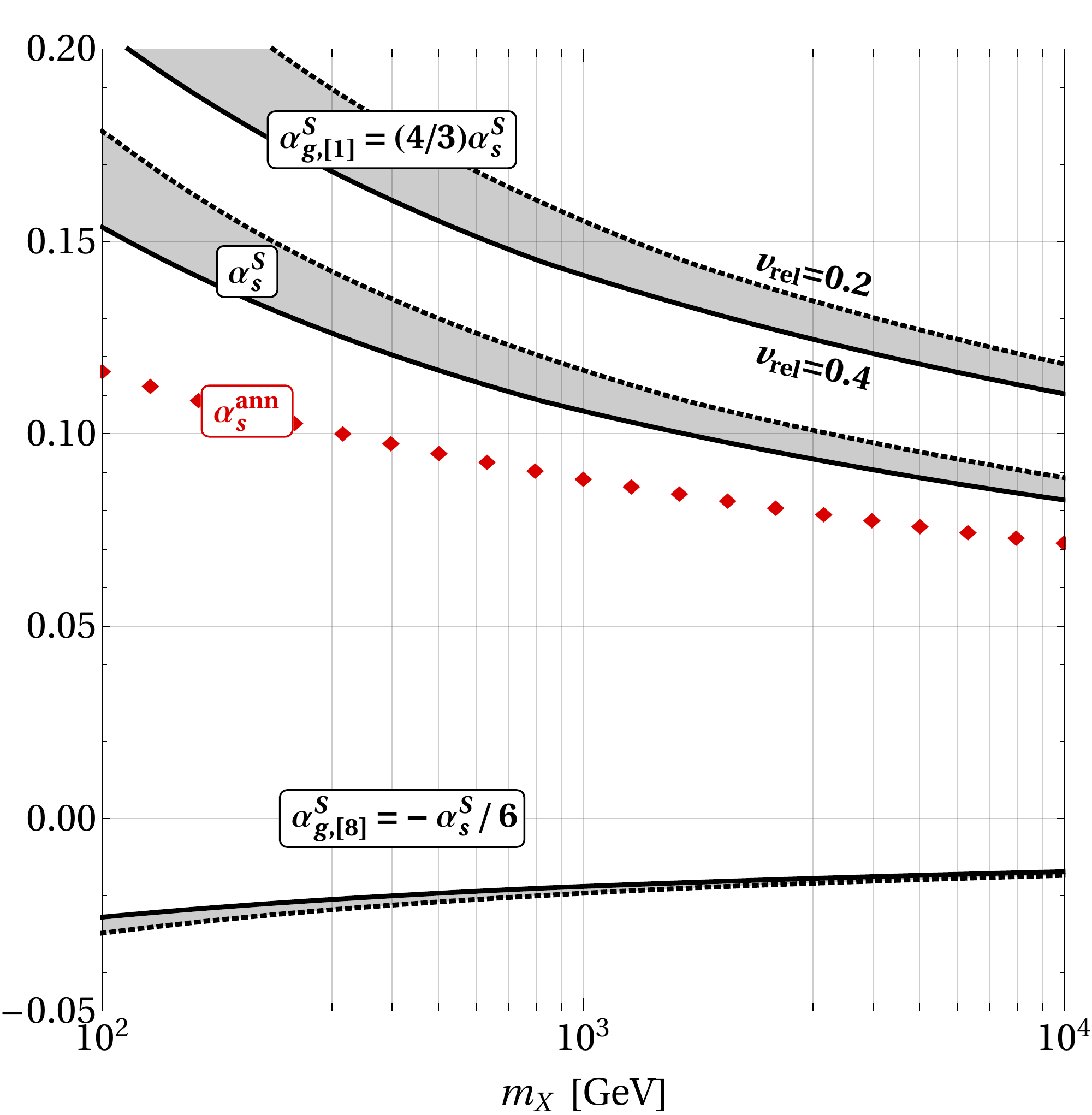}~~~
\includegraphics[height=0.44\textwidth]{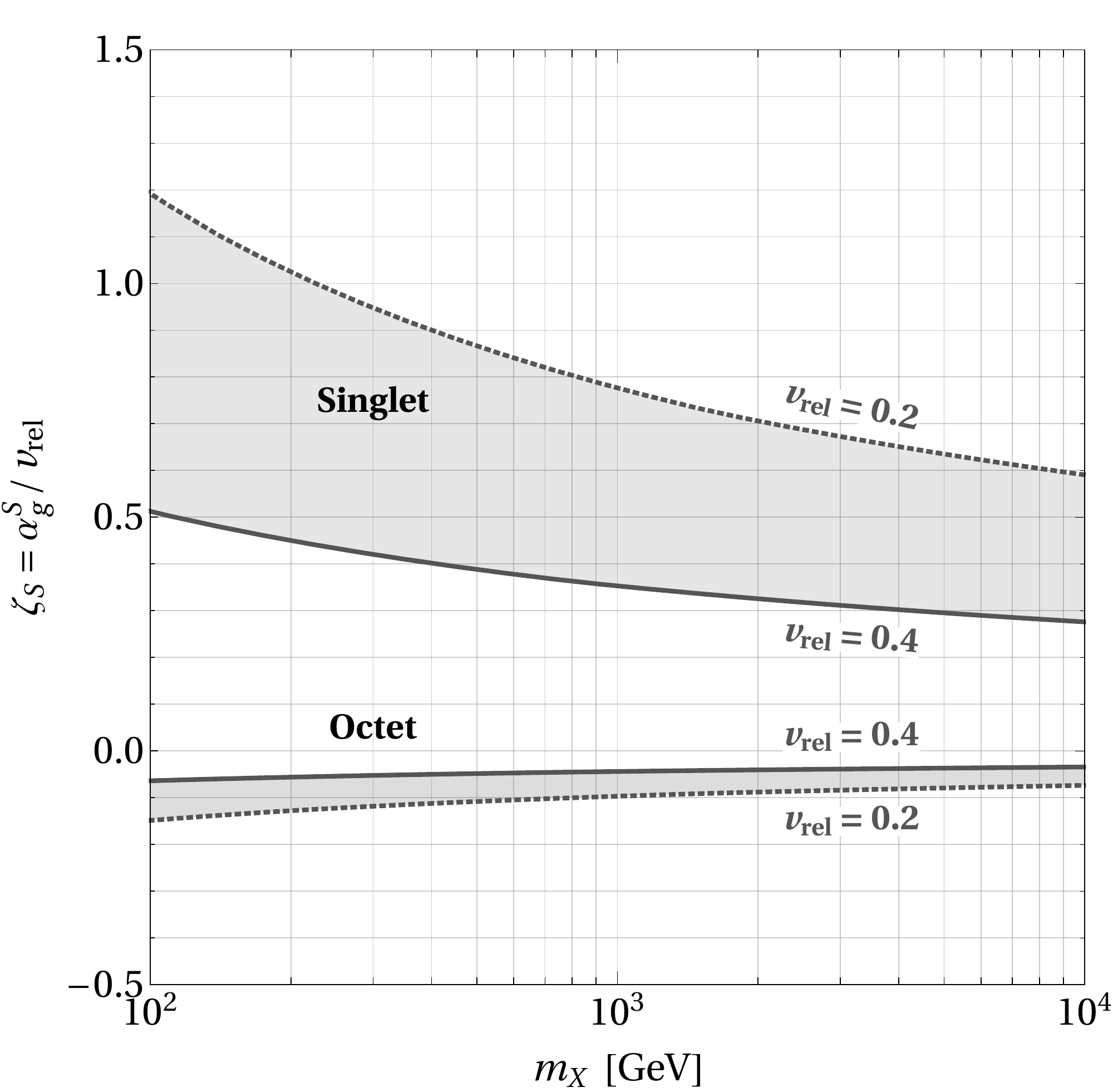}
\\
\includegraphics[height=0.44\textwidth]{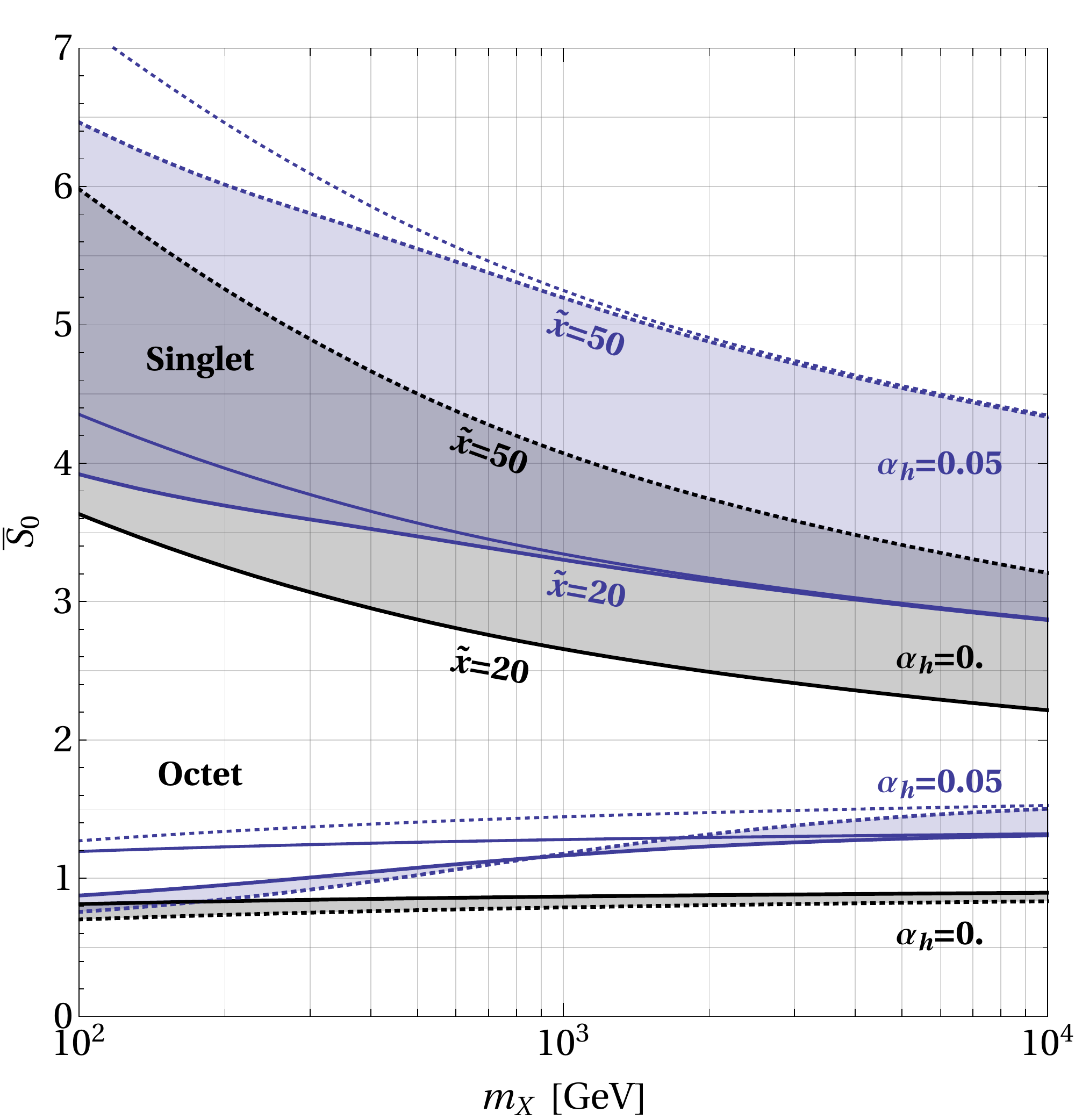}~~~~~~
\includegraphics[height=0.44\textwidth]{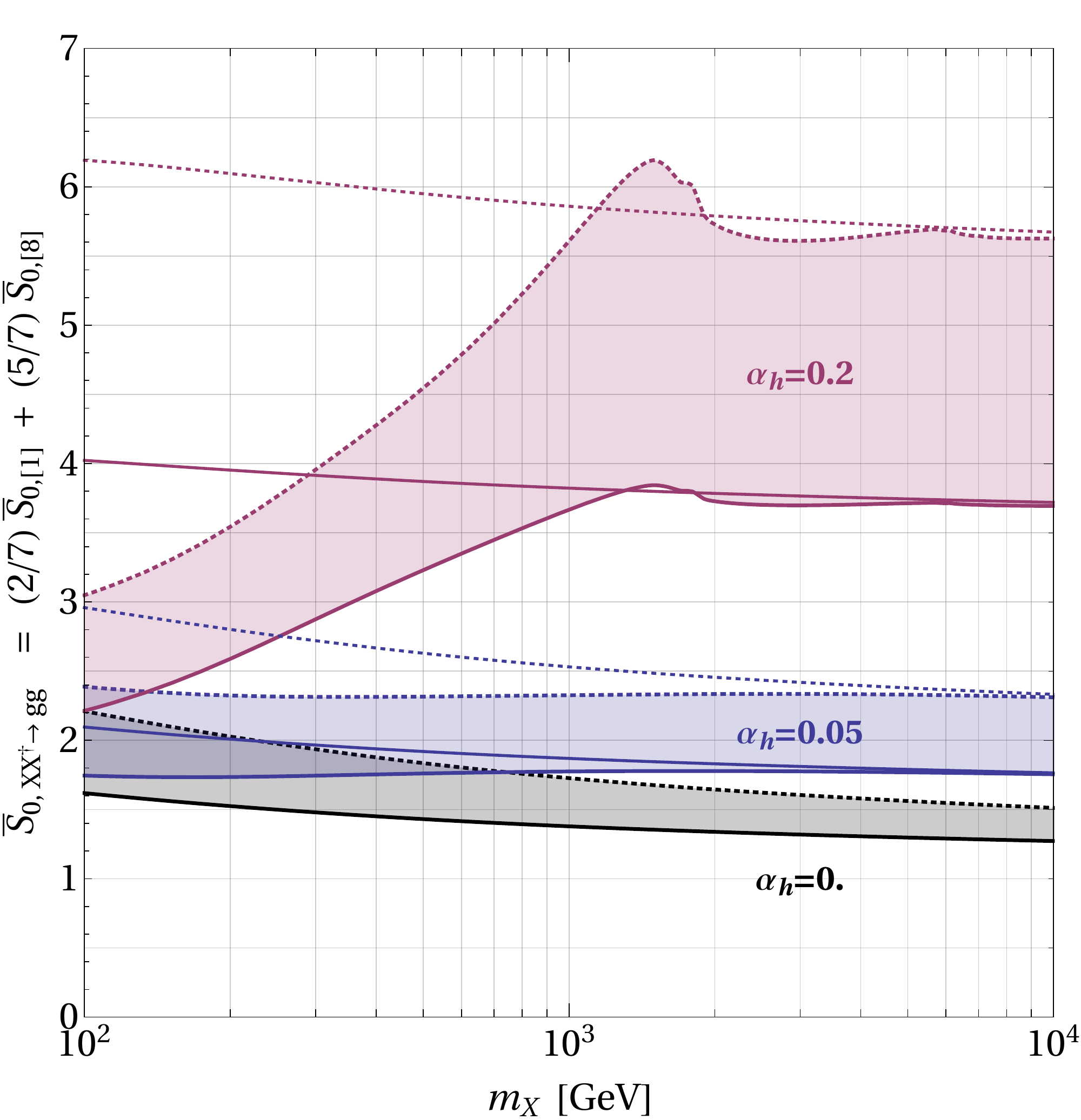}

\caption[]{\label{fig:AlphaRunning_ScattStates}
The effect of the running of $\alpha_s$ on the scattering states and the Sommerfeld enhancement/suppression of the annihilation processes. \\[1ex]
\emph{Upper left:} $\asScatt$ evaluated at the average momentum exchange in the scattering state, $Q = \mu \vrel$; for the colour-singlet state $\agScattSinglet = (4/3)\asScatt$, while for the colour-octet state $\agScattOctet = -\asScatt/6$ within the indicative velocity range $0.2 < v_{\rm rel} < 0.4$ that is typical during the DM freeze-out. For comparison, we show $\asAnn$ occurring in the perturbative annihilation processes taken at a momentum transfer $Q = \mX$ (\emph{red diamonds}).\\[1ex]
\emph{Upper right:}
The parameter $\zeta_g = \agScatt / v_{\rm rel}$ that determines the Sommerfeld effect due to gluon exchange only, for the singlet and octet configurations.\\[1ex]
\emph{Lower left:} 
The coloured bands show the thermally-averaged Sommerfeld factor for $s$-wave annihilation processes, $\bar{S}_0$, evaluated within the indicative range $20 < \tilde{x} \equiv \mX / T < 50$, during which the DM relic density is mostly determined. The Higgs exchange enhances the attraction / reduces the repulsion between the scattering-state particles  (\emph{blue bands: $\alpha_h = 0.05$}) with respect to gluon-only exchange (\emph{gray bands: $\alpha_h=0$}). The thin solid and dotted lines mark the Coulomb limit, $m_h \to 0$, which becomes a good approximation for large enough $\mX$.\\[1ex]
\emph{Lower right:} 
The thermally-averaged total Sommerfeld factor for annihilation into gluons. The Higgs exchange leads to a significant enhancement (\emph{blue band: $\alpha_h = 0.05$, purple band: $\alpha_h = 0.2$}) with respect to gluon-only exchange (\emph{gray band: $\alpha_h=0$}). 
}
\end{figure}

\begin{figure}[t!]
\centering
\includegraphics[height=0.48\textwidth]{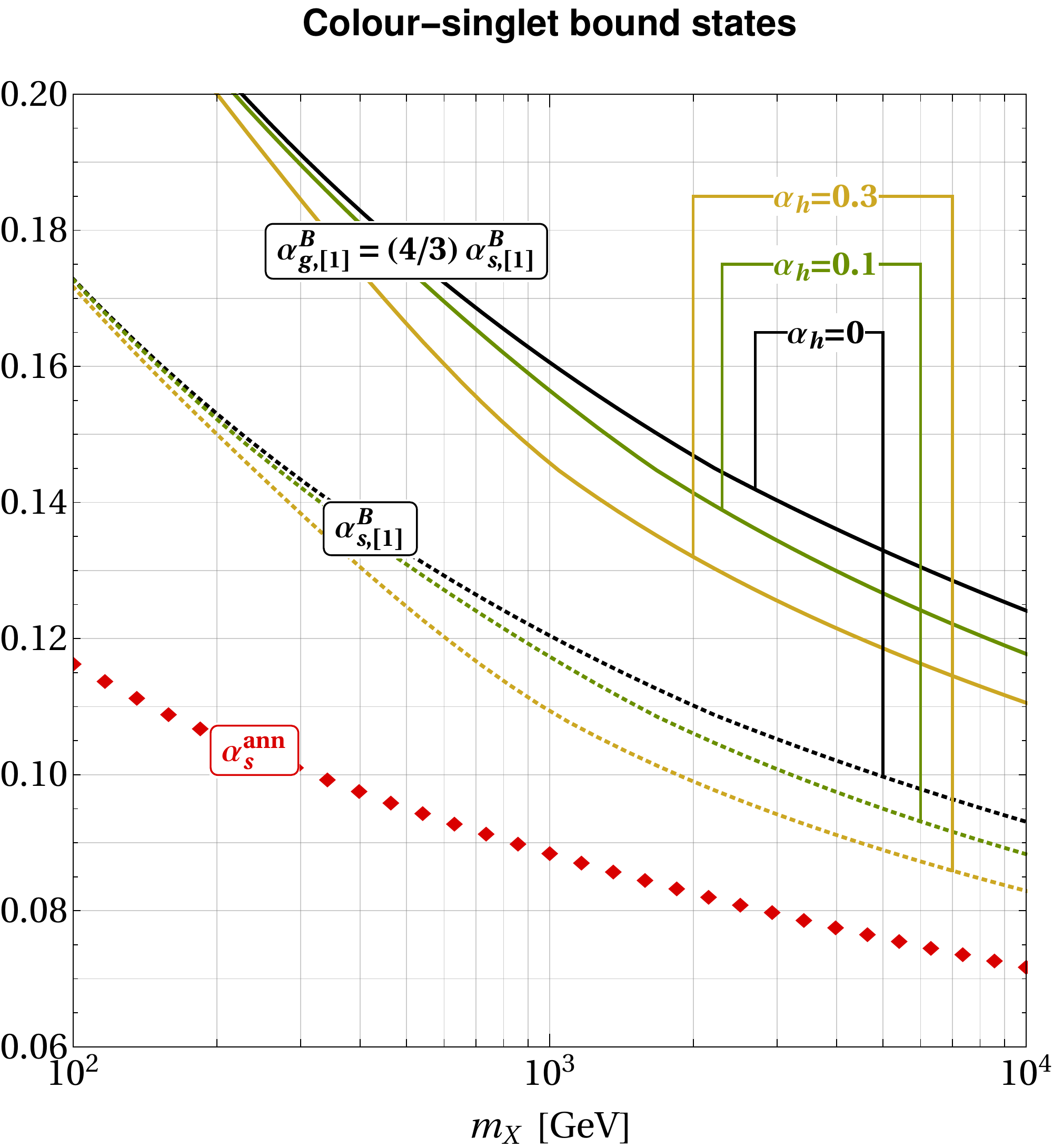}~~~
\includegraphics[height=0.48\textwidth]{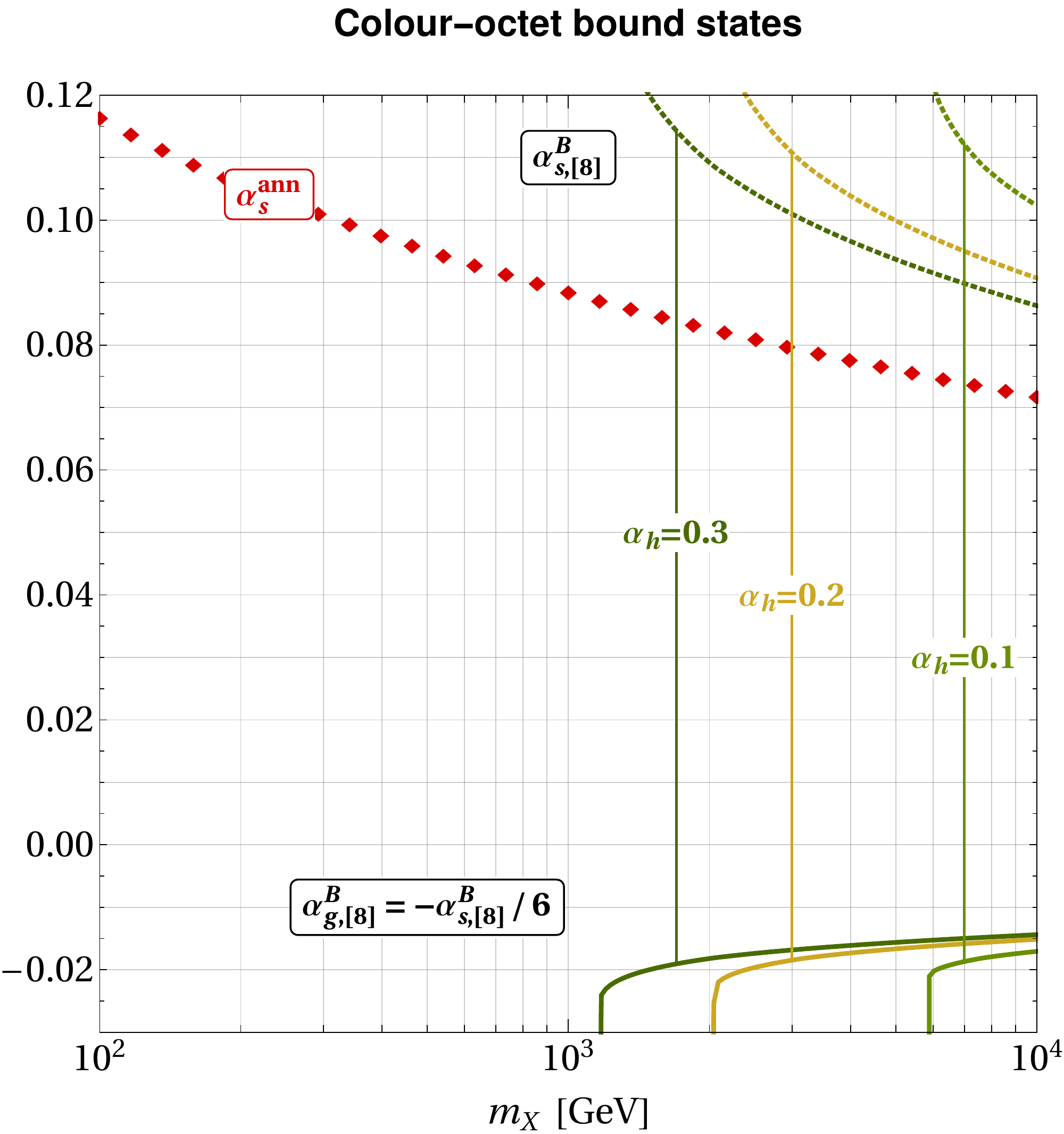}
	
\caption[]{\label{fig:AlphaRunning_BoundStates}
The strong couplings $\asBoundSinglet$ and $\asBoundOctet$ and the corresponding  $\agBoundSinglet = 4\asBoundSinglet/3$ and $\agBoundOctet = -\asBoundOctet/6$ that determine the colour-singlet (\emph{left}) and the colour-octet (\emph{right}) bound-state wavefunctions. The coupling to the Higgs, $\alpha_h$, increases the average momentum transfer within the bound states, thereby suppressing $\alpha_s$. While the effect is modest in case of the colour singlet, a large enough $\alpha_h$ is required for the existence of the octet bound state.
}
\vspace{2em}
\textbf{\small Colour-octet bound states}
\\
\includegraphics[height=0.44\textwidth]{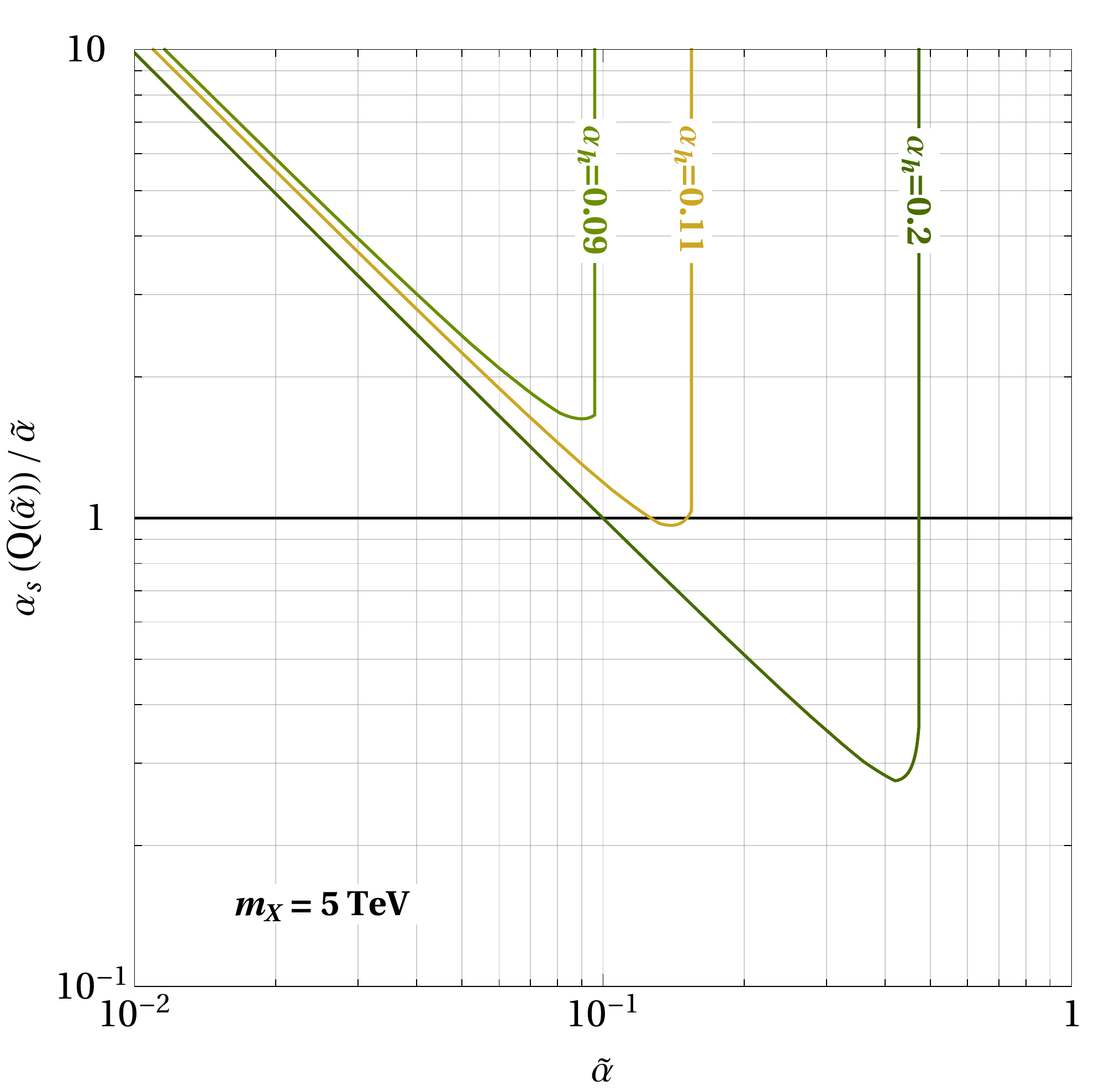}~~~
\includegraphics[height=0.44\textwidth]{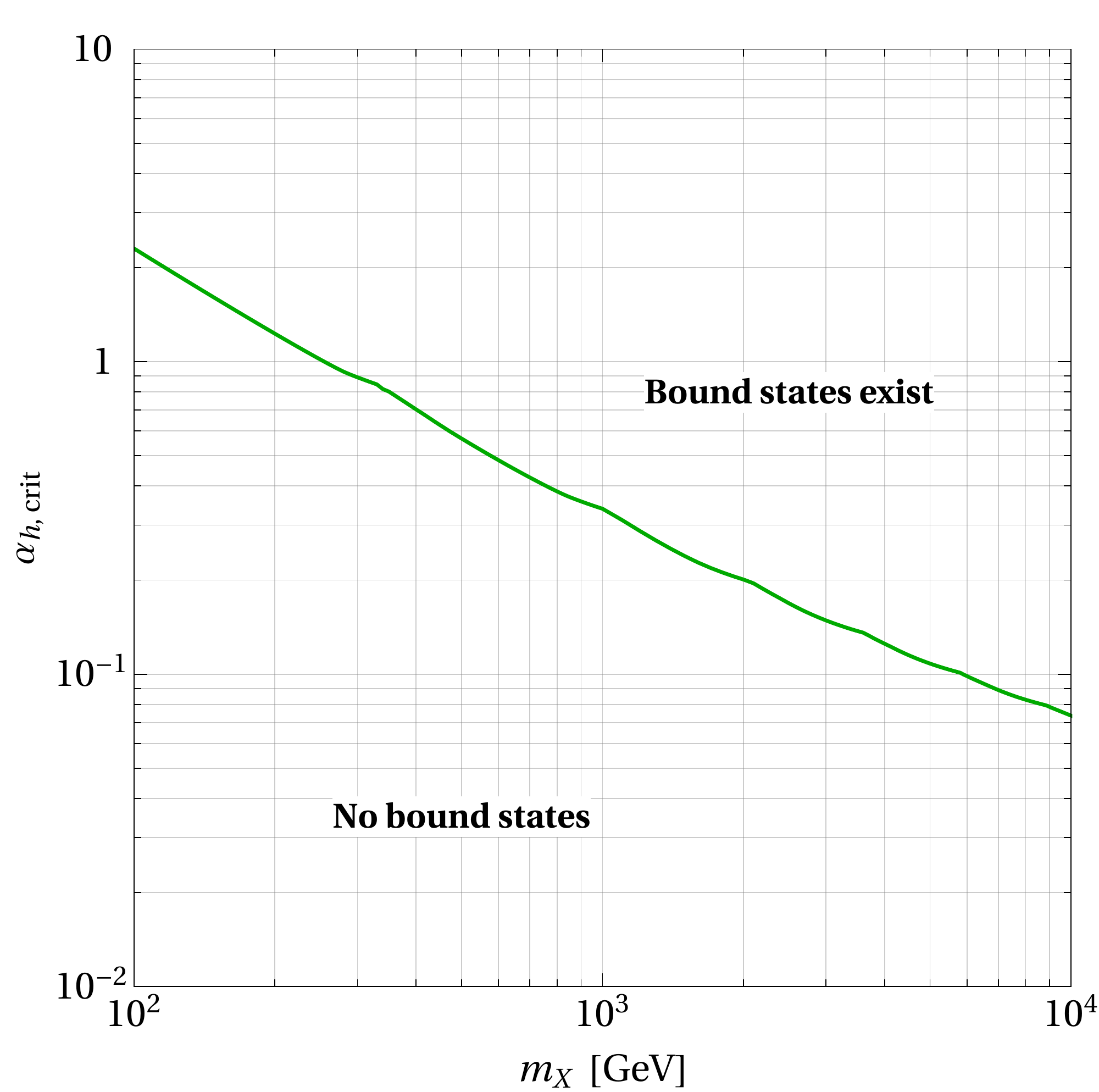}
\caption[]{\label{fig:alphaHcritical_Octet}
A colour-octet bound state exists only if the coupling $\alpha_h$ that gives rise to the attractive Yukawa potential is sufficiently strong to overcome the repulsive gluon potential.\\[1ex]
\emph{Left:} 
The equation $\alpha_s (Q (\tilde{\alpha}))  \ = \ \tilde{\alpha}$ has a solution when the coloured lines of fixed $\alpha_h$ intersect the horizontal black line. See text for discussion.\\[1ex]
\emph{Right:} The minimum coupling $\alpha_{h, \, \rm crit}$ for the octet bound states to exist as a function of $\mX$.
}
\end{figure}

\begin{figure}[t]
\centering
\includegraphics[height=0.46\textwidth]{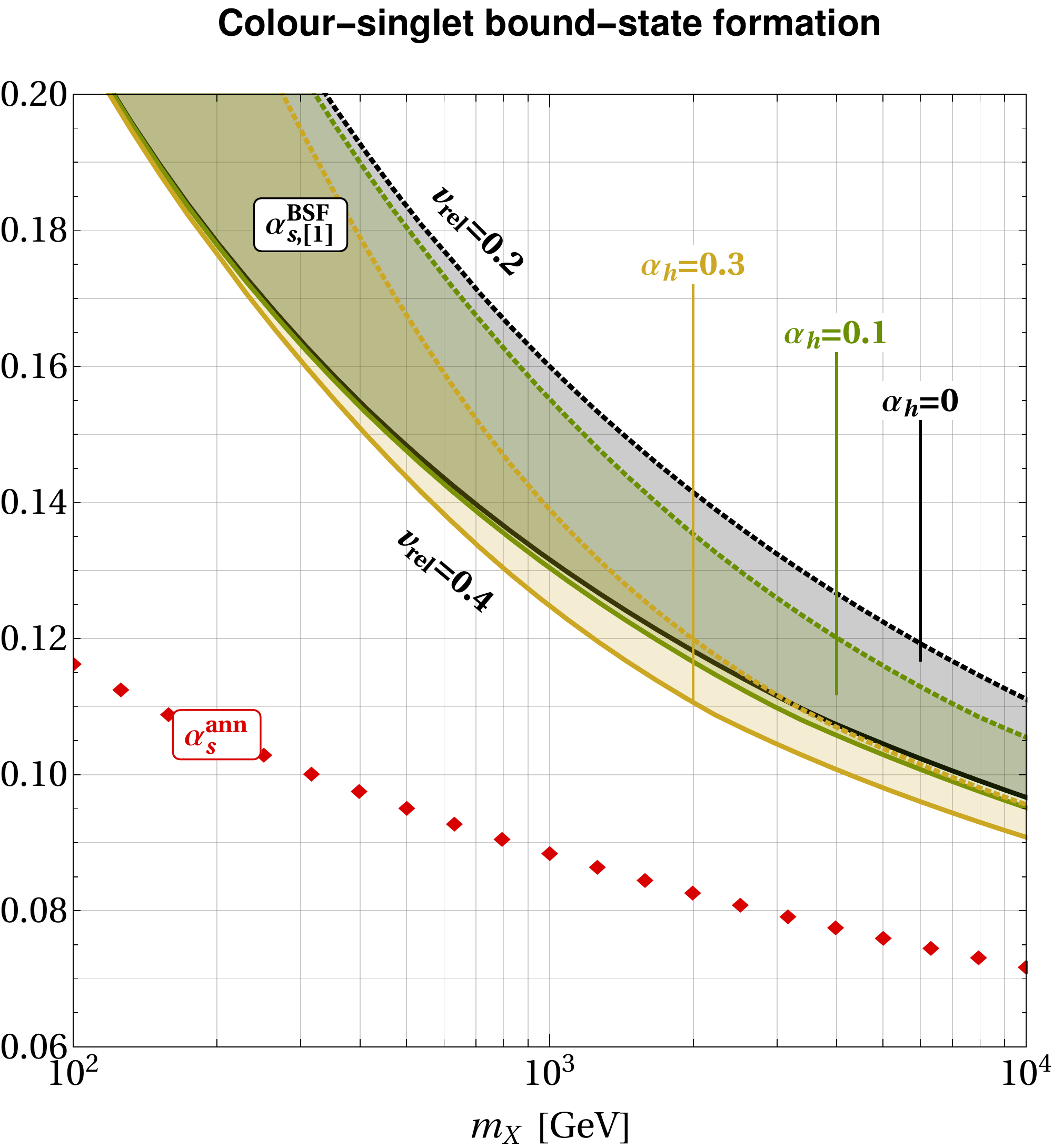}~~~
\includegraphics[height=0.46\textwidth]{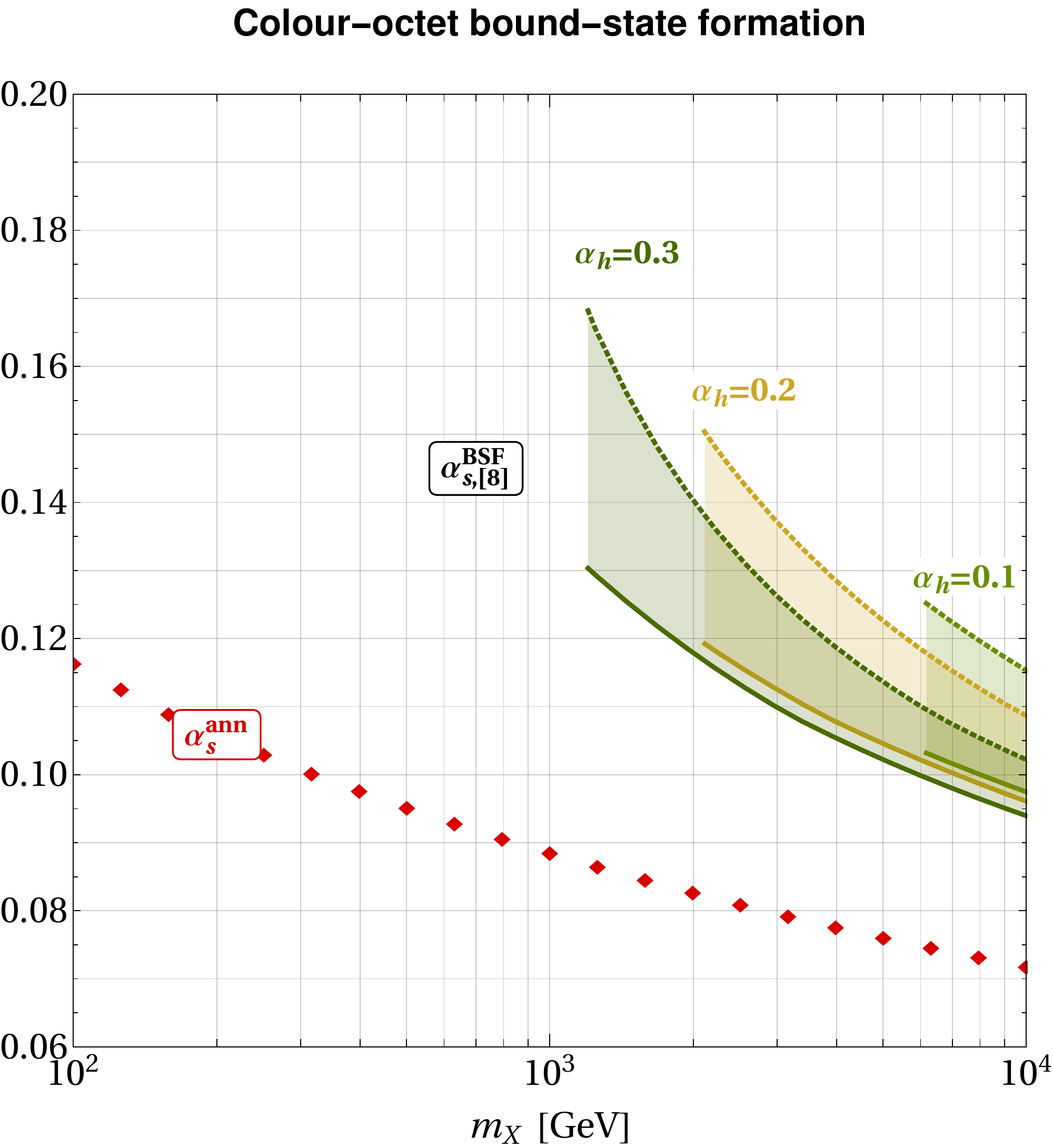}

\caption[]{\label{fig:AlphaRunning_BSF}
The strong coupling $\asBSFSinglet$ and $\asBSFOctet$ at the gluon emission vertex during the radiative capture of a colour-singlet (\emph{left panel}) or colour-octet (\emph{right panel}) bound state. The emitted gluon carries away the binding energy of the bound state plus the kinetic energy of the scattering state (see \cref{tab:MomentumTransfers}) that depends on the relative velocity $\vrel$. As before, we show $\asBSFSinglet$ and $\asBSFOctet$ in the typical velocity range $0.2 \leqslant \vrel \leqslant 0.4$ during freeze-out. As $\vrel$ decreases, the momentum transfer at the vertex drops and $\asBSF$ increases.  In the right panel, the cut-off of the bands for low $\mX$ reflects the fact that colour-octet bound states exist only for large enough $\mX$ and $\ah$ (cf.~\cref{fig:alphaHcritical_Octet}).  

}
\end{figure}

As it has been pointed out throughout this paper, the various vertices where the strong coupling appears in the annihilation and BSF diagrams are characterised by different momentum transfers $Q$.\footnote{
In fact, the smallness of the momentum transfer along the mediators in the ladder diagrams is responsible for the emergence of the non-perturbative phenomena, the Sommerfeld effect and bound states.}
Since the strong coupling is sensitive to the momentum transfer, $\as = \as(Q)$, it is important to carefully account for its running. 
An overview of the various scales is given in \cref{tab:MomentumTransfers}.\footnote{
An improved treatment would incorporate the running of the coupling in the Schr\"odinger equations for the scattering and bound states, by setting $Q=1/r$. The approximations of \cref{tab:MomentumTransfers} are sufficient for our purposes. We refer to~\cite{Pineda:2011aw,Pineda:2001ra,Manohar:2000kr} for the computation of the $\as$ running in NRQCD and pNRQCD.} 
In the occasions that the momentum transfer itself depends on the strong coupling, $Q = Q(\as)$, we solve numerically the following equation for $\tilde{\alpha}$,
\beq 
\as (Q(\tilde{\alpha}) ) = \tilde{\alpha} . 
\label{eq:alpha_s_determination} 
\eeq 
The effect of the running of $\as$ is depicted in \cref{fig:AlphaRunning_ScattStates,fig:alphaHcritical_Octet,fig:AlphaRunning_BoundStates,fig:AlphaRunning_BSF}, with related discussion in the captions.
A few remarks are in order here.

\paragraph{Scattering states.} The average momentum transfer $Q = \mu \vrel$, implies that $\asScatt$ becomes suppressed for large $\mX$. Consequently, for large $\mX$, the attraction due to the Higgs can significantly ameliorate or even overcome the repulsion due to the gluons in the colour-octet state. Due to the large multiplicity of the latter, (cf.~\cref{eq:sigma_XXbarTogg}), this enhances significantly the total annihilation rate, even for moderate values of $\ah$.\footnote{
This holds provided that $\vrel$ is not too low. See related discussion in \cref{sec:Results_Properties}.}
This is depicted in \cref{fig:AlphaRunning_ScattStates}.

\paragraph{Bound states.}
Because the momentum transfer depends on the strong coupling, we solve numerically \cref{eq:alpha_s_determination}. Generally, $\ah$ increases the average momentum transfer, thus suppressing $\asBound$, as shown in \cref{fig:AlphaRunning_BoundStates}. Overall though, the Bohr momentum $\kappa \equiv \mu (\agBound + \ah)$ increases with increasing $\ah$. 
If $\ah$  is sufficiently strong, it overcomes the gluon-mediated repulsion in the octet states, and gives rise to bound levels. In the octet ground state, 
\beq 
Q(\as) = \mu (\alpha_h -\as/6) \times 
\gamma_{1,0} \(-\frac{\as}{6\alpha_h}, \ d_h \) ,
\label{eq:Q8Bound}
\eeq
where we recall that $\dh = \mu \ah/\mh$, and the function $\gamma_{1,0}^{}(\lambda,\dh)$ has been determined numerically and is depicted in \cref{fig:gamma}. For this momentum transfer, we see graphically in the left panel of \cref{fig:alphaHcritical_Octet} that \cref{eq:alpha_s_determination} has a solution when the coloured lines of fixed $\alpha_h$ intersect the horizontal black line. Let us examine the scaling of $\alpha_s (Q (\tilde{\alpha})) / \tilde{\alpha}$ with $\tilde{\alpha}$.
\begin{itemize}
\itemsep 0pt	

\item
For sufficiently small $\tilde{\alpha}$ ($\tilde{\alpha} \ll 6\alpha_h$), the momentum transfer \eqref{eq:Q8Bound} is independent of $\tilde{\alpha}$, 
i.e.~$Q \simeq \mu \alpha_h \times \gamma_{1,0}^{} (0, d_h)$. Since $\alpha_s \sim 0.1$ for the $\mX$ values of interest, the ratio $\alpha_s (Q (\tilde{\a})) / \tilde{\alpha}$ is simply $0.1 / \tilde{\alpha}$, i.e.~starts from values $\gg 1$ and decreases with $\tilde{\alpha}$. 

\item
At larger $\tilde{\alpha}$ (but still $\tilde{\alpha} < 6\alpha_h$), this scaling changes. The factor $\gamma_{1,0}$ tends to 0, i.e.~$d_h$ approaches $d_{h, \, \rm crit} (\lambda = -\tilde{\alpha}/(6\alpha_h))$ [cf.~\cref{sec:Wavefunctions_BoundStates,fig:gamma}]. This, in turn, drives $Q$ to zero and consequently $\alpha_s (Q)$ to infinity.

\item
Between these two limits, $\alpha_s (Q (\tilde{\alpha})) / \tilde{\alpha}$ reaches the value 1 if $\ah$ is sufficiently large.
\end{itemize}
%
The minimum value of the Higgs coupling, $\alpha_{h, \, \rm crit}$, required for the octet bound states to exist is shown in the right panel of \cref{fig:alphaHcritical_Octet}.

\paragraph{Gluon-emission vertices in BSF.}
The momentum transfer --- which depends here as well on the strong coupling --- is the softer scale that enters our calculations, and thus yields the largest values of $\as$ (for fixed values of the other parameters). Since $\ah$ increases the binding energy, it suppresses $\asBSF$. This is seen in~\cref{fig:AlphaRunning_BSF}.

\begin{table}[b!]
\centering
\renewcommand{\arraystretch}{1}

\begin{tabular}{|c|c|c|c|} 
\hline
{\bf Vertices} 
&	\boldmath{$\as$}
&	\boldmath{$\ag$} 
&	{\bf Average momentum transfer}~\boldmath{$Q$}
\\ \hline  \hline 
\parbox[c]{15ex}{
\centering 
\smallskip
Annihilation: \\
gluon emission}
&	$\asAnn$
&	
&	$\mX$
\\[1.5ex] \hline
\multirow{2}{*}{
\hspace{-2ex}	
\parbox[c]{13ex}{
\centering 
\smallskip
Scattering state \\ 
wavefunction \\ 
(ladder)}
} 
&\multirow{2}{*}{$\asScatt$} 
&\parbox[c]{15ex}{
\centering 
\smallskip
Colour-singlet \\ 
\smallskip	
$\agScattSinglet = \dfrac{4\asScatt}{3}$}
&\multirow{2}{*}{$\mu \vrel$}
\\[3ex] \cline{3-3}
&
&\parbox[c]{15ex}{
\centering 
\smallskip
Colour-octet \\
\smallskip
$\agScattOctet = -\dfrac{\asScatt}{6}$}
&	
\\[3ex] \hline
\parbox[c]{15ex}{
\centering
Colour-singlet \\
bound-state \\ 
wavefunction \\ 
(ladder)} 
&	$\asBoundSinglet$ 
&	$\agBoundSinglet = \dfrac{4\asBoundSinglet}{3}$
&	\parbox[c]{46ex}{
\begin{multline} \nn
\kappaSinglet^{}  \ \gamma_{n\ell}^{} \(\l_{\bf \mathsmaller{[1]}}^{},\dh\) = 
\\
\mu \, \( \ah+ \dfrac{4\asBoundSinglet}{3} \) \times \gamma_{n\ell}^{} \(\dfrac{4\asBoundSinglet}{3\ah}, \ \dh\)
\end{multline}
}
\\[-8pt] \hline
\parbox[c]{15ex}{
\centering
Colour-octet \\
bound state \\ 
wavefunction \\ 
(ladder)} 
&	$\asBoundOctet$  
&	$\agBoundOctet = -\dfrac{\asBoundOctet}{6}$ 
&	\parbox[c]{46ex}{
\begin{multline} \nn
\kappa_{\bf \mathsmaller{[8]}}^{}  \ \gamma_{n\ell}^{} \(\l_{\bf \mathsmaller{[8]}}^{},\dh\) = 
\\
\mu \, \(\ah - \dfrac{\asBoundOctet}{6}\) \times \gamma_{n\ell}^{} \(-\dfrac{\asBoundOctet}{6\ah}, \ \dh\)
\end{multline}
}
\\[-8pt] \hline
\parbox[c]{15ex}{
\centering 
\smallskip
Formation of \\ 
colour-singlet \\
bound states: \\ 
gluon emission}
&	$\asBSFSinglet$
&	
&\parbox[c]{45ex}{
\beq 
\frac{\mu}{2} \left[ \vrel^2 
+ \(\ah + \dfrac{4\asBoundSinglet}{3}\)^2 
\gamma_{n\ell}^2 \left(\dfrac{4\asBoundSinglet}{3\ah}, \ \dh \right) \right] \nn 
\eeq}
\\[5ex] \hline
\parbox[c]{15ex}{
\centering 
\medskip	
Formation of \\ 
colour-octet \\
bound states: \\ 
gluon emission }
&	$\asBSFOctet$
&	
&\parbox[c]{45ex}{
\beq 
\frac{\mu}{2} \left[ \vrel^2 
+ \(\ah - \dfrac{\asBoundOctet}{6}\)^2 
\gamma_{n\ell}^2 \left(-\dfrac{\asBoundOctet}{6\ah}, \ \dh \right) \right] \nn 
\eeq}
\\[5ex] \hline
\end{tabular}
	
\caption[]{\label{tab:MomentumTransfers} 
The momentum transfer $Q$ at which the strong coupling $\as(Q)$ is evaluated. For the bound states, the functions $\gamma_{n\ell}(\lambda,\dh)$ are computed numerically (cf.~\cref{sec:Wavefunctions_BoundStates,fig:gamma}). We recall that $\lambdaR \equiv \agR/\ah$ with ${\bf R = 1,8}$, and $\dh = \mu\ah / \mh$.}
\end{table}

\begin{figure}[t!]
\centering
\includegraphics[height=0.46\textwidth]{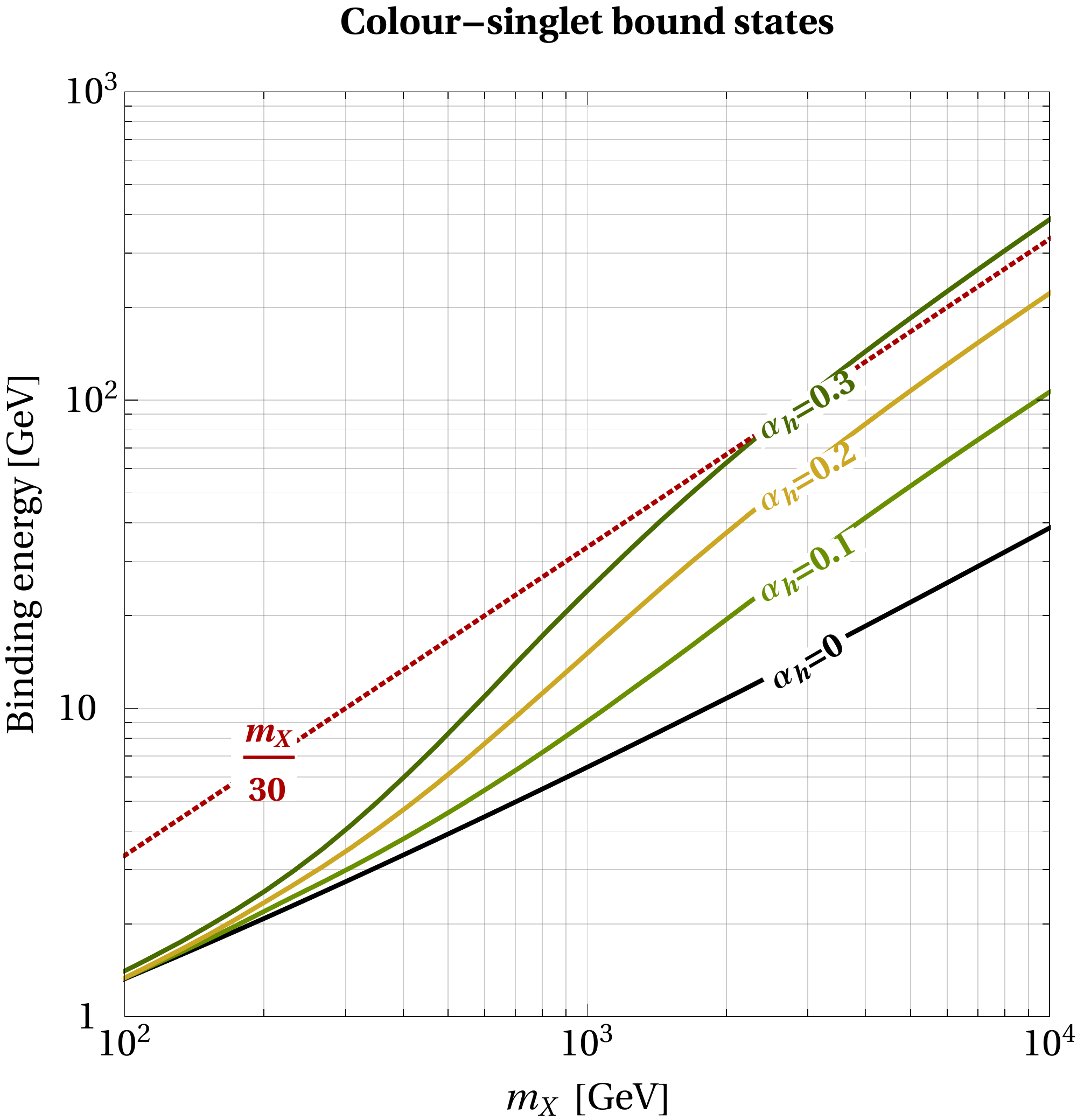}~~~
\includegraphics[height=0.46\textwidth]{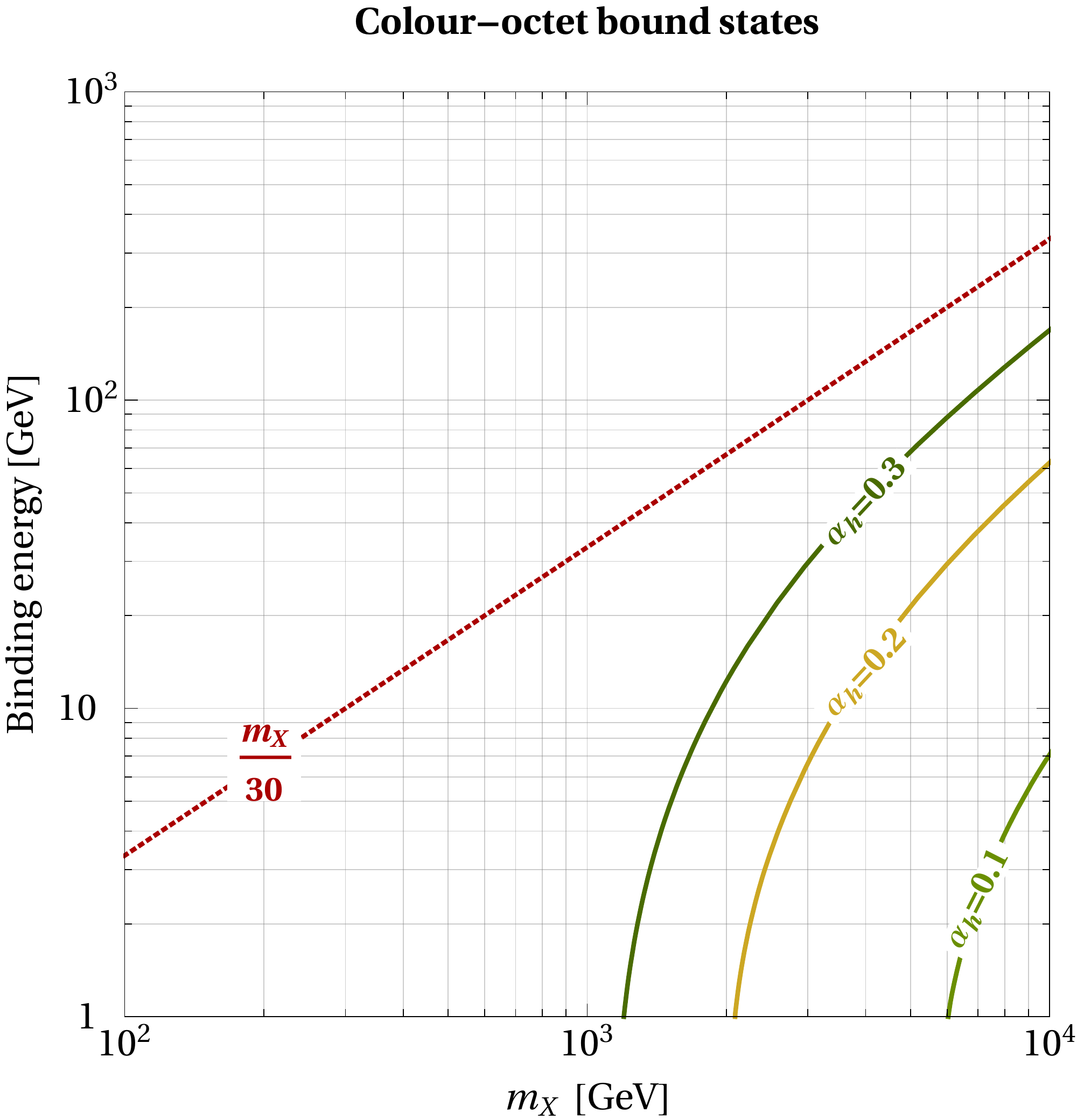}
\caption[The binding energies of the colour-singlet and octet states, for different values of the coupling to the Higgs]{\label{fig:BindingEnergy}	
The binding energy of the singlet and octet bound states, $B_{\bf [1]}$ and $B_{\bf [8]}$, for various values of the coupling to the Higgs, $\alpha_h$ (\emph{solid lines}). The \emph{dotted red lines} show the approximate temperature of freeze-out, $T_\mathsmaller{\rm FO} \approx \mX/30$.
}
\end{figure}


\clearpage 

\section{The long-range effect of the Higgs \label{Sec:Results}}

\subsection{Scattering states, bound states, annihilation and BSF rates \label{sec:Results_Properties}}

Based on the computations of previous sections, we now present and summarise the impact of the Higgs-mediated force on the properties of the scattering and bound states, and on the various cross-sections of interest.

\paragraph{Scattering states and Sommerfeld effect.} 

The Higgs-mediated potential enhances the attraction in the colour-singlet state, and suppresses or overcomes the repulsion in the octet state. Notably, the effect of the Higgs potential is influenced by the presence of the gluon-mediated potential, as was already shown in ref.~\cite{Harz:2017dlj}. In particular:
\begin{itemize}
\itemsep 0pt
\item
In the singlet state, the Higgs enhancement becomes sizeable --- and potentially reaches its Coulomb limit for a given coupling strength --- for lower masses (more generally, for lower $\dh \equiv \mu \ah / \mh$) than in the absence of the attractive gluon-mediated force~\cite[fig.~2]{Harz:2017dlj}.

\item
Conversely, in the octet state, a larger $\dh$ is required for the Higgs to affect the long-range interaction. However, the gluon-mediated repulsion is suppressed by a colour factor with respect to the full strength of the strong coupling, $\agOctet = -\asOctet/6$, suggesting that even a modest $\ah$ can potentially counteract it. This holds so long as $\vrel$ is not too low; at low $\vrel$, the gluon-induced repulsion becomes exponential [cf.~the Sommerfeld factor in the Coulomb limit~\eqref{eq:S0Coul} with $\zeta = \zetagScattOctet \lesssim-1$], and cannot be surmounted by a finite-range attractive force (cf.~ref.~\cite[fig.~3, left panel]{Harz:2017dlj}). However, the DM freeze-out occurs while $\vrel$ is fairly large and the gluon-induced repulsion is milder; this implies that the Higgs-mediated force has a significant effect.

\end{itemize}

\paragraph{Tighter bound states.}

\begin{figure}[t]
\centering
\includegraphics[width=0.46\textwidth]{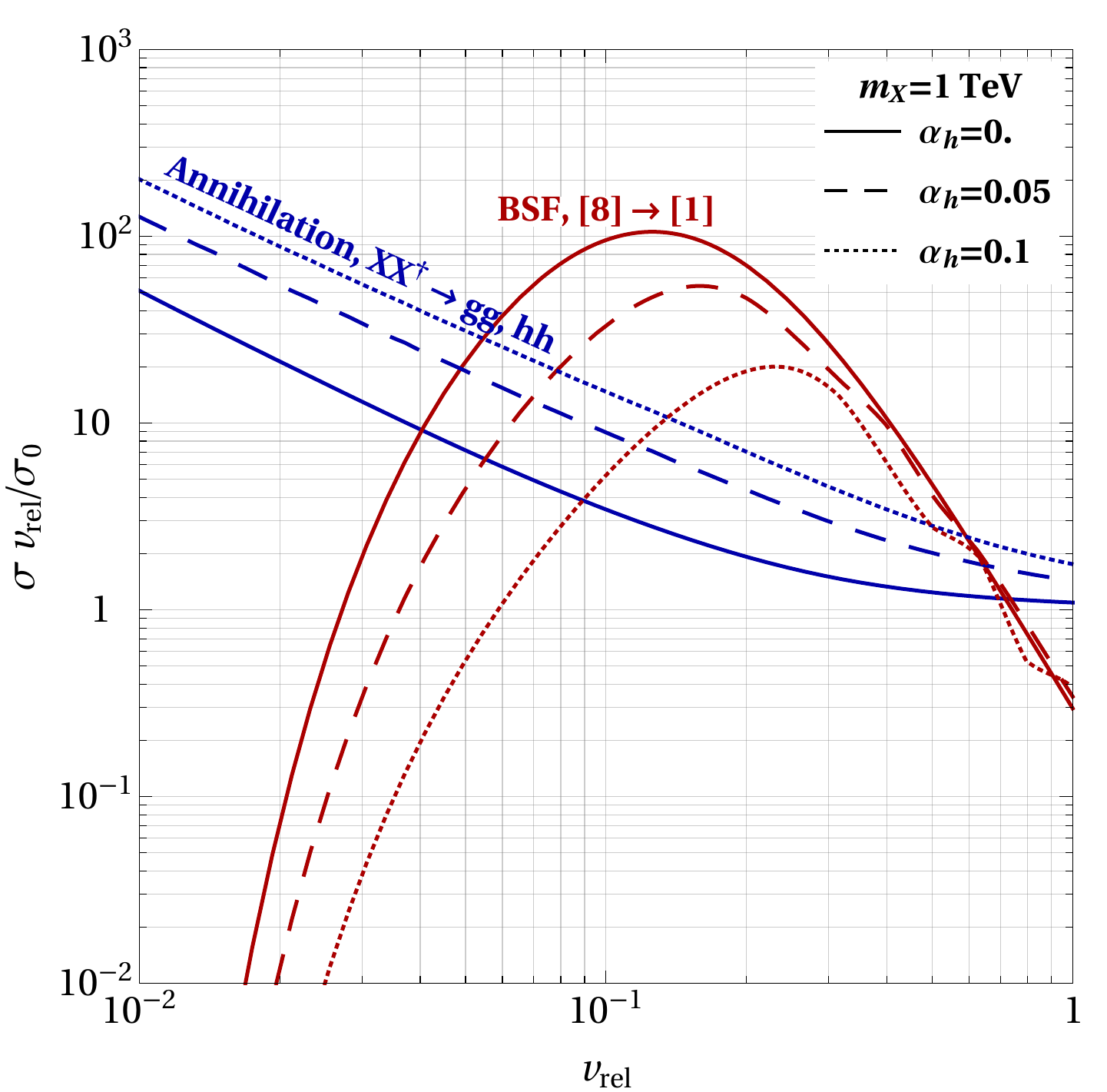}~~~
\includegraphics[width=0.46\textwidth]{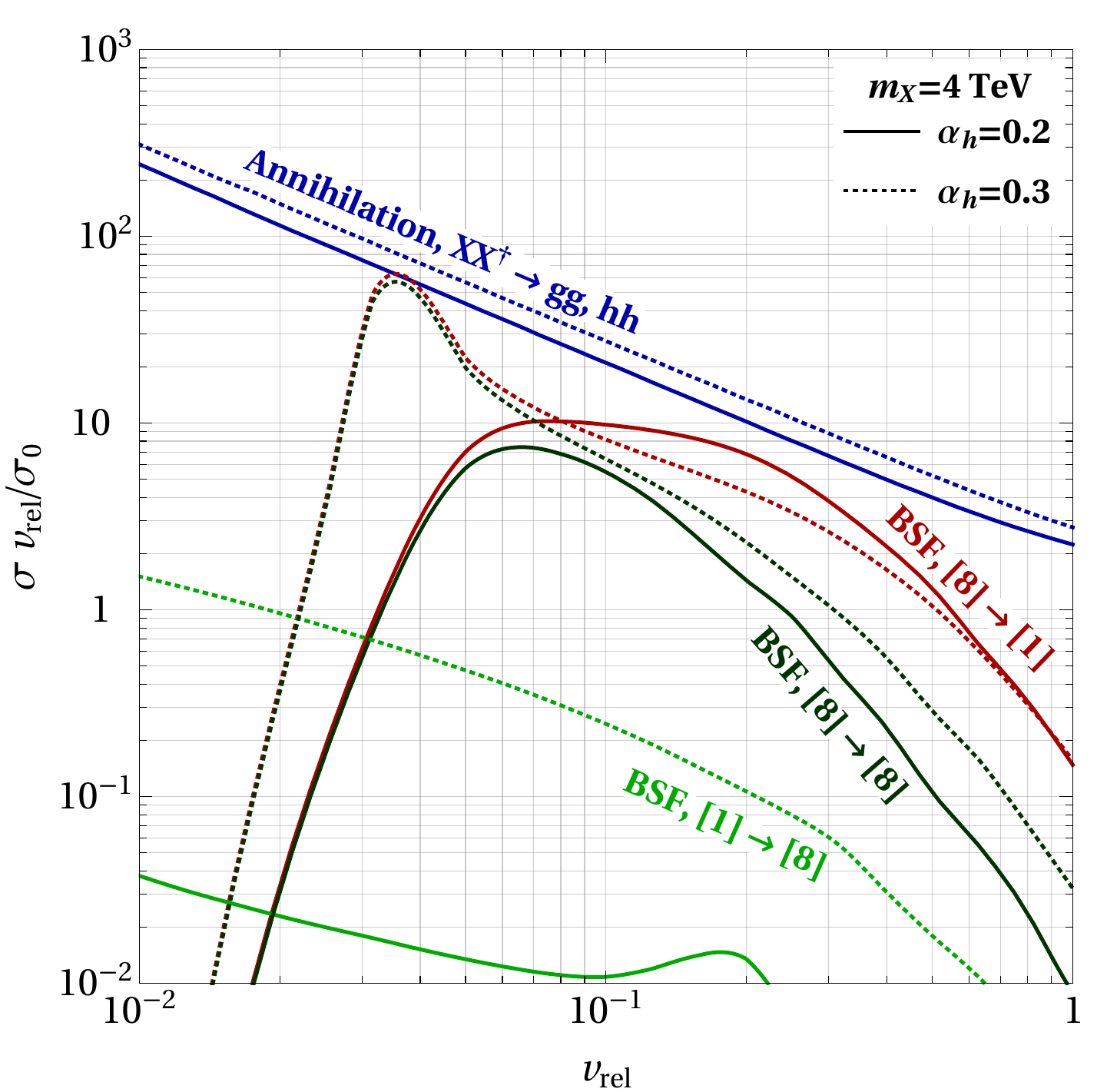}
\caption[]{\label{fig:BSFvsANN_numerical}
$\sigma \vrel$ for annihilation into gluons and Higgs bosons, $XX^\dagger \to gg, hh$ (\emph{blue lines}), and for the capture into the ground states via gluon emission, for the different colour transitions: $\bf{[8]\rightarrow [1]}$ (\emph{red lines}), $\bf{[1]\rightarrow [8]}$ (\emph{light green lines}), and $\bf{[8]\rightarrow [8]}$ (\emph{dark green lines}). All $\sigma \vrel$ are normalised to the perturbative value for $s$-wave annihilation, 
$\sigma_0 \equiv \sigma^{\mathrm{pert}}_{XX^\dagger\rightarrow gg} \vrel +\sigma^{\mathrm{pert}}_{XX^\dagger\rightarrow hh} \vrel$. 
Lines of different dashing correspond to different values for $\ah$, as denoted in the legends. 
\\[1ex]
\emph{Left:} For $\mX=1~\mathrm{TeV}$ and the values of $\ah$ considered, only colour-singlet bound states exist.   
\emph{Right:} For $\mX=4~\mathrm{TeV}$ and larger values of $\ah$, there exist also colour-octet bound states.
}

\vspace{2em}

\includegraphics[width=0.46\textwidth]{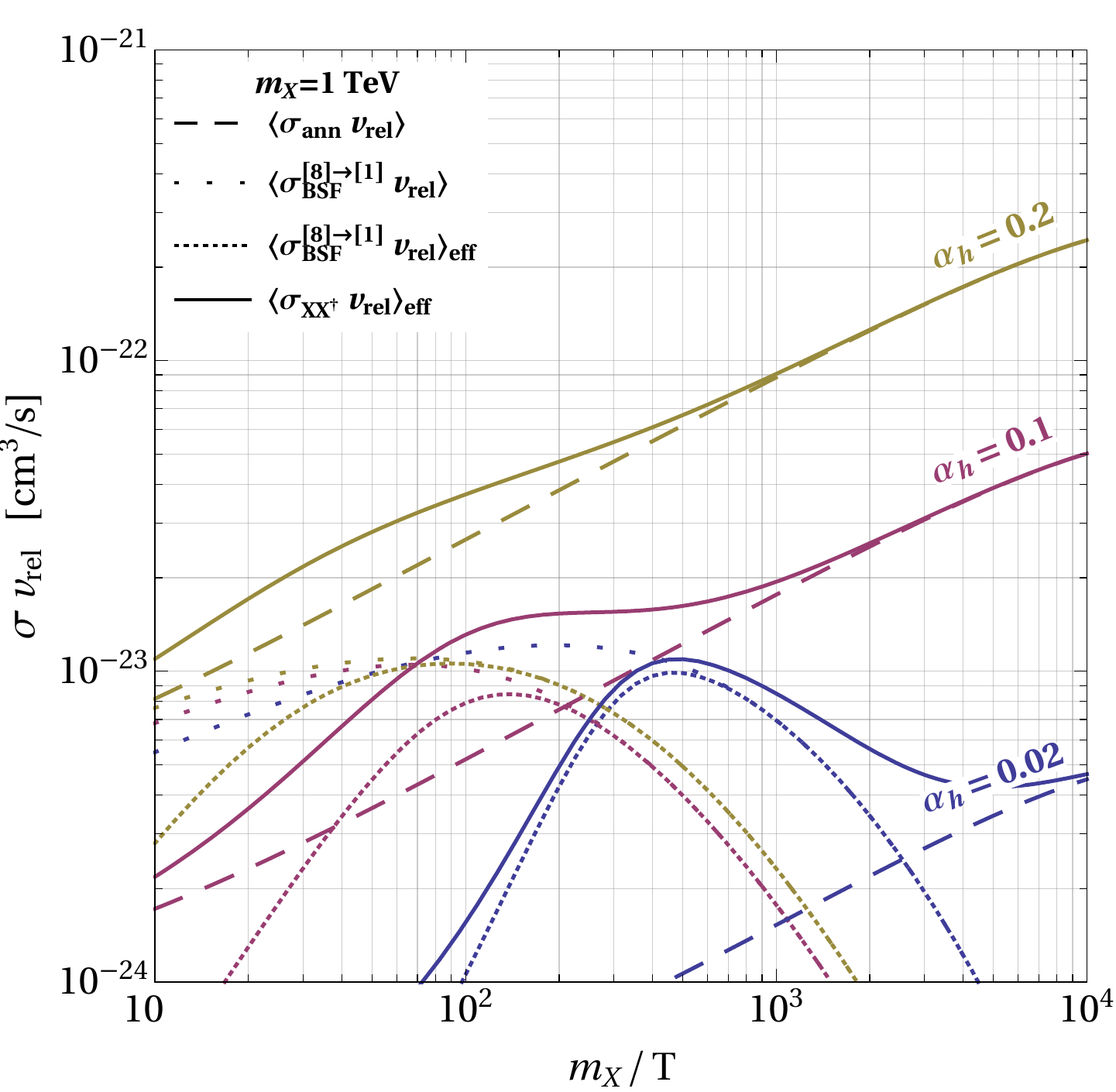}
\includegraphics[width=0.46\textwidth]{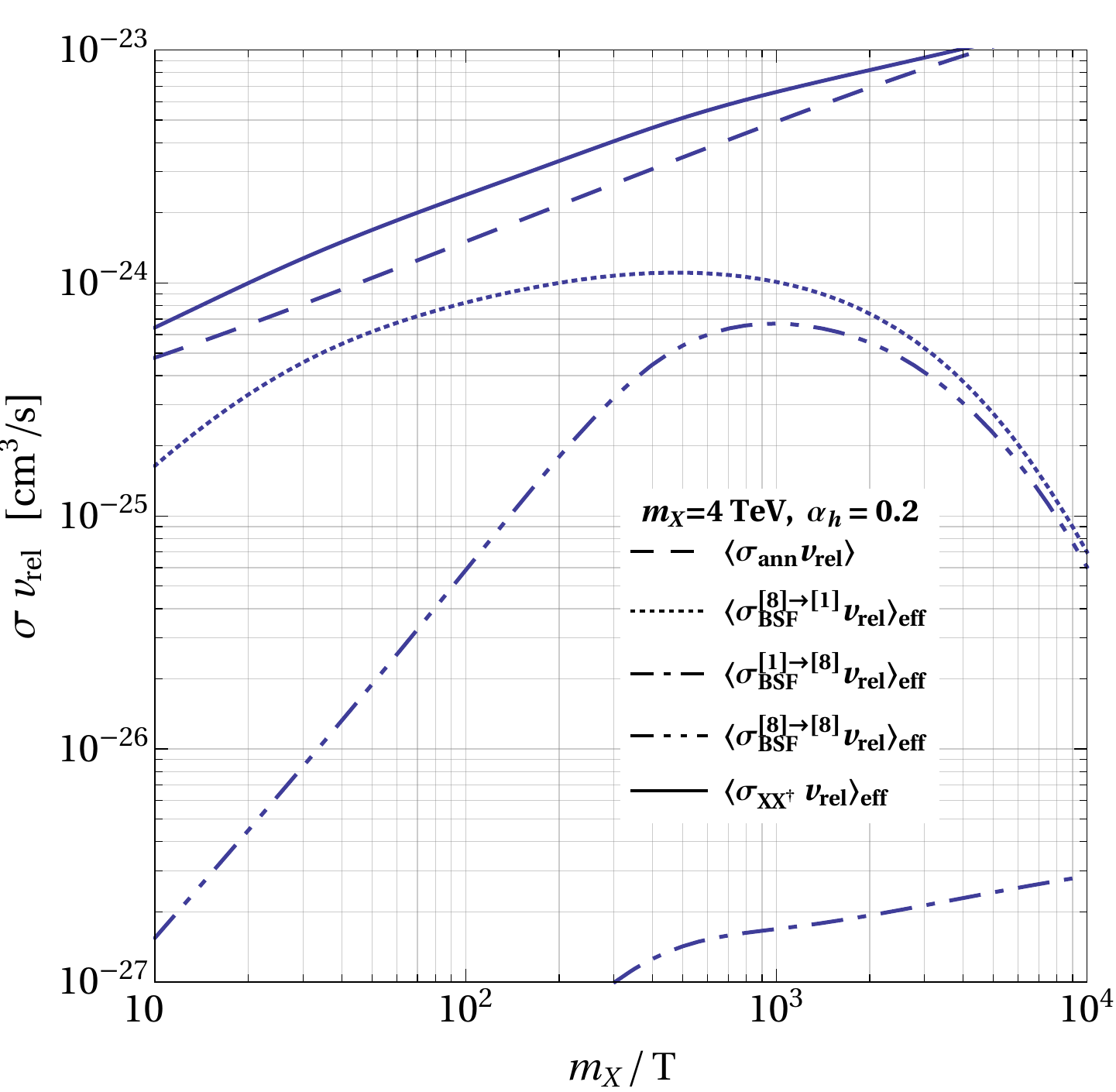}

\caption[]{\label{fig:BSFvsANN_ThermAver}
Comparison of thermally averaged cross-sections vs.~the time parameter $\tilde{x}=\mX/T$.\\[1ex]
\emph{Left:} We fix $\mX=1$~TeV, and consider different $\ah$ as denoted on the plot. The lines of different dashing denote 
annihilation \emph{(dashed)}, 
bound state formation $\bf{[8]\rightarrow [1]}$ (\emph{wide-spaced dotted}),
effective bound state formation $\bf{[8]\rightarrow [1]}$ weighted by the fraction that decay rather than being ionized  (\emph{densely-spaced dotted}), and
total effective cross-section that depletes DM (\emph{solid line = dashed + densely-spaced dotted}). For the $\mX$ and $\ah$ considered, only colour-singlet bound states exist.   
\\[1ex]
\emph{Right:} The values $\mX=4$~TeV and $\ah=0.2$ allow also for colour-octet bound states. We show the cross-sections for  
annihilation \emph{(dashed)} and  
effective bound state formation for different colour transitions, 
$\bf{[8]\rightarrow [1]}$ (\emph{dotted}), 
$\bf{[1]\rightarrow [8]}$ (\emph{double-dot dashed}),  
$\bf{[8]\rightarrow [8]}$ (\emph{dot dashed}). 
}
\end{figure}

The Higgs attraction increases the absolute value of the binding energies, as shown in \cref{fig:BindingEnergy}. This renders the bound-state dissociation processes inefficient earlier, when the DM density is greater, thus enhancing the efficacy of BSF in depleting DM. In \cref{fig:BindingEnergy}, we compare the binding energy with the typical temperature around DM freeze-out, $T_{\FO} \approx \mX/30$. If the binding energy equals or exceeds $T_{\FO}$, then the dissociation of bound states is inhibited already at freeze-out, and the efficiency of BSF in depleting DM is maximal. While this occurs only for very large values of $\ah$, we emphasise that the effective BSF rate can become comparable to or exceed the annihilation rate even at temperatures that are a factor of a few larger than the binding energy, even if it is not maximal (cf.~\cref{fig:BSFvsANN_ThermAver}).

\paragraph{\bf Additional bound states.} 
As already discussed in \cref{sec:BSF_AlphaRunning} and shown in \cref{fig:alphaHcritical_Octet}, for sufficiently large masses and couplings, the Higgs-mediated attractive force implies the existence of colour-octet bound states. However, these bound states are considerably looser than the singlet states, as seen in \cref{fig:BindingEnergy}, and have a limited effect on the DM density (except perhaps for $\ah \gtrsim 0.3$).

\paragraph{Impact on the strong coupling.}
The Higgs-mediated force increases the momentum transfer in the bound states and in the gluon-emission vertices of the radiative capture processes. Because of the running of the strong coupling, this suppresses the corresponding values of $\as$, as discussed in~\cref{sec:BSF_AlphaRunning} and depicted in \cref{fig:AlphaRunning_ScattStates,fig:AlphaRunning_BoundStates,fig:AlphaRunning_BSF}. However, the Bohr momentum and the binding energy increase with $\ah$, as seen in \cref{fig:BindingEnergy}.

\paragraph{Cross-sections.}

In \cref{fig:BSFvsANN_numerical}, we illustrate the velocity dependence of the BSF cross-sections, and compare them with the direct annihilation processes. 
\begin{itemize}
\itemsep 0pt	
\item
For large $\vrel$, the BSF cross-sections are rather suppressed and subdominant to the annihilation. This is due to the small overlap of the wavefunctions: the average momentum in the scattering states, $k = \mu \vrel$, is much greater than that in the bound states, which in the Coulomb approximation is the Bohr momentum, $\kappa = \mu (\agBound + \ah)$.  
In this regime, the BSF cross-sections scale as $\sigma_{\BSF} \vrel \propto (\kappa/k)^4 \approx [(\agBound + \ah)/\vrel]^4$. 

\item
At lower velocities, when $k\sim\kappa$ or $\vrel \sim \agBound + \ah$, the overlap of the wavefunctions is nearly maximal (although the precise value of $\vrel$ at which this occurs depends also on the Bohr momentum of the scattering state, hence on $\agScatt+\ah$). In this regime, the cross-section for capture into the tightest bound state (colour singlet) dominates over annihilation. Note that since $\as \sim 0.1$, this velocity range is relevant for freeze-out. 

\item
At $\vrel \lesssim \agScatt + \ah$, the Sommerfeld effect becomes important. 

Because of the gluon-mediated repulsion in the octet scattering states, the ${\bf [8] \to [1]}$ and ${\bf [8] \to [8]}$ capture processes become suppressed with decreasing $\vrel$. The coupling to the Higgs ameliorates this suppression, and large $\ah$ can even make these BSF cross-sections increase temporarily as the velocity drops; this can be seen in the right panel of \cref{fig:BSFvsANN_numerical}. However, at sufficiently low $\vrel$, the exponential Coulomb repulsion cannot be overcome by the finite-range Higgs-mediated attraction. 

On the other hand, the ${\bf [1] \to [8]}$ cross-section increases steadily with decreasing $\vrel$, due to the gluon and Higgs-mediated attractive forces in the colour-singlet scattering state. Similarly, the annihilation cross-section becomes dominated by the colour-singlet contribution and increases with decreasing $\vrel$.

\end{itemize}
Note that in \cref{fig:BSFvsANN_numerical}, we show the various cross-sections normalized to the perturbative annihilation cross-section, such that the direct dependence on $\mX$ cancels out. Nevertheless, $\mX$ affects the scale of $\as$ and thus has indirect impact on the various lines.

Let us now focus on the effect of the Higgs. As expected, the annihilation cross-section increases with larger $\ah$.  This is both due to the $XX^\dagger \to hh$ channel, and because the Higgs-mediated attractive potential enhances the annihilation cross-sections for all final states. In part because of the increase in the (perturbative) annihilation cross-section, the relative strength of BSF appears to diminish with increasing $\ah$. The suppression of $\asBSF$ at larger $\ah$ due to the running of the strong coupling (cf.~\cref{fig:AlphaRunning_BSF}) contributes to this trend. However, the larger Bohr momentum of the bound states implies that BSF peaks at a larger $\vrel$. This is very important for the DM depletion via BSF, as we discuss next.

\paragraph{Thermally averaged  cross-sections.} 
In \cref{fig:BSFvsANN_ThermAver}, we depict the impact of the Higgs on the thermally averaged annihilation and BSF rates. The features of $\<\sigma_{\ann} \vrel\>$ follow from the discussion above. For BSF, we present both the actual and the effective thermally averaged cross-sections, $\< \sigma_{\BSF} \vrel\>$ and $\< \sigma_{\BSF} \vrel\>_{\eff}$. Several comments are in order.
\begin{itemize}
\itemsep 0pt
	
\item
Despite the suppression of $\sigma_{\BSF} \vrel$ at large $\vrel$ discussed above, $\< \sigma_{\BSF} \vrel\>$ appears to be large even at high $T$; this is due to the Bose-enhancement factor [cf.~\cref{eq:sigmaBSF_averaged}]. 

\item
Because of the rapid ionisation processes, $\< \sigma_{\BSF} \vrel\>_{\eff}$ is suppressed at high $T$, and in fact becomes independent of $\sigma_{\BSF} \vrel$, as discussed in \cref{sec:Model_BoundStates}.  However, $\< \sigma_{\BSF} \vrel\>_{\eff}$ increases as $T$ drops. 

For capture from an octet state, $\< \sigma_{\BSF} \vrel\>_{\eff}$ peaks at $T\sim$~binding energy. The reason is three-fold: 
(i)~This condition is the thermally-averaged equivalent of $k\sim \kappa$, which maximises the overlap of the bound and scattering-state wavefunctions, as discussed above. 
(ii)~At and below this temperature, the ionisation rate becomes exponentially suppressed, therefore $\< \sigma_{\BSF} \vrel\>_{\eff}$ saturates to $\< \sigma_{\BSF} \vrel\>$.
(iii)~At even lower temperatures, the gluon repulsion in the octet state suppresses $\< \sigma_{\BSF}^{\mathsmaller{\bf [8]\to[1]}} \vrel\>$ and $\< \sigma_{\BSF}^{\mathsmaller{\bf [8]\to[8]}} \vrel\>$.

\item
At least for $0.02 \lesssim \ah \lesssim 0.1$,  $\< \sigma_{\BSF}^{\mathsmaller{\bf [8]\to[1]}} \vrel\>_{\eff}$ overcomes $\<\sigma_{\ann} \vrel\>$, even before it saturates to $\< \sigma_{\BSF}^{\mathsmaller{\bf [8]\to[1]}} \vrel\>$, i.e.~at temperatures higher than the binding energy of the singlet bound state.  Overall, BSF yields a sizeable contribution to the total effective annihilation cross-section $\< \sigma_{\mathsmaller{XX^\dagger}} \vrel\>_{\eff}$ for the range of couplings considered. 

\end{itemize}

\subsection{Dark matter relic density \label{sec:Results_RelicDensity}}

\begin{figure}[th!]
\centering
\includegraphics[width=0.45\textwidth]{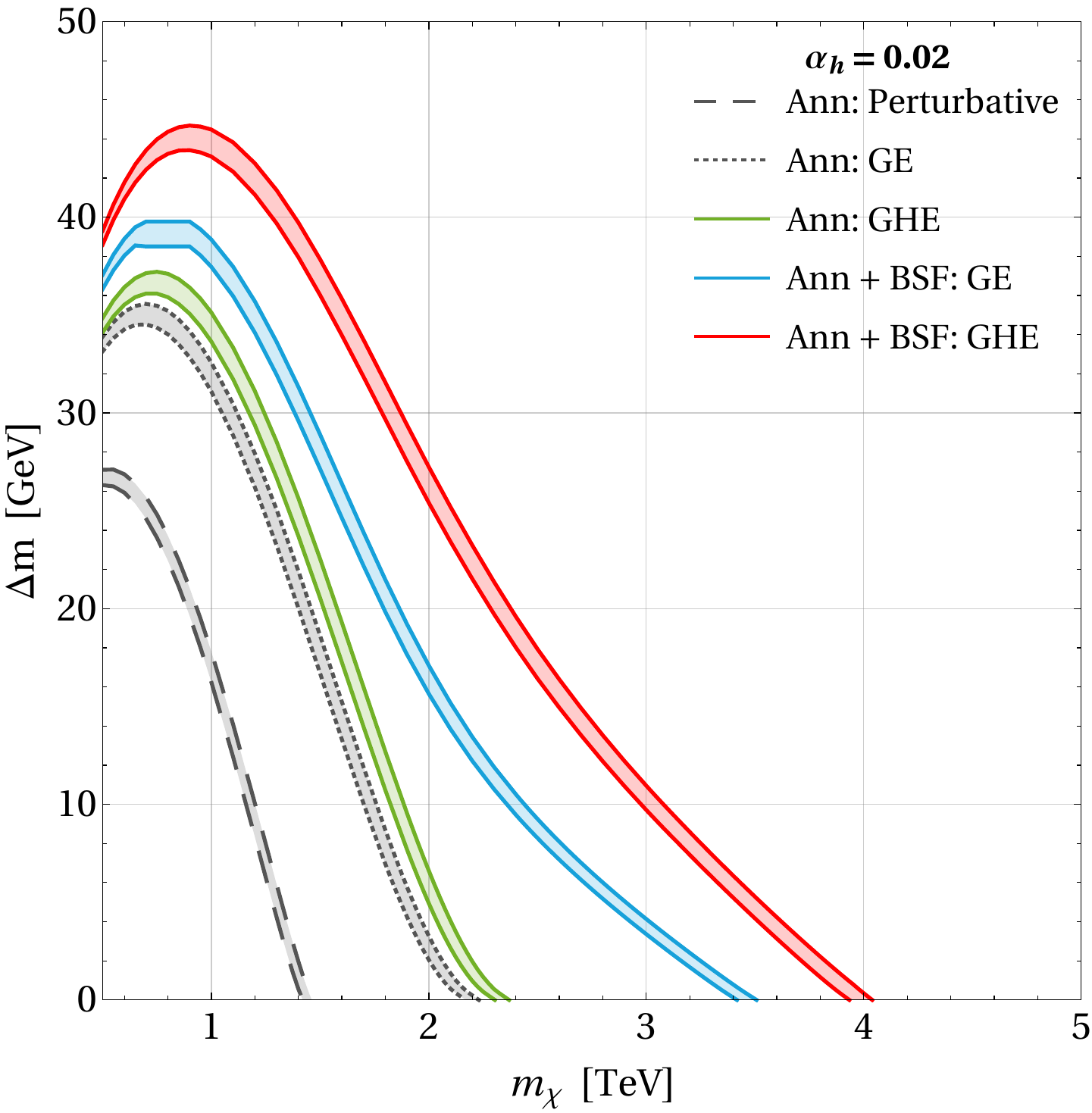}
\includegraphics[width=0.45\textwidth]{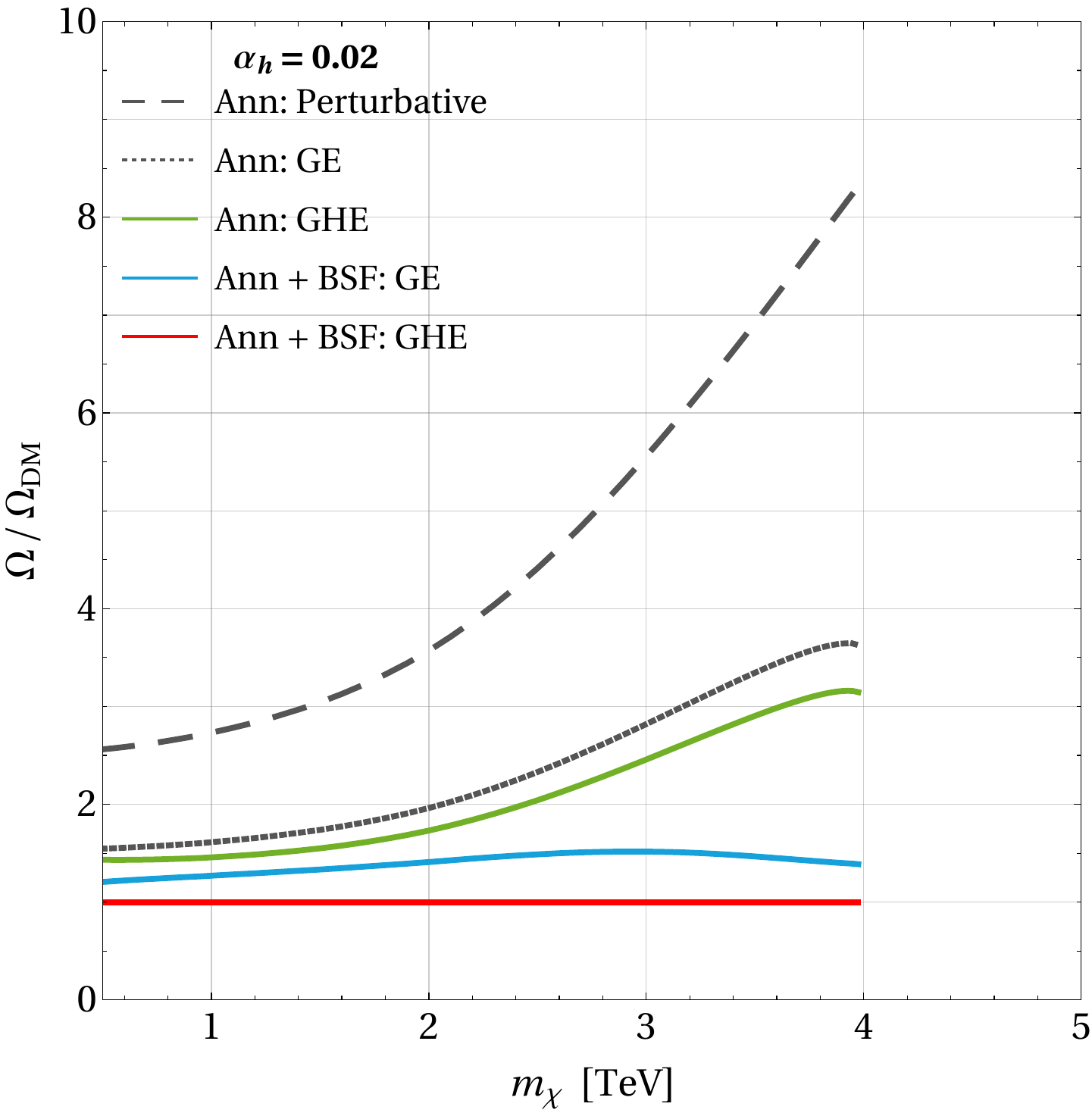}\\[1ex]
\includegraphics[width=0.45\textwidth]{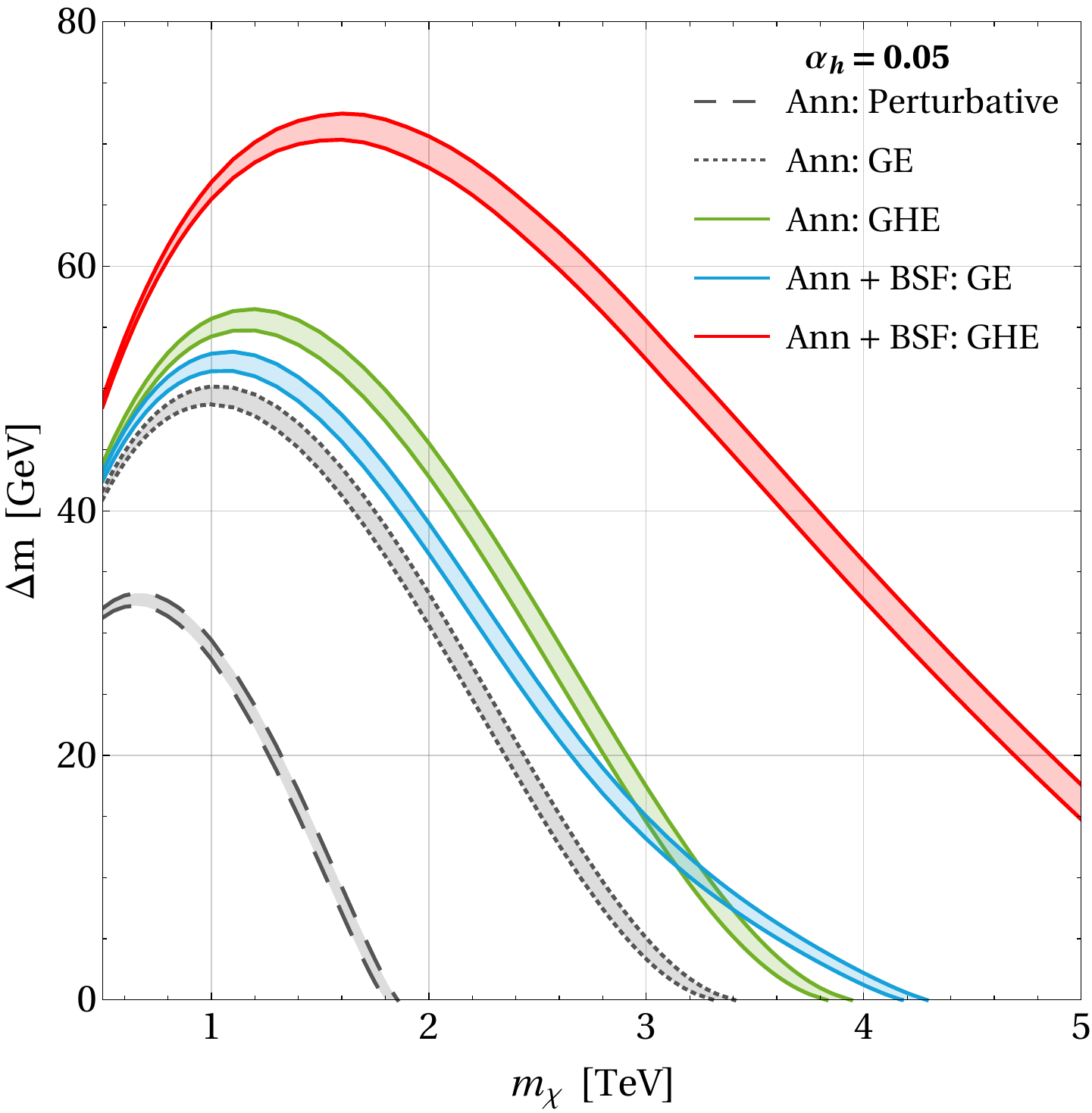}
\includegraphics[width=0.45\textwidth]{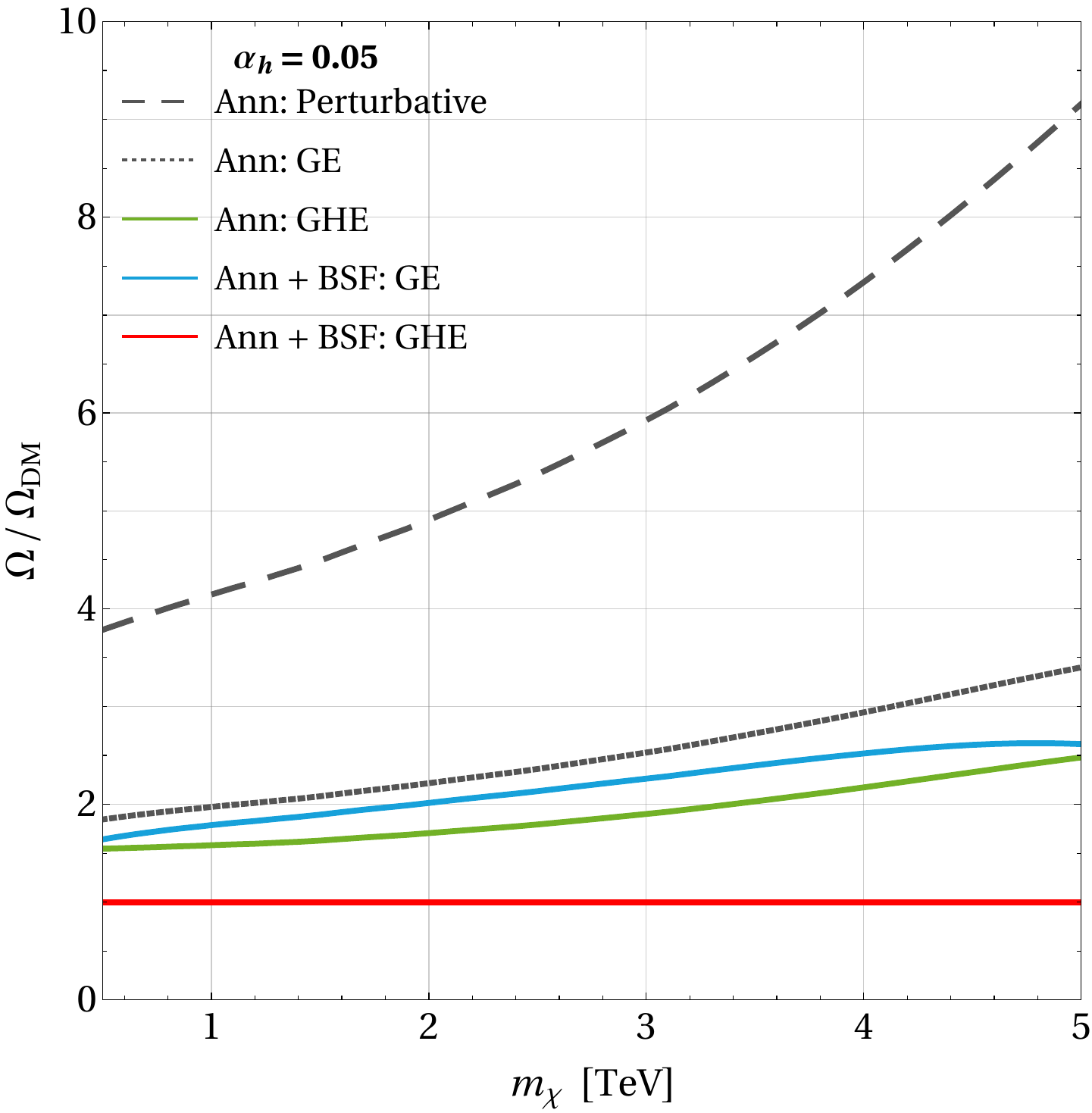}

\caption[]{\label{fig:RelicResults1}
\emph{Left panels}: The mass difference $\Delta m$ versus the DM mass $\mx$ that reproduce the observed DM density when taking into account:  
perturbative annihilation only (\emph{grey dashed}), 
annihilation with Sommerfeld effect due to gluon exchange only (\emph{grey dotted}), 
annihilation with Sommerfeld effect due to gluon and Higgs exchange (\emph{green}), 
annihilation with Sommerfeld effect and bound state formation with gluon exchange only (\emph{blue}), 
annihilation with Sommerfeld effect and bound state formation with gluon and Higgs exchange (\emph{red}). 
The width of the bands corresponds to the $3\sigma$ experimental uncertainty on the DM abundance, using the \emph{Planck} 2018 results. 
\\[1ex]
\emph{Right panels}: 
For the values of $\Delta m$ predicted by the full calculation (red bands on the left panels), we show the ratio of the relic density from each partial calculation to the observed DM abundance. The colour coding is the same as on the left panels.
\\[1ex]
We present the results for $\ah=0.02$ (\emph{up}) and $\ah=0.05$ (\emph{down}).
}
\end{figure}

\begin{figure}[th!]
\centering 
\includegraphics[height=0.45\textwidth]{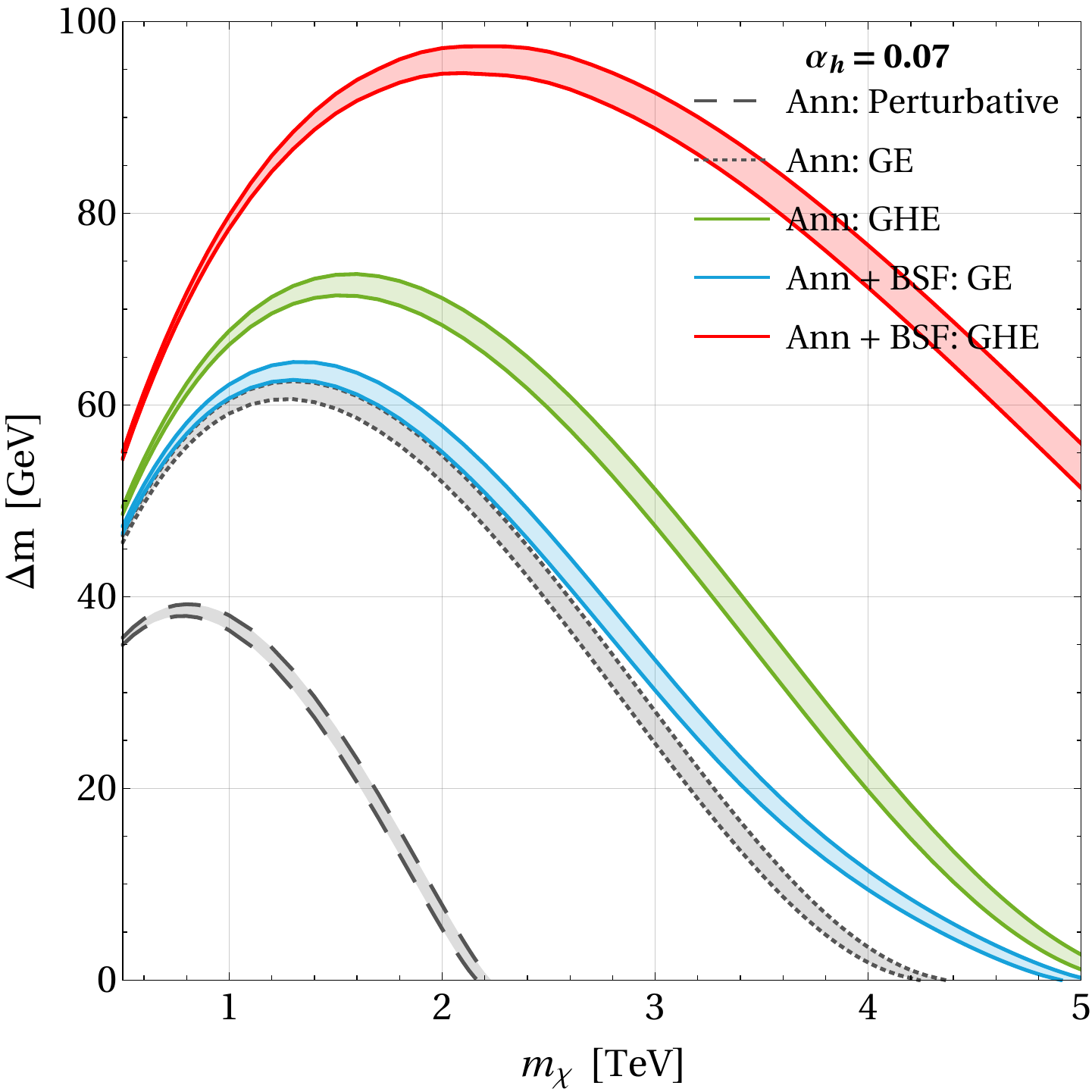}
\includegraphics[height=0.45\textwidth]{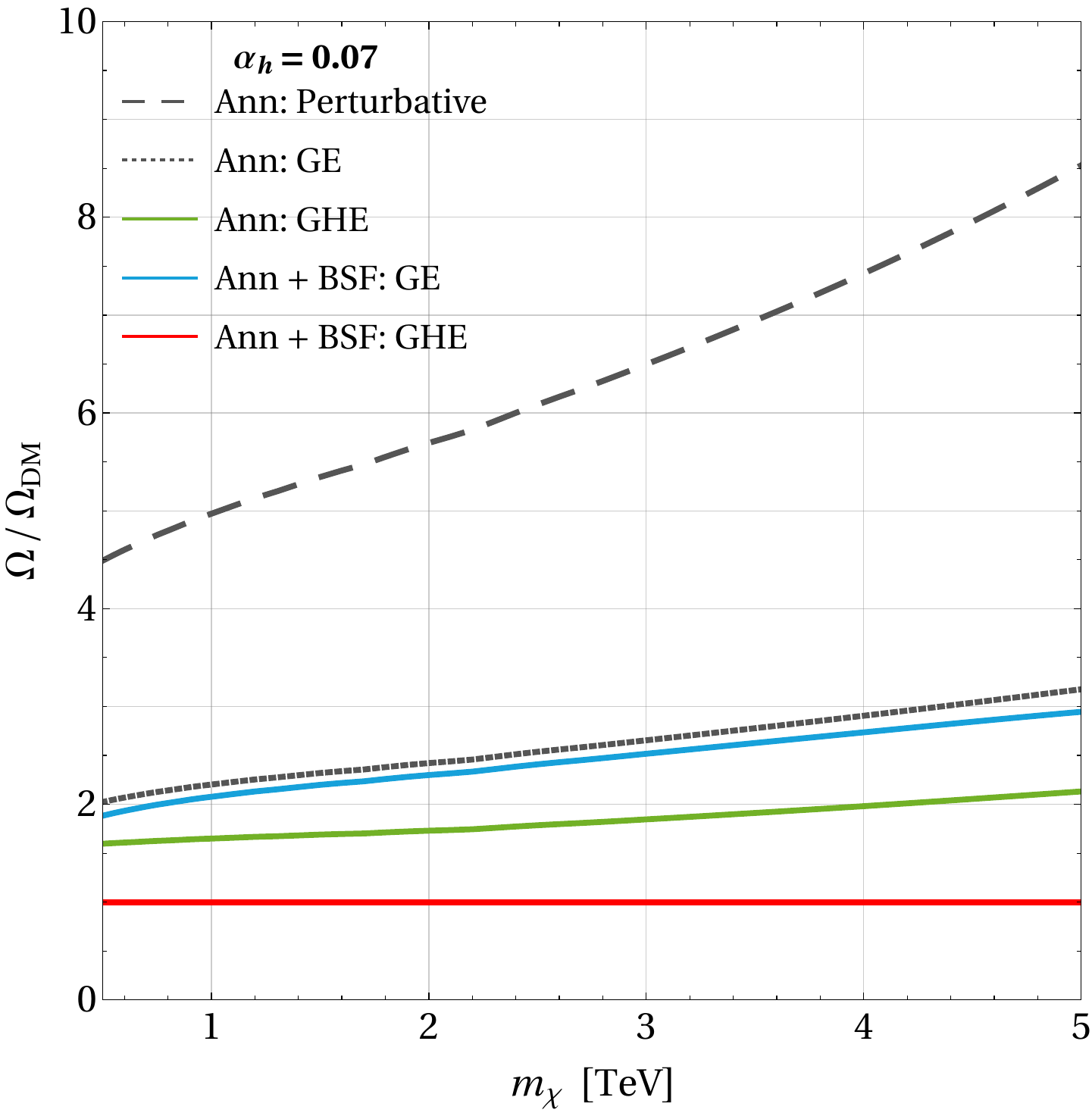}
\\[1ex]
\includegraphics[height=0.45\textwidth]{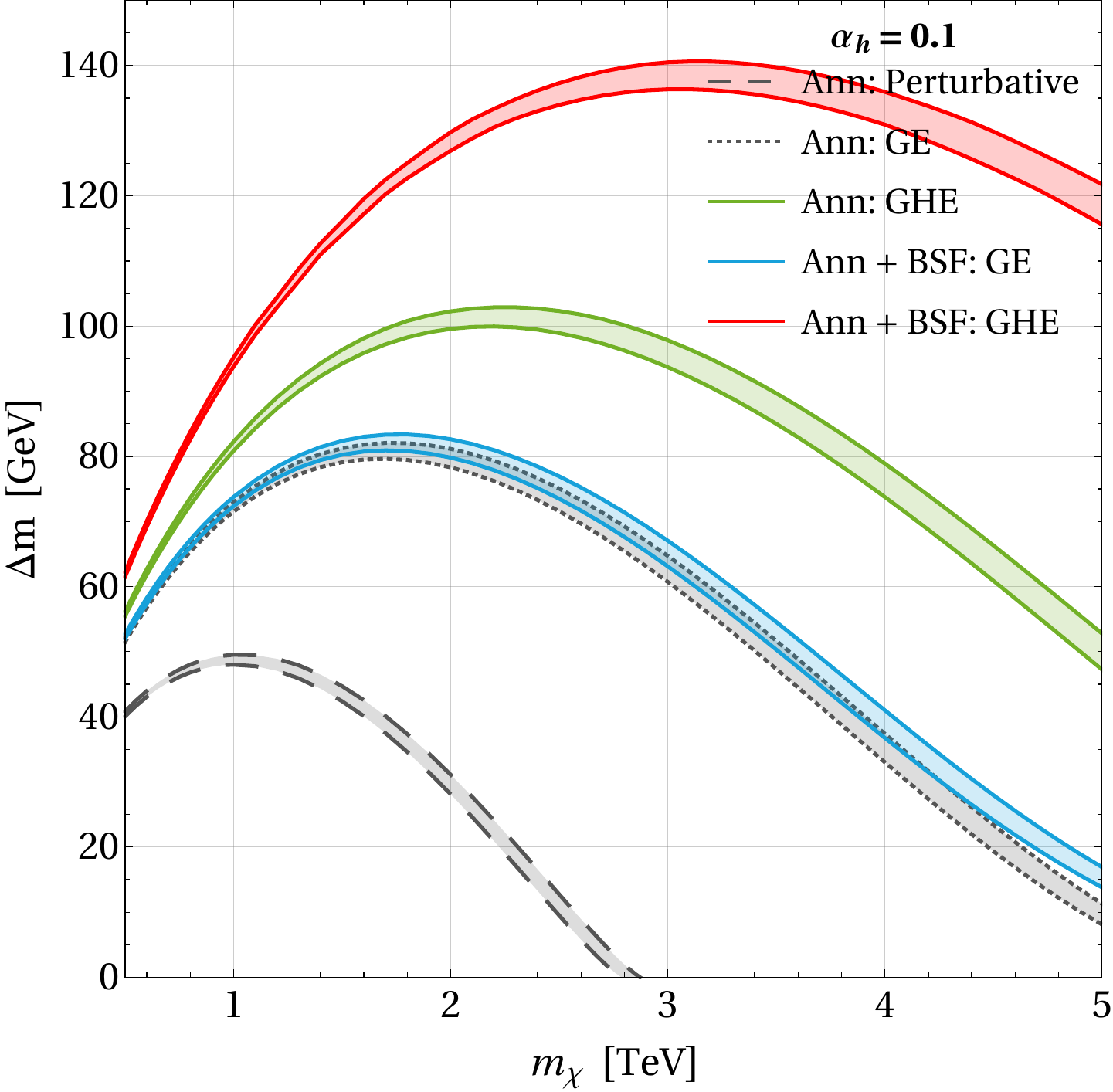}
\includegraphics[height=0.45\textwidth]{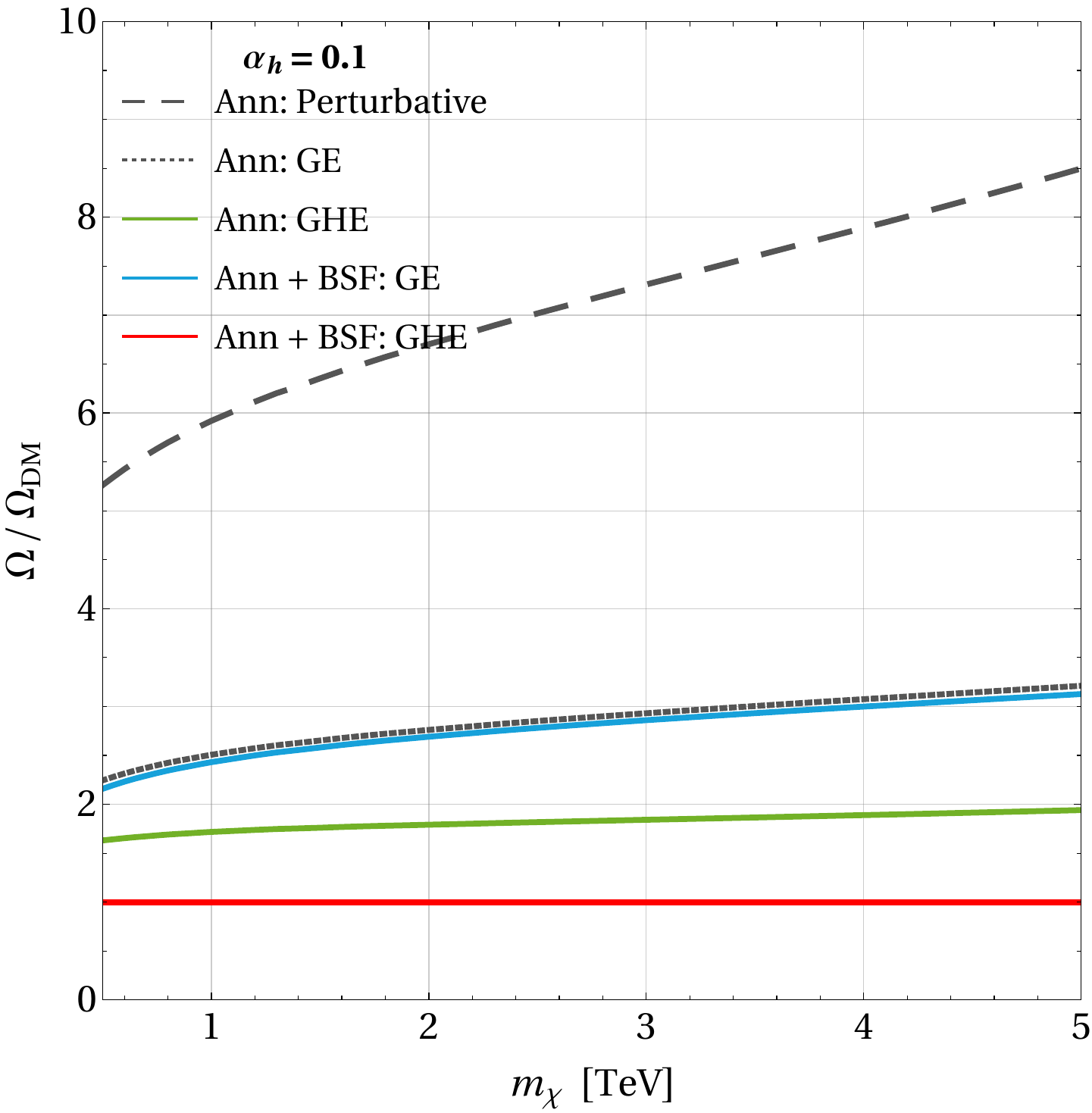}
\\[1ex]
\includegraphics[height=0.45\textwidth]{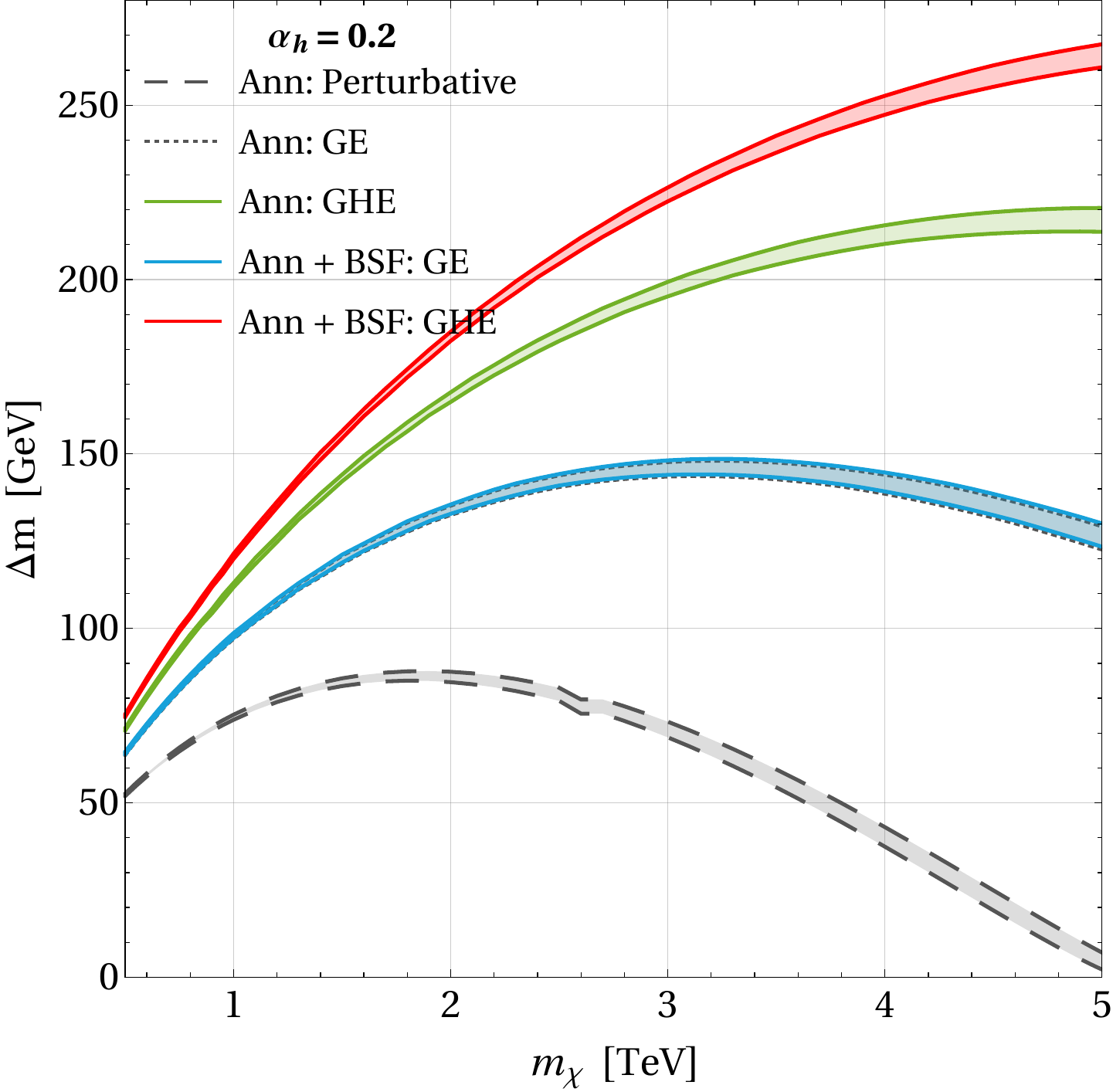}
\includegraphics[height=0.45\textwidth]{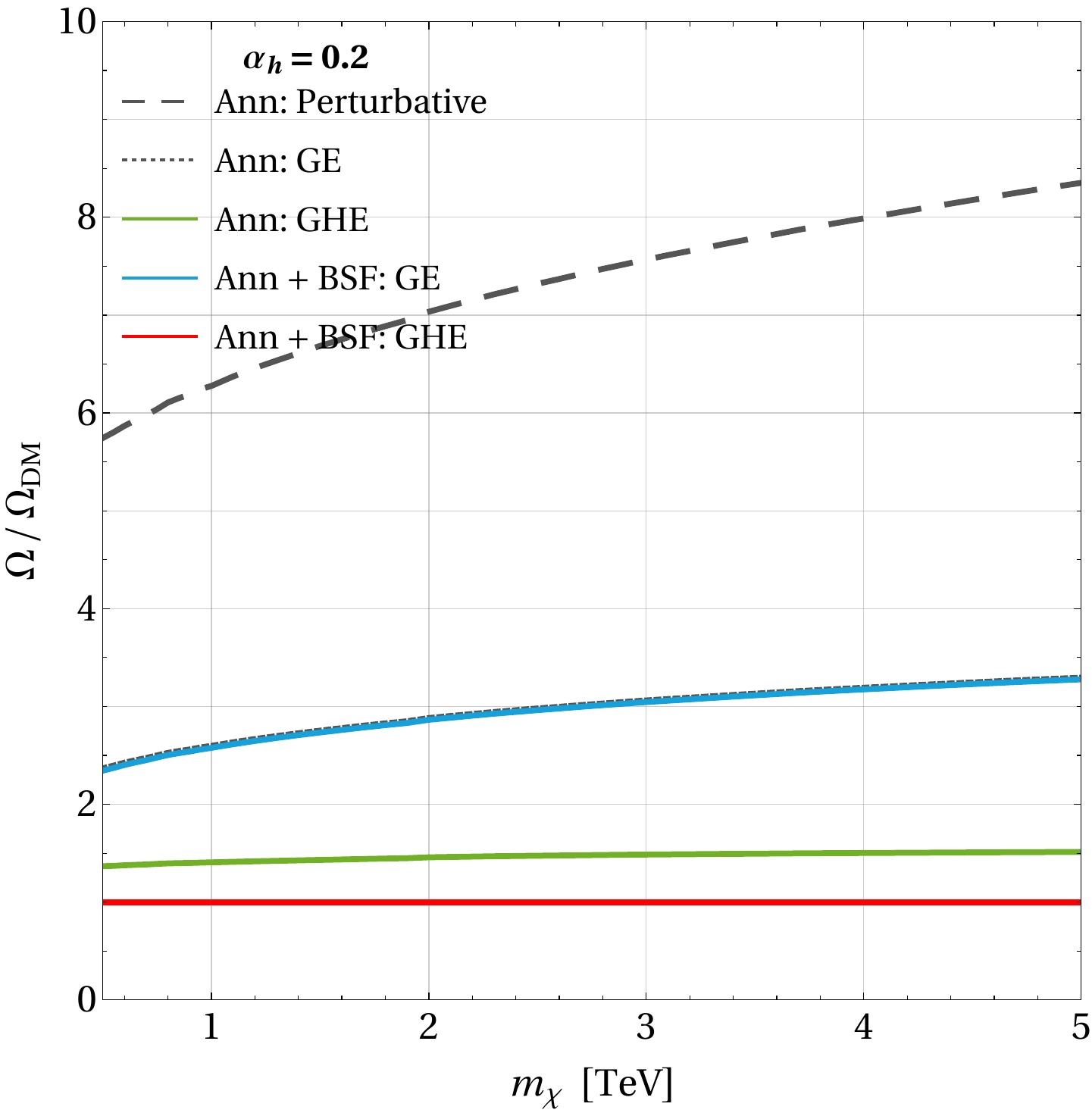}
\caption{As in \cref{fig:RelicResults1}, but for $\ah=0.1$ and $\ah=0.2$.
\label{fig:RelicResults2}
}
\end{figure}

We now turn to the impact of the Higgs-mediated force on the relic density. We perform the calculation as described in \cref{Sec:Model}, using the BSF cross-sections of \cref{Sec:BSF}. In the left panels of \cref{fig:RelicResults1,fig:RelicResults2}, we show, for different values of $\ah$, the NLP-LP mass difference $\Delta m$ versus the DM mass $m_{\chi}$ that give rise to the observed DM abundance. The width of the bands reflects the uncertainty in the measurement of the DM abundance by Planck. We compare different computations that incorporate combinations of the various non-perturbative effects discussed in this work: the Sommerfeld effect on the direct annihilation processes and BSF, due to the gluon and Higgs mediated potentials. We present the impact of these effect on the relic density in the right panels of \cref{fig:RelicResults1,fig:RelicResults2}.

For $\ah=0.02$, the Higgs enhancement of the direct annihilation gives only a moderate correction on the predicted relic density. However, the full calculation that includes annihilation and BSF with gluon and Higgs exchange, changes the predicted mass gap by up to $10~\mathrm{GeV}$, with respect to considering gluon exchange only. Given that the usual tools for computing the relic density include a perturbative calculation only, their estimation can be off by a factor of $2.5-8.5$ depending on the DM mass, even for a small value of $\ah$.

With increasing $\ah$, the effect of BSF with gluon exchange only appears to become less significant. This is because it has remained unaffected by $\ah$, while the direct annihilation into two Higgs bosons has become very rapid. However, including the Higgs-mediated potential gives rise to a very sizeable effect. This is true when considering direct annihilation only, and even more so when including BSF. We point out indicatively that for $\ah=0.1$ and $\ah=0.2$, $\Delta m$ is predicted to be larger by up to 40~GeV and 70~GeV respectively.

The above clearly demonstrate that in coloured co-annihilation scenarios, both the Sommerfeld effect and BSF, as well as both the gluon-exchange and Higgs-exchange potentials must be considered in order to obtain a reliable estimation of the relic density.


\subsection{Caveats \label{sec:Results_Caveats}}

The present work along with refs.~\cite{Harz:2017dlj,Harz:2018csl} aimed to demonstrate the long-range effect of the Higgs boson and the impact of BSF on the DM abundance. In order to not obfuscate these goals, various issues were not addressed, but would need to be properly considered in a comprehensive analysis. We briefly mention the main ones.

Our simplified model does not specify the LP-NLP interactions that are hypothesised to establish equilibrium between the two in the early universe. 
Departure from equilibrium during the NLP freeze-out would imply that the NLP annihilation (direct or via BSF) depletes DM less efficiently than implicitly assumed by the set of \cref{eq:Ytilde,eq:gstareff,eq:Boltzmann,eq:Y_equil,eq:sigma_eff,eq:sigma_XXdagger}. This in turn would shift the contours of \cref{fig:RelicResults1,fig:RelicResults2} to lower values of $\Delta m$ and $\mx$. 

The LP-NLP interactions, along with their mass difference, determine also the NLP decay rate. The latter must be smaller than the decay rate of the NLP-NLP bound states in order for BSF to deplete DM. This potentially implies that BSF cannot affect the DM density if the LP-NLP mass difference exceeds a threshold that depends on the strength of the LP-NLP interactions. Such a condition can be specified within complete models, e.g.~MSSM scenarios, where the couplings of the UV theory determine on the one hand the NLP decay rate, and on the other hand the entirety of the annihilation processes that control the bound-state decay rate. Moreover, these couplings determine the value of $\ah$. 

Including all significant annihilation channels in the analysis of course affects also the total effective annihilation rate --- both directly, and via the bound-state decay rate that determines the efficiency of BSF in depleting DM.  This could have a considerable impact on the results shown in \cref{fig:RelicResults1,fig:RelicResults2}, and therefore on the interpretation of the experimental constraints. 
For very large $\ah$, the radiative capture via Higgs emission that was neglected here, may also become significant, as noted in \cref{sec:Model_BoundStates}.

Ultimately, the precise computation of the DM density requires taking into account the thermal effects that were briefly discussed in \cref{sec:Model_NRpotential}. This could be potentially done using the formalism of ref.~\cite{Binder:2018znk}, which can account for the transition of the BSF processes from in-  to out-of equilibrium.  However, this formalism must be first extended to encompass radiative processes that involve ultrasoft modes, which are currently neglected.

Finally we note that sizeable couplings of the coloured particles to the Higgs endanger the stability of the $SU(3)_c$ symmetric vacuum. The determination of vacuum stability requires a dedicated study for any specific scenario under consideration. However, it has also been argued that this kind of dynamics could imply the emergence of a new phase in the MSSM, where the standard treatment of the vacuum stability does not apply~\cite{Giudice:1998dj}.

\clearpage
\section{Conclusions \label{Sec:Conclusions}}

While our searches for DM coupled to the electroweak gauge interactions have returned null results until now, it is plausible that the Higgs constitutes the portal to the dark sector. The discovery of the Higgs boson and the measurement of its properties in collider experiments urge the comprehensive assessment of its implications for DM.

If TeV-scale DM or its co-annihilating partners couple to the 125~GeV Higgs, then this interaction may have a sizeable long-range effect. In recent work, we showed that the DM relic abundance in coloured co-annihilation scenarios is affected significantly by the enhancement of the direct annihilation processes due to the Higgs-mediated force~\cite{Harz:2017dlj}, and by the formation and decay of bound states due to gluon exchange~\cite{Harz:2018csl}. Here we considered the effect of the Higgs-generated potential on the formation of bound states, and demonstrated the impact on the DM abundance. With respect to available tools that include perturbative calculations only, we have found that the predicted DM density may differ by up to almost one order of magnitude. While we focused on a particular class of models, we expect that the long-range effect of the Higgs boson is important in a variety of scenarios where the Higgs constitutes the portal to the dark sector.

Altogether, in the set-up we considered, we have shown that the Higgs-mediated force affects the DM density in a variety of ways that exhibit some salient features. In summary:
\begin{itemize}
\itemsep 0pt	

\item
The attractive interaction in the scattering states due to the Higgs exchange enhances both the direct annihilation cross-sections, as well as the capture into bound states.

\item
The Higgs-mediated potential increases the binding energy of the colour-singlet bound states. This implies that the capture rate becomes maximal at larger velocities, and consequently at higher temperatures in the early universe. Moreover, the bound-state dissociation becomes insignificant starting at higher temperatures. As a result, BSF begins to deplete DM efficiently at earlier times, when the DM density is higher, and therefore has a larger impact on the relic abundance.
	
\item	
The Higgs counteracts the gluon-mediated repulsion in the octet states. For large enough mass of the interacting particles and coupling to the Higgs, colour-octet bound states exist. Their formation and decay contributes to the DM depletion rate in the early universe, even if only modestly due to their small binding energy. 

\item
The Higgs exchange increases the momentum transfer in the bound states and at the gluon emission vertex in the capture process, hence driving the strong coupling to smaller values. This somewhat quells the strength of the BSF processes at large $\ah$. 

\item
The interference of the (gluon-generated) Coulomb potential and the (Higgs-generated) Yukawa potential influences the long-range effect of the latter, both in the scattering and the bound states.

\end{itemize}

The enhancement of the effective DM annihilation rate due to the Higgs-mediated force, implies that the LP-NLP mass difference and/or the DM mass must be larger than previously predicted, in order for DM to attain the observed density via freeze-out. This is particularly important for collider searches, since a larger mass gap yields harder jets that are easier to probe. Similarly, the increased values of the predicted DM mass --- which can lie even beyond the range considered here --- strengthen the motivation for indirect searches in the multi-TeV regime. 
However, the accurate interpretation of the experimental constraints necessitates that the effects discussed here are considered within complete models, where various relevant technicalities can be treated properly, as discussed in \cref{sec:Results_Caveats}.

\section*{Acknowledgements}

We thank Tobias Binder, Simone Biondini, Sacha Davidson, Adam Falkowski, Mikko Laine, Pasquale Serpico, Pietro Slavich and Michel Tytgat for useful discussions. J.H.~was supported by the Labex ILP (reference ANR-10-LABX-63) part of the Idex SUPER, and received financial state aid managed by the Agence Nationale de la Recherche (ANR), as part of the programme Investissements d'avenir under the reference ANR-11-IDEX-0004-02. J.H.~was further supported by the DFG Emmy Noether Grant No.~HA 8555/1-1. K.P.~was supported by the ANR ACHN 2015 grant (``TheIntricateDark" project), and by the NWO Vidi grant ``Self-interacting asymmetric dark matter".

\bibliography{Bibliography.bib}

\end{document}